\documentclass[twocolumn]{aastex631}

\usepackage{amsmath}
\usepackage{url}
\usepackage{graphicx}
\usepackage[caption=false]{subfig}
\usepackage{comment}

\shorttitle{AASTeX v6.3.1 Sample article}
\shortauthors{MAGIC et al.}
\graphicspath{{./}{figures/}}

\begin{document}

\title{Time-Dependent Modeling of the Sub-Hour Spectral Evolution During the 2013 Outburst of Mrk~421}

\author{K.~Abe}
\affiliation{Japanese MAGIC Group: Department of Physics, Tokai University, Hiratsuka, 259-1292 Kanagawa, Japan}
\author[0000-0001-7250-3596]{S.~Abe}
\affiliation{Japanese MAGIC Group: Institute for Cosmic Ray Research (ICRR), The University of Tokyo, Kashiwa, 277-8582 Chiba, Japan}
\author[0000-0001-8215-4377]{J.~Abhir}
\affiliation{ETH Z\"urich, CH-8093 Z\"urich, Switzerland}
\author{A.~Abhishek}
\affiliation{Universit\`a di Siena and INFN Pisa, I-53100 Siena, Italy}
\author[0000-0001-8816-4920]{A.~Aguasca-Cabot}
\affiliation{Universitat de Barcelona, ICCUB, IEEC-UB, E-08028 Barcelona, Spain}
\author[0000-0002-3777-6182]{I.~Agudo}
\affiliation{Instituto de Astrof\'isica de Andaluc\'ia-CSIC, Glorieta de la Astronom\'ia s/n, 18008, Granada, Spain}
\author{T.~Aniello}
\affiliation{National Institute for Astrophysics (INAF), I-00136 Rome, Italy}
\author[0000-0002-5613-7693]{S.~Ansoldi}
\affiliation{Universit\`a di Udine and INFN Trieste, I-33100 Udine, Italy}\affiliation{also at International Center for Relativistic Astrophysics (ICRA), Rome, Italy}
\author[0000-0002-5037-9034]{L.~A.~Antonelli}
\affiliation{National Institute for Astrophysics (INAF), I-00136 Rome, Italy}
\author[0000-0001-9076-9582]{A.~Arbet Engels$^\star$}
\affiliation{Max-Planck-Institut f\"ur Physik, D-85748 Garching, Germany}
% ...
\collaboration{300}{(MAGIC Collaboration)}
\author[0000-0001-6640-0179]{M.~Petropoulou$^\star$}
\affiliation{Department of Physics, National and Kapodistrian University of Athens, University Campus Zografos, GR 15784, Athens, Greece}\affiliation{Institute of Accelerating Systems \& Applications, University Campus Zografos, Athens, Greece}
\author[0000-0002-8889-2167]{M.~Polkas$^\star$}
\affiliation{Donostia International Physics Center, Paseo Manuel de Lardizabal 4, E-20118 Donostia-San Sebasti\'an, Spain}
\author[0000-0001-5217-4801]{A.~Mastichiadis}
\affiliation{Department of Physics, National and Kapodistrian University of Athens, University Campus Zografos, GR 15784, Athens, Greece}

\correspondingauthor{A.~Arbet-Engels, M.~Polkas, M.~Petropoulou, D.~Paneque}
\email{contact.magic@mpp.mpg.de}



\begin{abstract}

In April 2013, the TeV blazar Markarian~421 underwent one of its most powerful emission outbursts recorded to date. An extensive multi-instrument campaign featuring MAGIC, VERITAS, and \textit{NuSTAR} provided comprehensive very-high-energy (VHE; $E > 100$\,GeV) and X-ray coverage over nine consecutive days. The VHE flux peaked at approximately 15 times that of the Crab Nebula, with rapid variability detected on timescales down to 15 minutes in both X-ray and VHE bands. This rich dataset, characterized by its dense temporal coverage and high photon statistics, offers an unparalleled opportunity to probe the broadband emission dynamics in blazars. In this work, we perform a detailed spectral analysis of the X-ray and VHE emissions on sub-hour timescales throughout the flare. We identify several clockwise spectral hysteresis loops in the X-rays, revealing a spectral evolution more complex than a simple harder-when-brighter trend. The VHE spectrum extends beyond 10\,TeV, and its temporal evolution closely mirrors the behavior in the X-rays. Crucially, we report the first evidence of VHE spectral hysteresis occurring simultaneously with the X-ray loops. To interpret these findings, we apply a time-dependent leptonic model to 240 broadband spectral energy distributions (SEDs) binned on a 15-minute scale, allowing us to self-consistently track the particle distribution's history. Our modeling shows that the majority of the sub-hour flux and spectral variations are driven by changes in the luminosity and slope of the injected electron distribution. The required variations in the electron slope are difficult to reconcile with magnetic reconnection but are consistent with a shock-acceleration scenario where the shock compression ratio evolves by a factor of $\sim2$. The model also points to a relatively stable magnetic field and emitting region size, favoring a scenario where the emission originates from a stationary feature in the jet, such as a recollimation shock. However, this scenario requires a jet Lorentz factor that significantly exceeds values from VLBI measurements to account for the high minimum electron energy implied by the lack of variability in the optical band.

\end{abstract}



\section{Introduction} \label{sec:intro}
Markarian 421 (Mrk$\,$421; RA=11$^\text{h}$4$^{\prime}$27.31$^{\prime\prime}$, Dec=38$^{\circ}$12$^{\prime}$31.8$^{\prime\prime}$, J2000, $z=0.031$) is the closest and among the brightest TeV blazars, which are jetted active galactic nuclei (AGNs) that have their plasma jet oriented at a small angle relative to the line of sight. Blazars are identified by a spectral energy distribution (SED) dominated by nonthermal radiation from the jet, and the SED displays two broad components. The low-energy component ranges from radio wavelengths and can reach up to the X-ray band. The high-energy component is located at gamma-ray energies. The low-energy SED component of Mrk~421 peaks above $10^{15}$\,Hz, thus classifying it as a high synchrotron peaked blazar (HSP) following the nomenclature of \citet{2010ApJ...716...30A}. Thanks to its brightness, the flux evolution of Mrk~421 can be precisely characterized from radio wavelengths to very high energies (VHE; $E>100$,GeV) by current instruments down to (at least) daily timescale independently of the flux activity \citep[for recent works, see e.g.][]{2016ApJ...819..156B, 2021A&A...655A..89M, 2021MNRAS.504.1427A}. Because of that, Mrk\,421 is an ideal target to probe blazar radiation mechanisms and investigate particle dynamics within extragalactic jets.\par 

In April 2013, Mrk~421 underwent one of the brightest outbursts recorded to date \citep{paper1}. The VHE flux reached $\sim15$ times that of the Crab Nebula, a factor $\gtrsim30$ above its average activity \citep{2014APh....54....1A}. The flare lasted more than a week, and was covered over nine consecutive days by several instruments from radio bands to VHE. Notably, an extensive coverage in the X-ray and VHE regimes could be obtained thanks to the Nuclear Spectroscopic Telescope Array \citep[\textit{NuSTAR};][]{2013ApJ...770..103H}, and two Imaging Atmospheric Cherenkov Telescopes (IACTs) located in the northern hemisphere, the Major Atmospheric Gamma Imaging Cherenkov telescope \citep[MAGIC][]{2008ApJ...674.1037A}, and the Very Energetic Radiation Imaging Telescope Array System \citep[VERITAS;][]{2008AIPC.1085..657H}. A total of 43\,hours of simultaneous X-ray/VHE observation was gathered, making it one of the most extensive characterization of an HSP flare at those energies \citep[see][for details]{paper1}. In the context of theoretical modeling of HSP emission, the simultaneous X-ray/VHE data sample is of paramount importance since within a leptonic emission scenario the same electron population radiating in the X-ray band via synchrotron radiation is expected to emit in the VHE band via inverse-Compton scattering \citep[see for example][]{1992ApJ...397L...5M, 2011ApJ...736..131A}. The obtained data sample thus provides an ideal opportunity to test and constrain theoretical models.\par

\subsection{Summary of the April 2013 flare and the observed broadband evolution characteristics}

Initial investigations focused on the X-ray activity were reported in \citet{2015ApJ...811..143P, 2016ApJ...831..102K}. The radio-to-VHE flux variability and correlation patterns were subsequently studied in details in \citep[][hereafter Paper~1]{paper1}. We summarize below the main results before describing the scope of this work, which extends the study performed in Paper~1.\par 

Owing to the exceptionally high emission state, the fluxes throughout the April 2013 flare could be determined in 15\,min time bins in both the X-ray and VHE bands. Extreme variability was measured and the light curves displayed an overall variation amplitude of more than one order of magnitude. In particular, significant variability down to sub-hour timescale was detected. Interestingly, for several of the days, the X-ray and VHE light curves could be decomposed into a slow (multi-hour) and fast (sub-hour) component, hinting towards several emitting regions contributing to those bands. \par 

The X-ray and VHE correlation was quantified after splitting the fluxes into multiple sub-energy bands: 3-7\,keV, 7-30\,keV, 30-80\,keV for the X-rays, and 0.2-0.4\,TeV, 0.4-0.8\,TeV, $>0.8$\,TeV for the VHE. For each combination, the VHE emission correlated strongly with the X-ray one without any time-lag. However, the slope of the correlation was dependent on the exact energy bands considered. A linear relationship (slope $\approx1$ in a log-log flux plane) was found between the $>0.8$\,TeV and 3-7\,keV bands, while a sub-linear trend (slope $\approx0.4$ in a log-log flux plane) was detected between 0.2-0.4\,TeV and 30-80\,keV bands. This complexity in the correlation patterns is in line with leptonic models \citep{2005A&A...433..479K}.    

In the optical/UV and radio bands the variability is much more suppressed compared to the VHE and X-ray regimes. The fluxes varied by at most 30\%, and the corresponding flux level remained close to those measured in January-March 2013, when Mrk~421 was in very low VHE and X-ray activity \citep{2016ApJ...819..156B}. As for the MeV-GeV band, the flux variations were only by factors of a few. No correlation between the MeV-GeV band and the keV and TeV bands was found. 

\subsection{Scope of this work}

As follow-up study of Paper~1, we first present an investigation of the X-ray and VHE spectral behavior. While Paper~1 focused on the integrated flux variability, the sensitivity of \textit{NuSTAR} and MAGIC, coupled with the large photon statistics due to the high flux, offers the possibility to characterize spectral changes on sub-hour timescale and probe a higher degree of complexity than the common ``harder-when-brighter'' trend. An example of such complex spectral evolution is the so-called hysteresis pattern, that have been already measured in the X-rays, but not at VHE, likely due to a lack of statistics \citep{2004A&A...424..841R, 2017ApJ...834....2A}. Hysteresis patterns provide important clues on the dynamics of the underlying emitting particles since they relate to cooling losses or acceleration processes \citep{Kirk98}.\par 

In a second step, we build a time-dependent radiation model to describe the broadband evolution during the flare. Our work focuses primarily on the reproduction of the VHE and X-ray (flux and spectral) behavior, where the most pronounced variability happened. We applied and compared the model to multi-band light curves and broadband spectra obtained over 15-minute bins. Unlike most of the previous published work that uses \textit{time-independent} models \citep[i.e., where particles are \textit{not} evolved with time taking into account self-consistently radiative losses and/or escape, see e.g.][]{2017ApJ...834....2A, 2020A&A...633A.162H, 2020A&A...637A..86M, 2021A&A...647A.163M}, our aim is to make an essential step forward to test radiation and acceleration models. We take advantage of the densely sampled VHE and X-ray observations from MAGIC, VERITAS and \textit{NuSTAR} to constrain the evolution of the particle distribution up to the highest energies in the jet, which can then be compared with predictions from particle-in-cell (PIC) simulations.\par

The paper is structured as follows. In Sect.~\ref{sec:analysis} we summarize the observations together with the data processing methods. The results of the spectral analysis are presented in Sect.~\ref{sec:spectral_analysis}. The time-dependent modeling is shown in Sect.~\ref{sec:modelling}. The discussion of our findings is given in Sect.~\ref{sec:discussion}, while the final concluding remarks are drawn in Sec.~\ref{sec:summary}.

\begin{figure*}[ht!]
\gridline{\fig{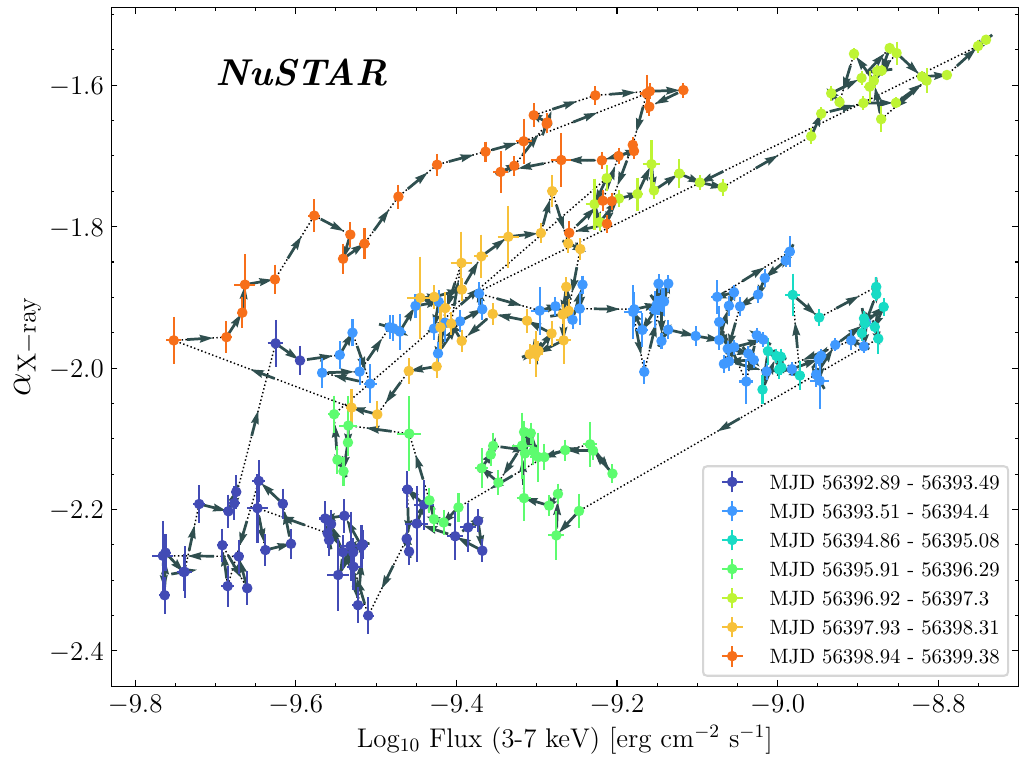}{0.7\textwidth}{}}
\caption{Spectral parameter $\alpha$ versus the 3-7\,keV flux as measured by \textit{NuSTAR} between MJD~56393 and MJD~56399. $\alpha_{\rm X-ray}$ is derived from a log-parabola fit with the curvature parameter $\beta_{\rm X-ray}$ fixed to 0.38 (see text in Sect.~\ref{sec:analysis} for more details). The data are binned in 15-min time intervals. Each day is plotted with a different color and black arrows indicate the direction of time.
\label{alpha_vs_flux_nustar_all_bins}}
\end{figure*}

\section{Observations and data processing} \label{sec:analysis}

The instruments involved in this work are the ones described in Paper~1, and cover the entire electromagnetic spectrum from radio wavelengths to VHE. As in Paper~1, we consider the time period from MJD~56393 (11 April 2013) until MJD~56401 (19 April 2013). In Appendix~\ref{sec:appendix_analysis} the data processing is summarized. The methods are similar to those of Paper~1, with the only difference that we reprocessed \textit{NuSTAR} and \textit{Fermi}-LAT data with updated instrument response functions. In the VHE band, we extended the MAGIC analysis to extract spectral parameters and SED points. To extend and maximize the temporal coverage at VHE, we also make use of the VERITAS data published in \cite{2017ICRC...35..641B}, which was also used in Paper~1. The multiwavelength light curves can be found in Fig.~\ref{light_curve_flare} in Appendix~\ref{sec:appendix_analysis}.\par

\section{X-ray \& VHE spectral behaviors} \label{sec:spectral_analysis}

We investigate the X-ray spectral evolution by fitting the \textit{NuSTAR} spectra with a log-parabola model (pivot energy fixed at 1\,keV) in the 3-79\,keV band in each of the 15-min time intervals (286 in total) of the light curve shown in Figure~\ref{light_curve_flare}. Compared to a simple power-law model, about 60\% of the bins show a $>3\sigma$ preference for a log-parabola, and therefore the latter model will be used throughout this work to characterize the spectral trends in the X-rays. After this first series of fits, we find that the curvature parameter $\beta_{\rm X-ray}$ does not exhibit a strong dependence on the flux. The relation between $\beta_{\rm X-ray}$ and the 3-7\,keV flux is shown in Figure~\ref{betavsflux} in Appendix~\ref{sec:VHE_spectral_analysis_app}. We thus perform a second series of fits in \texttt{Xspec} with $\beta_{\rm X-ray}$ fixed to $0.38$, which is the average of all the 15-min intervals. This procedure cancels any potential correlation between the $\alpha_{\rm X-ray}$ and $\beta_{\rm X-ray}$ spectral parameters and the parameter $\alpha_{\rm X-ray}$ can be used to directly quantify the hardness evolution. We stress that in the ``$\beta$-fixed'' model, the vast majority of the fits (93\%) is compatible with the data within $2\sigma$ based on a $\chi^2$ test. For only 2 fits (out of 286) the ``$\beta$-fixed'' model is incompatible with the data with a significance above $3\sigma$. Hence, our conclusions on the spectral evolution are not significantly biased by fixing $\beta_{\rm X-ray}$.\par 

In Figure~\ref{alpha_vs_flux_nustar_all_bins}, $\alpha_{\rm X-ray}$ as a function of the 3-7\,keV flux is displayed between MJD~56393 until MJD~56399, which is the period where the source was in its highest emission state during the flare. Each day is plotted in a different color, and black arrows show the direction of the time evolution. Significant spectral variability occurs on $\sim 15$\,min scales, and an overall harder-when-brighter behavior is detected at a significance level above $10\sigma$ (Pearson's coefficient of $0.60\pm0.04$), in line with previous \textit{NuSTAR} observations of Mrk~421 \citep{2016ApJ...819..156B, 2021A&A...655A..89M}, and other blazars \citep[e.g.][]{1998ApJ...492L..17P, 2004ApJ...601..151K}. A more complex behavior is however visible, beyond the simple harder-when-brighter trend. The source follows different paths in the parameter space depending on the day, and at different occasions spectral hysteresis loops are visible.\par

\begin{figure*}[ht]
\gridline{\fig{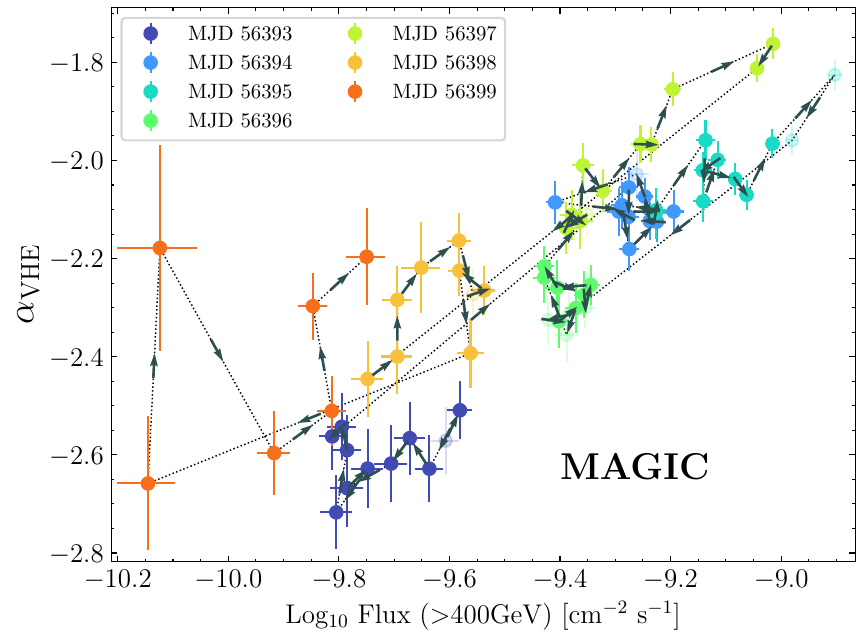}{0.49\textwidth}{(a)}
\fig{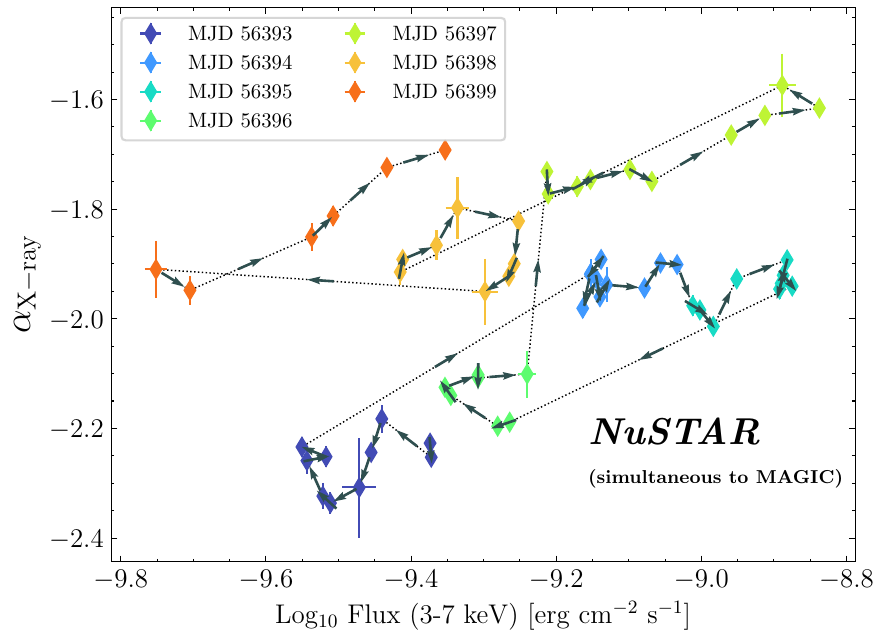}{0.498\textwidth}{(b)}
          }
\caption{Spectral parameter $\alpha$ versus flux for (a) MAGIC and (b) \textit{NuSTAR} between MJD~56393 and MJD~56399. The \textit{NuSTAR} fluxes are evaluated in the 3-7\,keV bands, while for MAGIC they are computed above 400\,GeV. In each panel the data are binned over 30\,min, strictly simultaneous intervals. Each day is plotted with a different color and black arrows indicate the direction of time. The MAGIC points plotted with transparent markers are temporal bins that are not accompanied by a strictly simultaneous \textit{NuSTAR} measurement.
\label{alphavsflux_magic}}
\end{figure*}

A first hysteresis loop in clockwise direction is observed from MJD~56393 to MJD~56396. The flux is initially in a decaying phase on MJD~56393 (violet points) until a local minimum is reached. The emission is rising again and shows an abrupt hardening, from $\alpha_{\rm X-ray}\approx-2.2$ to $\approx-2.0$ between MJD~56393 and MJD~56394. On MJD~56396 (green markers), the emission is decreasing and softening, thus drawing an hysteresis pattern.\par

Two additional hysteresis loops in clockwise direction are visible over intraday timescales on MJD~56398 (light orange markers) and MJD~56399 (orange markers). It is interesting to note that the emission on MJD~56399 is significantly harder than the rest of the days despite covering a comparable flux range.\par 

For the last two days discussed in this work, MJD~56400 and MJD~56401 (see Fig.~\ref{light_curve_flare}), the flux is significantly lower and $\alpha_{\rm X-ray}$ lies between  $\approx-2.2$ and $\approx-2.4$. For visibility purposes, these two days are not included in Fig.~\ref{alpha_vs_flux_nustar_all_bins} in order to focus on the MJD~56393 - MJD~56399 period where most of the complex patterns are measured. They are nevertheless presented (together with the rest of the flare) in Fig.~\ref{alpha_vs_flux_nustar_alldays} in Appendix~\ref{sec:VHE_spectral_analysis_app}.\par

We performed a spectral analysis in the VHE band using the MAGIC observations. Instead of using 15-min time intervals, as done for \textit{NuSTAR}, we fit the MAGIC spectra in 30-min intervals to increase the photon statistics, which is necessary given the sensitivity of MAGIC. The vast majority of the MAGIC spectra are well described with the log-parabola model given in Eq.~\ref{eq:logparabola_MAGIC} and the fit p-values are all above $10^{-2}$. As in the X-ray band, we find that the curvature parameter $\beta_{\rm VHE}$ shows no strong dependence on the flux. The lack of correlation between $\beta_{\rm VHE}$ and the flux above 400\,GeV is shown in Figure~\ref{betavsflux} in Appendix~\ref{sec:VHE_spectral_analysis_app}. Given this lack of correlation, and for the same arguments mentioned previously in the case of the \textit{NuSTAR} spectral analysis, a second series of fit were performed with $\beta_{\rm VHE}$ fixed to $0.40$ (the average from all the individual time intervals). We also note that based on a $\chi^2$ test the ``$\beta$-fixed" model is consistent with the data within $3\sigma$ for all fits (73 in total), and only two fits are inconsistent with the data above $2\sigma$. No significant bias is thus introduced in our interpretation of the spectral evolution when using a ``$\beta$-fixed'' model. Finally, we ran all fits above a common minimum energy of 250\,GeV such that $\alpha_{\rm VHE}$ can be fairly compared between all time intervals. The energy threshold of MAGIC increases with the observing zenith angle, and using a threshold at 250\,GeV instead of the minimum achievable by MAGIC ($\sim$75\,GeV) allows us to include time bins with an observing zenith angle up to $\approx52^\circ$ while preserving sufficient statistics. Only a minority of the bins were taken at zenith angles above $52^\circ$, and they are removed from the spectral analysis.\par 

In Figure~\ref{alphavsflux_magic} (a), we show $\alpha_{\rm VHE}$ as function of the flux above 400\,GeV. For comparison purposes, in Figure~\ref{alphavsflux_magic} (b) the spectral index $\alpha_{\rm X-ray}$ is plotted as function of the 3-7\,keV flux based on 30\,min time intervals strictly simultaneous to MAGIC. Overall, the MAGIC spectra reveal a significant harder-when-brighter trend at VHE. This is in agreement with previous work on Mrk~421 \cite[e.g.][]{2021MNRAS.504.1427A, 2021A&A...655A..89M}. The correlation significance is $\approx 11\sigma$ and the Pearson coefficient is $0.86\pm0.03$. The significance is mostly dominated by the day-by-day variations, however, we note a significant ($\gtrsim 5\sigma$) intranight harder-when-brighter behavior measured on MJD~56397.\par 

Similar to what was observed in the hard X-rays, the VHE spectra populate different regions in the parameter space depending on time. Figure~\ref{alphavsflux_magic} unveils an evident similarity between the behavior observed in MAGIC and \textit{NuSTAR}, and $\alpha_{\rm VHE}$ draws similar pattern in the parameter space as $\alpha_{\rm X-ray}$ throughout the entire flare. On MJD~56398 (orange data points), the evolution of $\alpha_{\rm VHE}$ strongly indicates an intraday hysteresis loop in clockwise direction simultaneous to the one observed in \textit{NuSTAR}.\par

\subsection{Daily VHE spectral analysis} \label{sec:daily_averaged_magic_spectra}

In this section, we investigate the nightly averaged MAGIC spectra, which will be used as input for the time-dependent modeling of the flare to constrain parameters such as the Doppler factor or the magnetic field within the source. In a leptonic scenario, the highest energies (above a few TeVs) are affected by the Klein-Nishina suppression of the inverse-Compton cross-section. Averaging VHE spectra on each night provides a precise characterization of the multi-TeV spectrum, allowing us to probe the presence (or not) of a cut-off.\par

For most of the days, the nightly MAGIC spectra show a $>3\sigma$ statistical preference for a log-parabola model with an exponential cut-off (dubbed as ELP) instead of a pure log-parabola model (as applied to the 30-min intervals; see previous section). We parametrize the ELP as following:
\begin{equation}
\label{eq:elp_MAGIC}
        \frac{dN}{dE} = f_0 \left(\frac{E}{E_0}\right)^{\alpha-\beta \log_{10}{\left(\frac{E}{E_0}\right)}} e^{-\frac{E}{E_{\rm cut-off}}}
\end{equation}
where $E_0$ is the normalization energy fixed to 500\,GeV. The best-fit parameters are listed in Table~\ref{tab:spectral_parameters_56395_56397}, together with the $\chi^2/{\rm d.o.f.}$ and the statistical preference for an ELP model ($\sigma_{ELP}$). For days MJD~56398 to MJD~56401, the preference for an ELP model is not significant ($<3\sigma$), thus we only quote the parameters of the best-fit log-parabola function. The measured spectra extend to $\sim$10\,TeV, at which the EBL absorption becomes significant for Mrk~421 ($\tau_{EBL}\sim 1$). We thus tried different available EBL models \citep{2012MNRAS.422.3189G, 2008A&A...487..837F,2018A&A...614C...1F} to search for possible systematic effects, but all results were compatible with fits using the \citet{2011MNRAS.410.2556D} model.

\begin{figure}[ht!]
\plotone{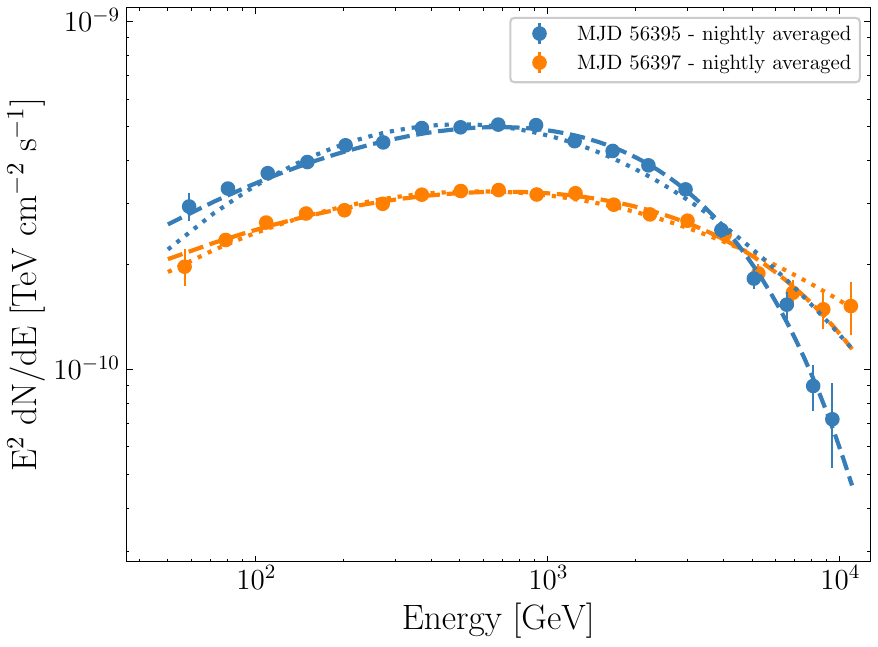}
\caption{Nightly MAGIC spectra from MJD~56395 and MJD~56397, which show the highest VHE flux of the April 2013 flare. The dotted and dashed lines are the best-fit function using a log-parabola and log-parabola with cut-off model, respectively. For both days, the log-parabola model with a cut-off is preferred at a significance level above $3\sigma$. The SED points and spectral model curves are corrected for the EBL absorption using the model of \citet{2011MNRAS.410.2556D}.
\label{nightly_vhe_spectra}}
\end{figure}

\begin{table*}[h]
\centering
\resizebox{0.9\textwidth}{!}{\begin{tabular}{ccccccc}
\hline \hline
   Date  &  $f_0$ & $\alpha_{\rm VHE}$ & $\beta_{\rm VHE}$ & $E_{\rm cut-off}$  & $\chi^2/{\rm d.o.f.}$ & $\sigma_{ELP}$\\
   MJD &  TeV cm$^{-2}$ s$^{-1}$ &  &  & TeV  &  &\\
\hline
 56393 & $(0.69\pm0.05)\times10^{-9}$ & $-2.36\pm0.09$ & $0.31\pm0.08$ & $2.8\pm1.1$ &17.0/14 & 3.0\\ 
56394 & $(1.49\pm0.04)\times10^{-9}$ & $-1.92\pm0.04$ & $0.03\pm0.04$ & $2.9\pm0.4$ & 15.0/16 & 8.6 \\ 
56395 & $(2.21\pm0.03)\times10^{-9}$ & $-1.83\pm0.02$ & $0.14\pm0.03$ & $4.7\pm0.7$ & 7.8/16 & 7.7 \\ 
56396 & $(1.27\pm0.05)\times10^{-9}$ & $-2.06\pm0.05$ & $0.14\pm0.05$ & $2.7\pm0.5$ & 14.5/15 & 6.4\\ 
56397 & $(1.34\pm0.01)\times10^{-9}$ & $-1.91\pm0.02$ & $0.12\pm0.03$ & $13\pm4$ & 22.3/16 & 3.2\\ 
56398 & $(0.65\pm0.01)\times10^{-9}$ & $-2.30\pm0.01$ & $0.36\pm0.02$ & $----$ & 13.7/16 & 2.1 \\ 
56399 & $(0.38\pm0.01)\times10^{-9}$ & $-2.34\pm0.03$ & $0.40\pm0.05$ & $----$ & 23.9/15 & 0.0 \\
56400 & $(0.30\pm0.01)\times10^{-9}$ & $-2.58\pm0.04$ & $0.69\pm0.10$ & $----$ & 19.5/15 & 1.8\\ 
56401 & $(0.15\pm0.01)\times10^{-9}$ & $-2.96\pm0.12$ & $0.96\pm0.39$ & $----$ & 10.5/12 & 0.0 \\

\hline
\end{tabular}}
\label{tab:spectral_parameters_56395_56397}\caption{Spectral parameters obtained by fitting a the nightly averaged MAGIC spectra. An ELP model (see Eq.~\ref{eq:elp_MAGIC}) is fitted for days MJD~56393 to MJD~56397. For days MJD~56398 to MJD~56401, a simple log-parabola model is adopted since the statistical preference for an ELP is $<3\sigma$. The statistical preference for a ELP is given in the last column.}
\end{table*}

Days MJD~56395 and MJD~56397 are characterized by the hardest spectra and the highest cut-off energies, located at $\sim5$\,TeV and $\sim13$\,TeV for MJD~56395 and MJD~56397, respectively. The spectra from these two nights are overlaid in Fig.~\ref{nightly_vhe_spectra}, where we compare a log-parabola fit (dotted lines) with an ELP model (dashed lines). For MJD~56397, we set the lower limit on $E_{cut-off}$ at the 95\%  confidence level (C.L.) by performing a $\chi^2$ scan around the best-fit $E_{cut-off}$. During the scan, all parameters are left free except $E_{cut-off}$. The 95\% C.L. limit is given by the $E_{cut-off}$ value at which the $\chi^2$ increases by 2.71 with respect to the minimum ($\chi^2_{min}$). Following this procedure, the 95\% C.L. lower limit on $E_{cut-off}$ is 8.3\,TeV. \par

\section{Time-dependent modeling of the flare} \label{sec:modelling}

Our time-dependent model assumes a purely leptonic scenario for the non-thermal emission. The optical-to-X-ray fluxes originate from synchrotron radiation by relativistic electrons/positrons within the jet (hereafter both particle types are denoted as electrons for simplicity), while the gamma-ray flux is produced via electron inverse-Compton scattering off the synchrotron photons emitted by the same population of electrons. This model is commonly referred to as the synchrotron self-Compton model \citep[SSC; see for example][]{1992ApJ...397L...5M, 1998ApJ...509..608T, 2004ApJ...601..151K}, and is usually able to reproduce well the SED of Mrk~421 \citep[see e.g.][]{2011ApJ...736..131A, 2017ApJ...834....2A, 2021A&A...655A..89M}. The SSC scenario is supported by the tight correlation between the X-ray and VHE fluxes (see results in Paper~1), and the similar simultaneous spectral evolution patterns in those bands (see Sect.~\ref{sec:analysis}). This strongly indicates a single underlying particle population responsible for the two SED components (here, the electrons), differently from hadronic models \citep{1993A&A...269...67M, 2001APh....15..121M} that additionally invoke protons to explain the gamma-ray emission.\par  

The model consists of two spatially separated, non-interacting emitting regions inside the jet. The first zone, more compact and energetic, dominates the emission in the X-ray and VHE bands. The second zone occupies a larger volume and is populated with less energetic electrons such that it is dominantly responsible for radio-to-optical and MeV/GeV emission. At radio frequencies, the fluxes are treated as upper limits given that they likely receive a substantial contribution from broader regions located much farther downstream the jet. Such a multiple component approach is suggested by the strong correlation between the X-ray \& VHE bands, while the X-ray/VHE emissions do not correlate with the radio/optical/MeV-GeV fluxes. Moreover, the optical/UV fluxes vary by at most 30\%, while X-ray and VHE fluxes display flux changes by about an order of magnitude (see Paper~1 and Fig.~\ref{light_curve_flare}). \par 

In the first zone (hereafter dubbed as the ``fast'' zone) the parameters and the electron distribution are evolved on $\lesssim15$\,min timescales to account for the X-ray \& VHE variability on sub-hour timescales (see previous sections). As for the second zone (hereafter dubbed as the ``slow'' zone), variations of the parameters are imposed on daily timescales since the radio/optical/MeV-GeV show no significant intranight variability. The modeling of the light curve and SEDs is performed for the nine days of the flare (see Fig.~\ref{light_curve_flare}), and solely within time ranges where strictly simultaneous X-ray \& VHE measurements are available.\par

\begin{figure*}[htb!]
\centering
\gridline{\fig{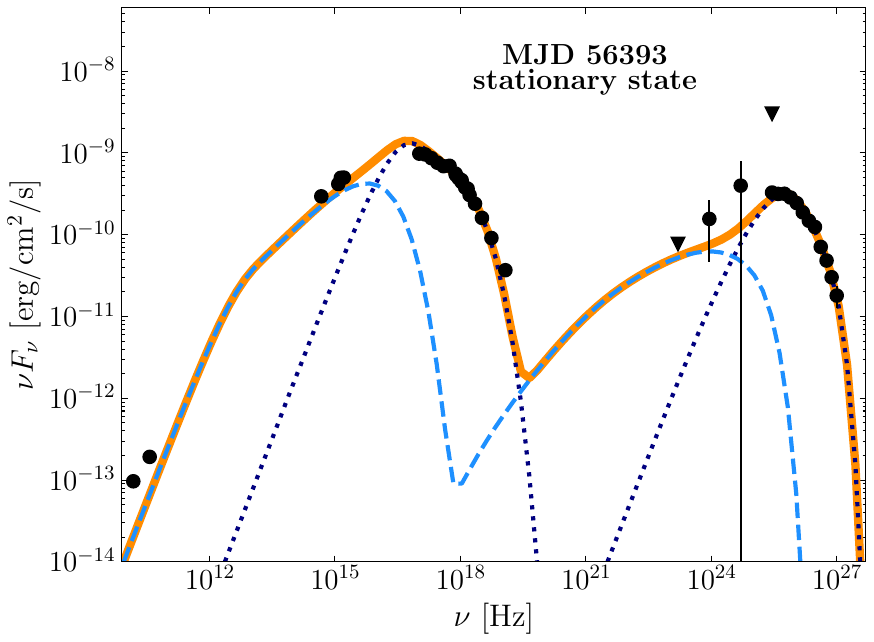}{0.48\textwidth}{(a) MJD~56393}
          \fig{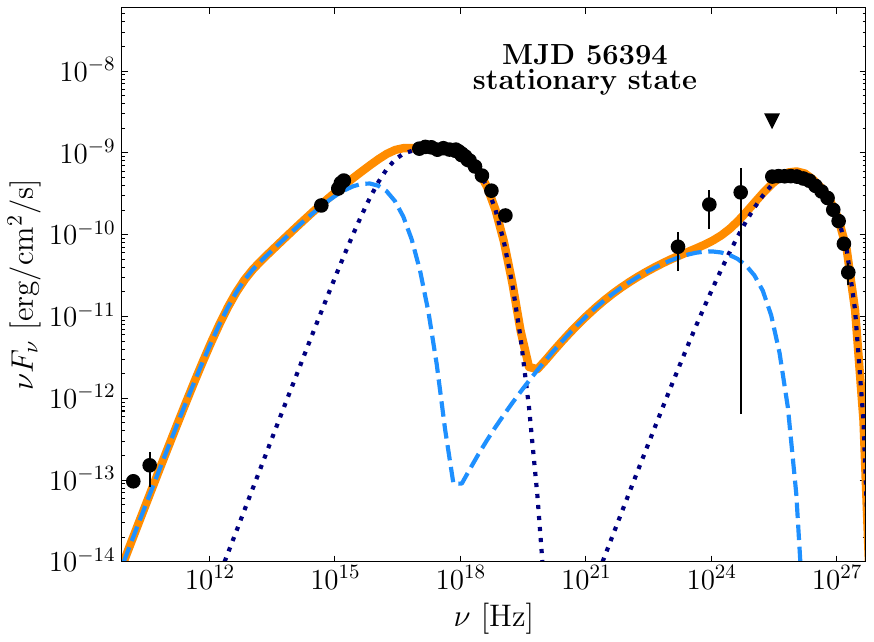}{0.48\textwidth}{(b) MJD~56394}}
\gridline{\fig{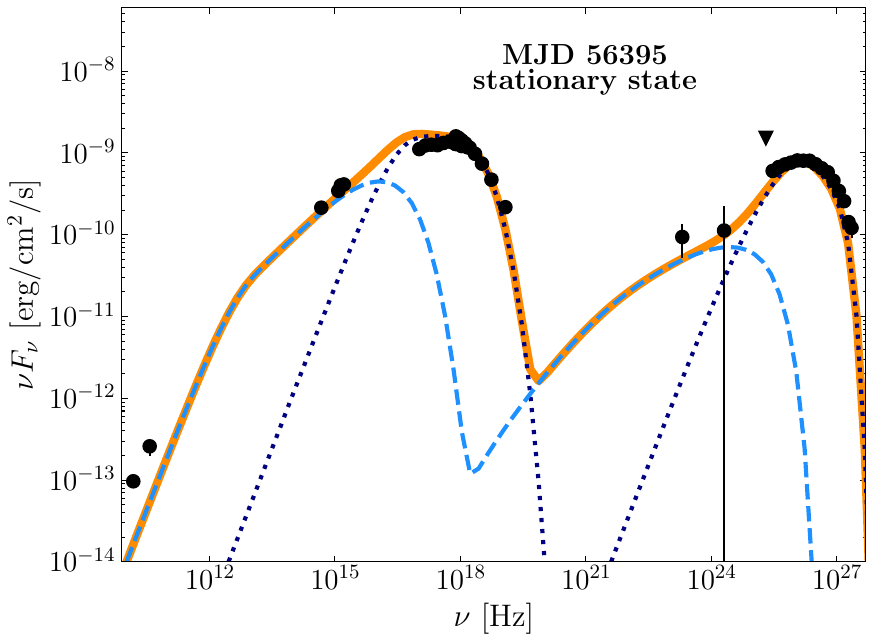}{0.48\textwidth}{(c) MJD~56395}
          \fig{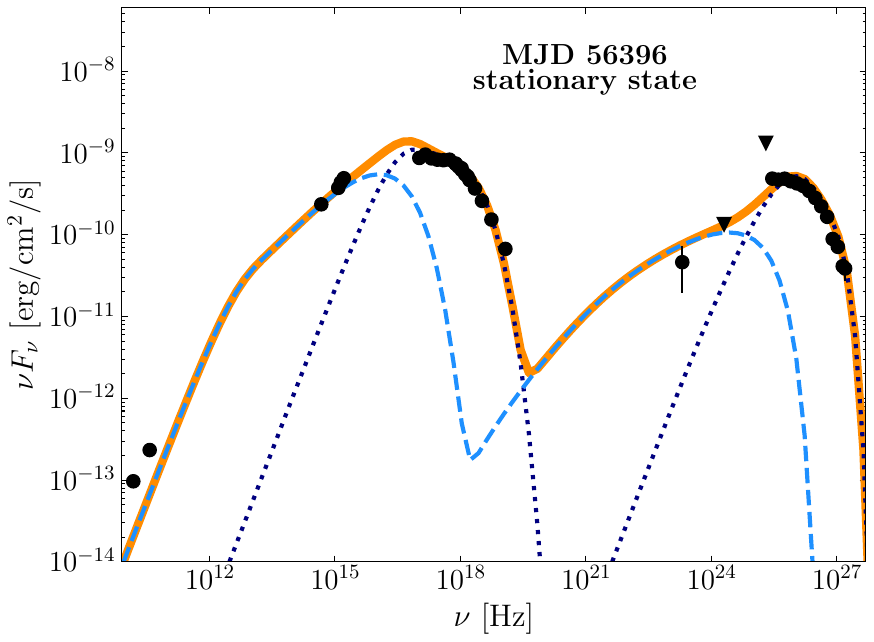}{0.48\textwidth}{(d) MJD~56396}}
\gridline{\fig{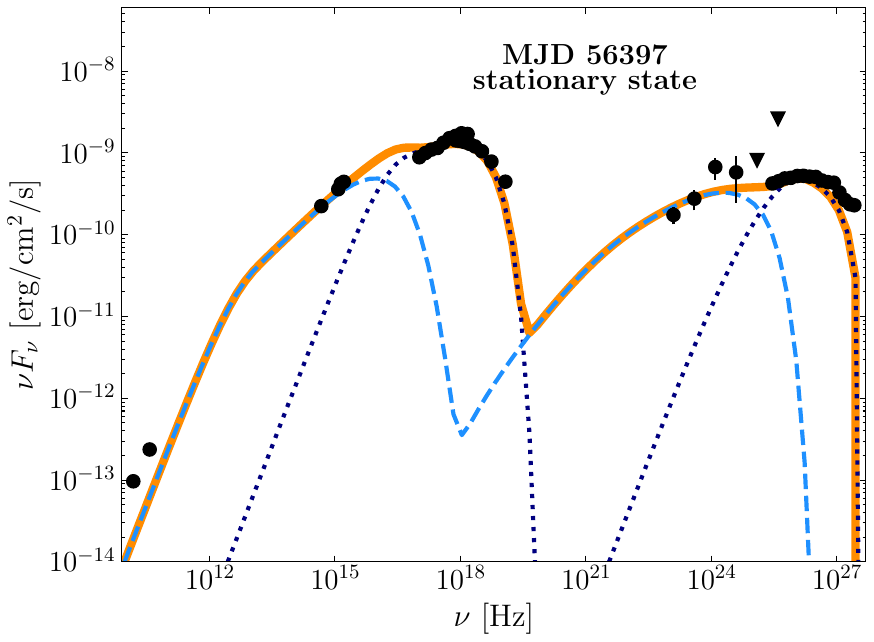}{0.48\textwidth}{(e) MJD~56397}
          \fig{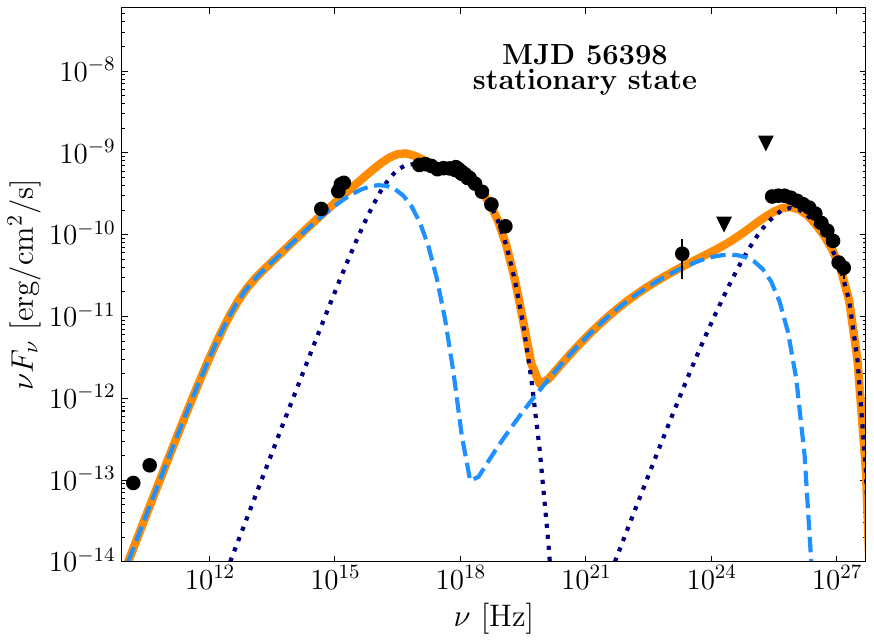}{0.48\textwidth}{(f) MJD~56398}}
\caption{Stationary states fitted to nightly averaged SEDs, from MJD~56393 to 56398. The light blue dashed line represent the emission from the ``slow'' zone, the dark blue dotted line is the emission from the ``fast'' zone and the solid line is the sum of the two components. Data are depicted with dark points. The corresponding model parameters are listed in Tab.~\ref{table:steady}.
\label{stationary_states}}
\end{figure*}
\begin{figure*}[htb!]
\centering
\gridline{\fig{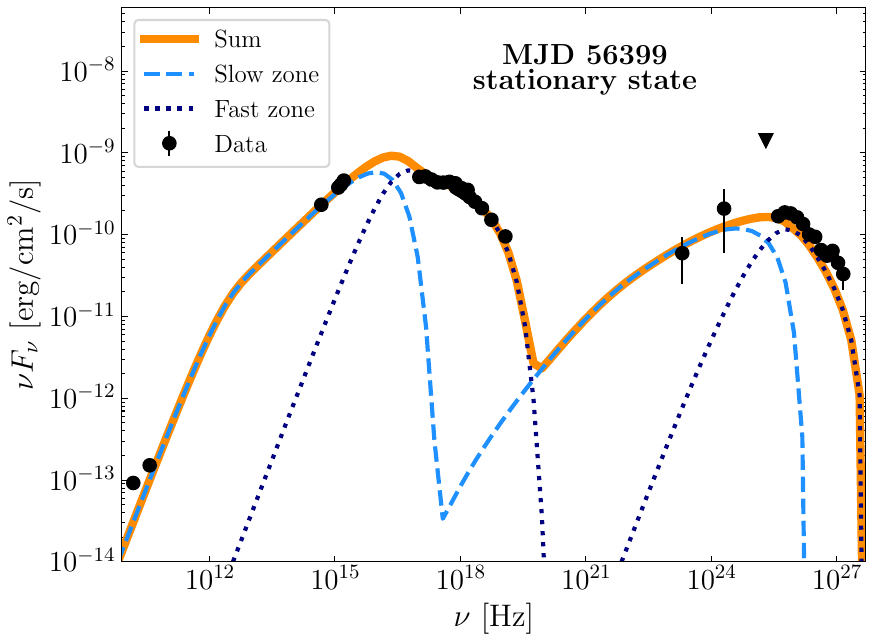}{0.48\textwidth}{(g) MJD~56399}
          \fig{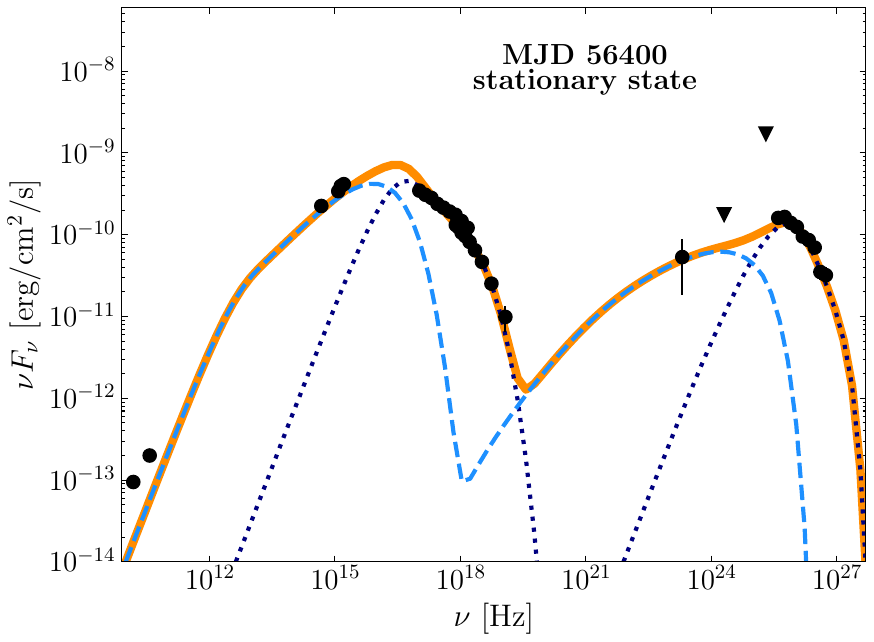}{0.48\textwidth}{(h) MJD~56400}}
\gridline{\fig{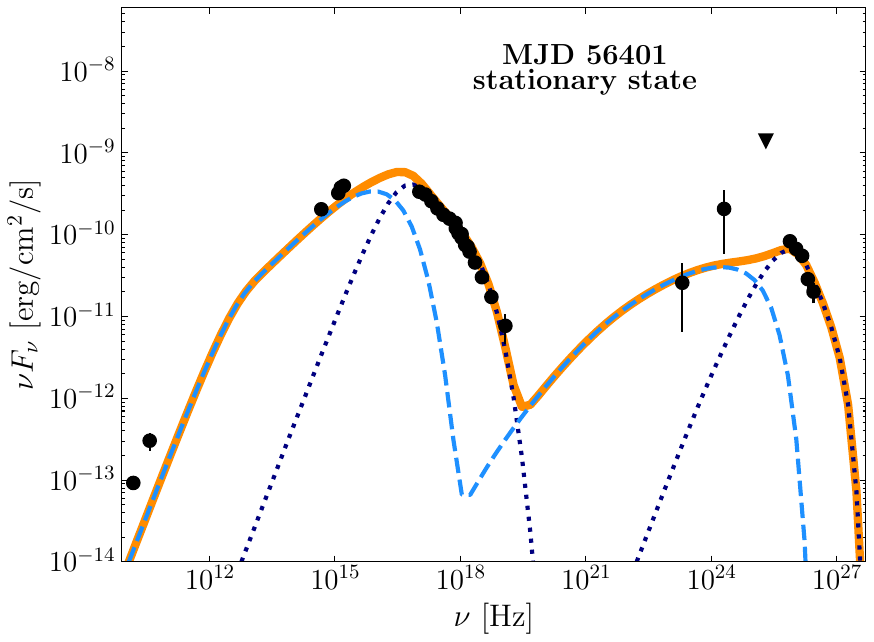}{0.48\textwidth}{(i) MJD~56401}}
\caption{Same as Fig.~\ref{stationary_states}, from MJD~56399 to MJD~56401.
\label{stationary_states_2}}
\end{figure*}

\begin{table*}[h]
\centering
\resizebox{0.9\textwidth}{!}{\begin{tabular}{ccccccc}
\hline \hline
   Day  &   $\log_{10}(l_e) +5$ & $p$    &  $\log_{10}(\gamma_{\rm min})$ & $\log_{10}(\gamma_{\rm max})$  & $B$ [G] & $R$ [$10^{16}$cm]  \\ 
\hline \hline
  ``Slow'' zone  & & &    &  $\delta=30$ &  & \\  
  \hline
  $56393$ & $ 0.10 $ & $ 2.0 $ &   $2.5 $ & $4.4  $ &$0.21  $   & $ 1 $ \\    
  $56394$ & $  0.10 $ & $ 2.0 $ &   $2.5 $ & $ 4.4 $ & $0.21  $   & $ 1 $ \\    
  $56395$ & $ 0.20 $ & $ 2.0 $ &   $2.5 $ & $ 4.6 $ &$0.21  $   & $  1$ \\    
  $56396$ & $ 0.10 $ & $ 2.0 $ &   $2.5 $ & $4.6  $ &$0.21  $  & $ 1 $ \\    
  $56397$ & $ 0.70 $ & $ 2.0 $ &   $2.5 $ & $ 4.5 $ &$0.21  $   & $ 0.5 $ \\   
  $56398$ & $ 0.00  $ & $ 2.0 $ & $2.5 $ & $ 4.6 $ & $0.21  $   & $ 1 $ \\   
  $56399$ & $  0.15 $ & $ 2.0 $ &   $ 2.5$ & $4.5  $ & $0.21  $  & $ 1  $ \\  
  $56400$ & $  -0.05 $ & $ 2.0 $ &   $ 2.5$ & $4.5  $ & $0.21  $  & $ 1  $ \\    
  $56401$ & $ 0.05 $ & $ 2.0 $ &   $ 2.5$ & $4.5  $ & $0.21  $  & $ 1  $ \\    

\hline 
  ``Fast'' zone   & & &   & $\delta=100$ &  & \\
\hline  
  $56393$ & $ -0.30 $ & $3.63  $ &   $4.2 $ & $5.2  $ & $ 0.160 $  & $ 0.122 $ \\   
  $56394$ & $ -0.10 $ & $ 2.94 $ &   $4.3 $ & $5.3  $ & $ 0.100 $  & $ 0.122 $ \\    
  $56395$ & $ -0.05 $ & $ 2.82  $ &  $4.3 $ & $5.3  $ & $ 0.110 $  & $ 0.122   $ \\    
  $56396$ & $ -0.10 $ & $ 3.20$ &   $4.3 $ & $5.3  $ & $ 0.100 $  & $  0.122  $ \\   
  $56397$ & $ -0.10 $ & $2.61 $ &   $4.2 $ & $5.4  $ & $ 0.100 $  & $ 0.122   $ \\    
  $56398$ & $ -0.30 $ & $ 3.11 $ &   $4.2 $ & $5.4  $ & $ 0.120 $  & $  0.122  $ \\   
  $56399$ & $ -0.45 $ & $ 3.35  $ & $4.2 $ & $5.5 $ & $ 0.130 $  & $  0.122  $ \\ 
  $56400$ & $ -0.30 $ & $ 3.96  $ & $4.3 $ & $5.4 $ & $ 0.086 $  & $  0.122  $ \\    
  $56401$ & $ -0.55 $ & $ 3.96  $ & $4.3 $ & $5.3 $ & $ 0.110 $  & $  0.122  $ \\    
\hline
\hline
\end{tabular}}
\label{table:steady}
\caption{Stationary state model parameters for each day for the ``slow'' and ``fast'' zones ($\delta_{\rm slow}=30$ and $\delta_{\rm fast}=100$ respectively). The day labeling refers to the MJD date entering during this segment. The values of $p$ for the ``fast'' zone are representative of the daily average state, and additional short-timescale variations will be applied to model the light curves (see Sect.~\ref{sec:p_le_variations_setup} and Fig.~\ref{p_le_vs_mjd}).}
\end{table*}

\subsection{Model set-up and initial conditions} \label{sec:modelling_setup}

Each emitting zone is modelled as a sphere (in the jet comoving frame) of radius $R$ permeated by a tangled magnetic field of strength $B$. The emitting plasma moves with a bulk velocity of $\Gamma_b$, leading to a relativistic beaming of the radiation with a Doppler factor $\delta$. Pre-accelerated electrons are injected at a rate $Q_e^{\rm inj}(\gamma)$ into each zone with a power-law distribution of Lorentz factor $\gamma$ (or equivalently kinetic energy $(\gamma-1)\,m_e c^2 \approx \gamma\, m_e c^2$) with a slope $p$ between $\gamma_{\rm min}$ and $\gamma_{\rm max}$, and followed by a modified exponential cut-off for Lorentz factors beyond $\gamma_{\rm max}$: 

\begin{equation}
\label{eq:pwl_cutoff}
Q_e^{\rm inj}(\gamma) \propto \begin{cases}
 \gamma^{-p}, \quad \gamma_{\rm min}<\gamma<\gamma_{\rm max}\\
 \gamma^{-p} \, \exp{\left(1-(\gamma / \gamma_{\rm max})^{a}\right)}, \quad \gamma > \gamma_{\rm max}\\
\end{cases}
\end{equation}
where $a=2$, which we find to yield the best description of the spectral hardness and the fluxes at all bands. In principle, within a shock acceleration process the exponent $a$ can be related to the type of diffusion/turbulence in the plasma \citep[see e.g.][]{2007A&A...465..695Z,2024PhRvD.109j3039B}, and $a=2$ may occur in a Bohm-type diffusion. Super-exponential high-energy cutoff are also derived from PIC simulations of magnetic reconnection events \citep[see e.g.][]{2016ApJ...816L...8W}, although recent simulations of large-scale systems seem to prefer a simple exponential cutoff \citep{2018MNRAS.481.5687P}. The normalization of the distribution is characterized by the electron compactness $l_e$. The electron compactness is a dimensionless measure of the electron luminosity $L_e$, and is defined as follows: 
\begin{equation}
L_e = m_e c^2 \int_{\gamma_{\rm min}}^{\gamma_{\rm max}} Q_e^{\rm inj}(\gamma) (\gamma-1)
\label{eq:le_luminosity}
\end{equation}
\begin{equation}
l_e = \frac{3 \sigma_T L_e}{4\pi m_e c^3 R}.
\label{eq:le}
\end{equation}
After injection, the subsequent electron and photon populations can be exactly determined by solving the Fokker-Planck equations (kinetic equations hereafter), taking into account all the gain/loss terms of all relevant radiative processes. These are the emission from synchrotron and inverse Compton scattering, the synchrotron self-absorption and the $\gamma \gamma $ pair-production. We note that for the inverse Compton scattering in the Klein-Nishina regime the full expression is used \citep{Blumenthal70}. We further assume that in both zones the particles escape the emitting region on a timescale $t_{\rm esc} = R/c$. When solving the kinetic equations, the Compton scattering of cold electrons ($\gamma \approx 1$) and their radiation is not considered\footnote{The emission from electrons with $\gamma \approx 1$ is very low, so we assume that once they cool down to such energies they simply escape the blob. Additionally, the code used in this work is not designed to treat both non-relativistic and relativistic particle distributions.}. For solving the kinetic equations we use a modified version of the code of \citet{Mastichiadis95}, as further developed in \citet{polkas21} for producing time-dependent SEDs by continuously varying one parameter of the SSC model. The main differences with their work and the current model are: i) we use high-cadence X-ray light curves (binned over 15-minute intervals) instead of gamma-ray light curves of 1-day resolution ii) we impose variations on the power-law slope of the electrons $p$ iii) we use $l_e$ variations as a correction motivated by the model-data residuals following the variations of $p$ (see later for more details) iv) no external photon fields are considered as seeds for the inverse-Compton scattering. The general scheme for transforming from observations to model parameters and the interpolation scheme applied remains the same. In our setup the energy resolution is fixed at $0.1$\,dex.\par

The kinetic equations for electrons and photons are solved in the rest frame of the blob. We therefore have to transform the properties of the photon field to the observer's frame using the Doppler factor $\delta$. The bolometric luminosities transform as $\delta^4$, the photon energies by multiplying by $\delta$ and time-contracts as $1/\delta$. The photon distribution is transformed to an observed SED, using the distance of Mrk~421 \citep[$z=0.031$;][]{1975ApJ...198..261U}. These transformations are not going to be mentioned again for the rest of this section.\par

\subsubsection{Stationary States}
\label{sec:sec_stationary}

The time-dependent modeling starts from stationary states determined for each day separately. For this, we first build daily average SEDs from the data, which we then model with our two-component scenario. We extract daily VHE spectra from the MAGIC observations as well as \textit{NuSTAR} SEDs averaged over time ranges that are strictly simultaneous to those from MAGIC. The \textit{NuSTAR} SEDs are complemented by \textit{Swift}-XRT observations that are the closest to the center of the MAGIC/\textit{NuSTAR} exposures. We also use \textit{Fermi}-LAT SEDs averaged over 18\,hours around the center of each MAGIC/\textit{NuSTAR} observation. Finally, we consider quasi-simultaneous radio/optical/UV observations (within less than $\sim10$\,hours, being smaller than the typical variability timescale observed at those frequencies).\par 

We compute the steady-state spectrum of each zone separately by running the code until $10$ light-crossing times without applying any parameter variations and then add their SEDs. The total spectrum is compared to the data. We update the input parameters and repeat the procedure until a satisfactory description of the measurement is achieved.\par

In order to limit the degrees of freedom, the Doppler factor of the ``slow'' zone is set to $\delta_{\rm slow}=30$ \citep[in line with previous modeling of this source in average state, see][]{2011ApJ...736..131A} and the electrons are injected with an index $p=2$ for all days, which is found to well describe the radio/optical/MeV-GeV data. We constrain the radius of the emitting blob to be consistent with the variability timescale using causality arguments ($R<\delta_{\rm slow} \, c \, \Delta t_{\rm obs}$, where $\Delta t_{\rm obs}$ is the observed variability timescale being $\sim1$\,day for the radio/optical/MeV-GeV bands). We find that using $B\approx 0.07$\,G provides a reasonable description of the data, and hence we fix it to the latter value for all days.\par

Similar considerations are made to constrain the degrees of freedom for the stationary states of the ``fast'' zone. First, we derive that $\delta_{\rm fast}\sim100$ is required to properly reproduce the VHE spectra up to the highest energies. As highlighted in Sect.~\ref{sec:daily_averaged_magic_spectra}, MAGIC nightly spectra extend over multi-TeV energies. Adopting a lower Doppler factor, $\delta_{\rm fast} \sim 50$ comparable to that of the ``slow'' zone, leads to an evident underestimation of the measurements at the highest energies due to the Klein-Nishina suppression of the inverse-Compton cross-section. Following \citet{1998ApJ...509..608T}, the Klein-Nishina effects fully set in and strongly affect the peak frequency (and luminosity) of the inverse-Compton component when the Doppler factor is below $\delta_{\rm KN}$ that is given by:
\begin{equation}
\label{eq:delta_limit}
    \delta_{\rm KN} = \sqrt{\frac{\nu_{\rm s} \nu_{\rm IC}}{(3/4) (m_e c^2 / h)^2}} ,
\end{equation}

where $\nu_{\rm s}$ and $\nu_{\rm IC}$ are the observed synchrotron and inverse-Compton peak frequencies, respectively, and $h$ the Planck constant. Taking as example MJD~56395, we have $\nu_{\rm s}\approx 1$\,keV and $\nu_{\rm C}\approx 1$\,TeV (see Fig.~\ref{nightly_vhe_spectra}), yielding $\delta_{\rm KN} \approx 70$. We stress that, even above this limit, the Klein-Nishina effects can have a non-negligible impact, and we find that increasing the Doppler factor to at least $\delta_{\rm fast}\sim100$ is necessary to obtain a satisfactory description of the VHE data up to highest energies for most days. In Appendix~\ref{sec:doppler_comp_steadystate}, we compare for illustration purposes the SED from the ``fast'' zone when $\delta_{\rm fast}=50$ and $\delta_{\rm fast}=100$ for MJD~56393 and MJD~56395. These tests confirm the expected underestimation of the VHE spectrum at the highest energies when a Doppler factor below 100 is used. As reported in detail in Sect.~\ref{sec:modelling_results}, MJD~56397 still suffers from an underprediction of the highest VHE spectral points, even when using $\delta_{\rm fast} = 100$. While we argue that it may also suggest the appearance of an extra zone filled with energetic electrons, one cannot exclude that, for this particular day, the Doppler could be even larger than 100.\par 

The radius of the ``fast'' zone is set to \mbox{$R=0.122 \times 10^{16}$\,cm} in agreement with the $\sim15$\,min variability detected in the X-ray \& VHE bands. Contrary to the ``slow'' zone, the magnetic field is kept as a free parameter which is a requirement to match the variation of the Compton dominance (ratio between the inverse-Compton and synchrotron peak luminosities) that is well constrained by the X-ray and VHE data. Finally, the injected electron slope ($p$) is determined directly from the X-ray data.\par 

\begin{figure}[t!]
\plotone{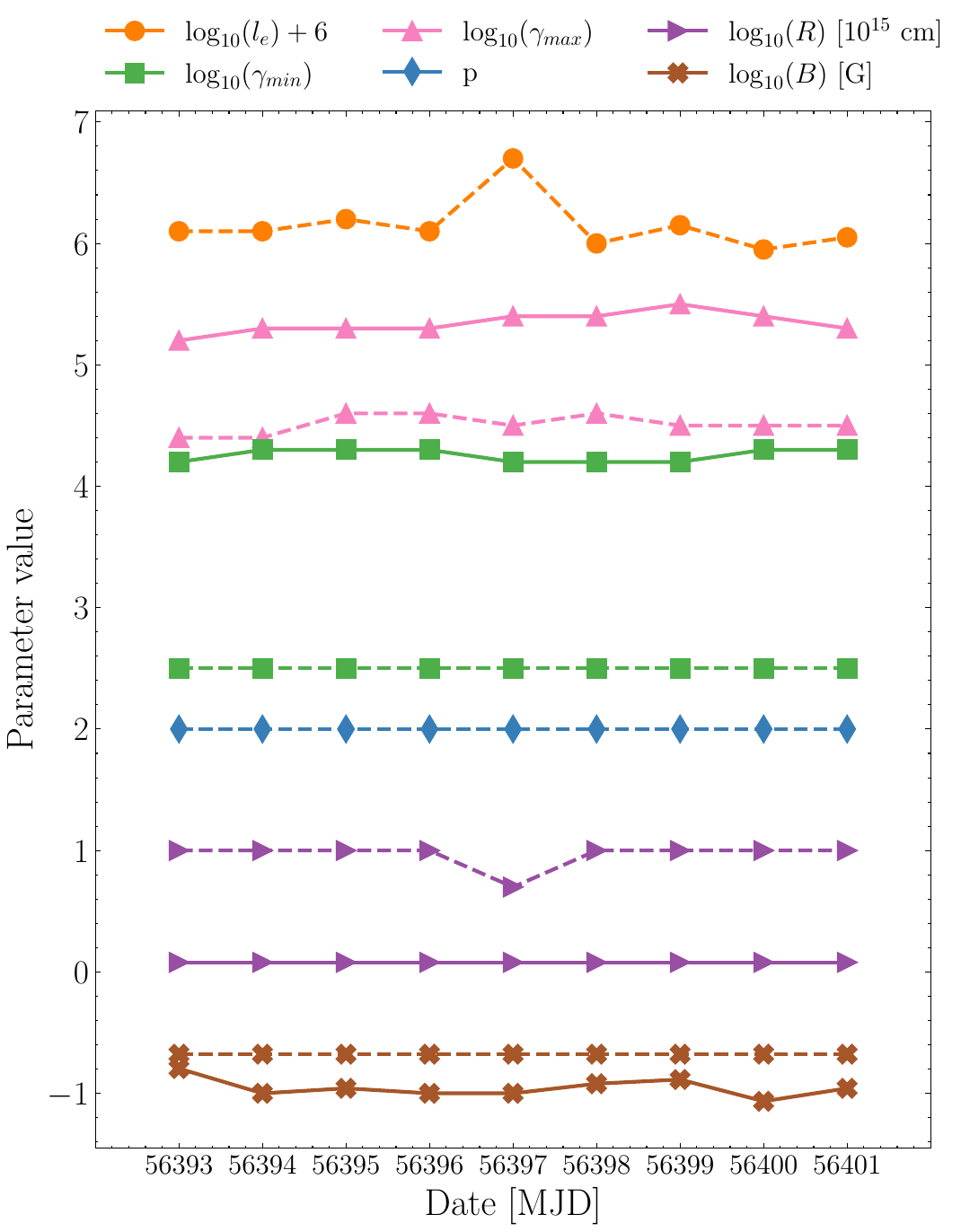}
\caption{Temporal evolution of the parameters from the theoretical model applied to broadband SEDs from single days. The solid and dashed lines depict the model parameters from the "fast" and "slow" zones, respectively.}
\label{stationary_state_evolution}
\end{figure} 

The model curves for these stationary states adjusted to the daily broadband SEDs are depicted in Fig.~\ref{stationary_states} and \ref{stationary_states_2}. The corresponding model parameters are listed in Tab.~\ref{table:steady}, and their temporal evolution is illustrated in Fig.~\ref{stationary_state_evolution}. Within the ``fast'' zone, the magnetic field $B$ varies by less than a factor of two. Regarding the ``slow'' zone, most of the parameters remain unchanged, in line with the low variability observed in the UV/Optical/MeV-GeV fluxes.\par 

\begin{figure*}[th!]
\centering
\includegraphics[width=2.2\columnwidth]{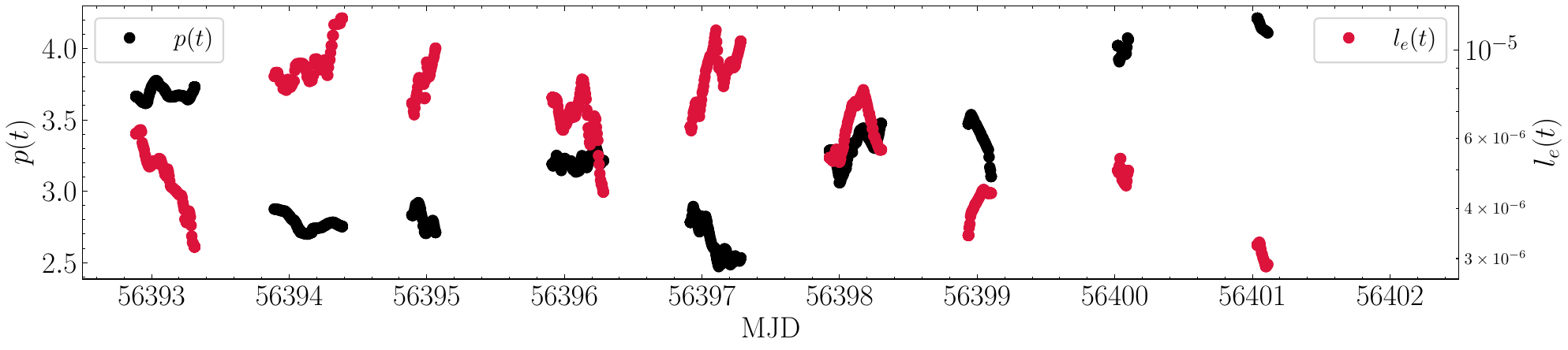}
\caption{Temporal evolution of the model parameters $p(t)$ (black markers) and $l_e(t)$ (red markers) that characterize the electron distribution injected in the ``fast'' zone. See text for more details on the procedure applied to determine the corresponding variations. 
\label{p_le_vs_mjd}}
\end{figure*}

The small variability of $R$ and $B$ throughout the days suggests that both components are almost stationary within the jet. In case the ``fast'' zone moves downstream (as in a blob-in-jet model) with a Lorentz factor $\Gamma_{\rm b, fast} \sim \delta_{\rm fast} /2 = 50$ (assuming a closely aligned jet), one can estimate a travelled distance of $\Delta d = \frac{\Gamma_{\rm b, fast}^2 \beta_{\rm fast} c \Delta T_{\rm obs}}{1+z} \sim 10^{20}$\,cm $\sim 30$\,pc within a window of $\Delta T_{\rm obs}=9$\,day in the observer's frame. Assuming that the blob started to form at a distance $d_0 \sim R/\Theta_d \sim R \, \Gamma_{\rm b, fast} \sim 10^{17}$\,cm from the jet's base (where $\Theta_d \sim 1/ \Gamma_{\rm b, fast}$ is the opening angle of the jet), this leads to a ratio $\Delta d/d_0 \sim 10^3$. Under the assumption where the magnetic field scales as $\sim R^{-1}$ \citep[due to flux freezing and energy conservation in a toroidal field;][]{1984RvMP...56..255B}, $B$ would evolve by several orders of magnitude between the start and the end of the flare, in clear contradiction with Table~\ref{table:steady}. We thus favor a scenario where the two zones are quasi-stationary features in the jet, possibly associated to two different re-collimation shocks \citep{1988ApJ...334..539D}. In this configuration, even considering stationary shocks the radiation is boosted by a factor $\delta^4$ \citep[see discussion in ][]{2021A&A...654A..96Z, 1997ApJ...484..108S}.\par

We neglect the interaction between the two regions, meaning that the emission from each zone is entirely computed based on the local physical conditions. In reality, particles in a given zone may interact with the radiation field from the other zone if these are close to each other, leading to an additional inverse-Compton component. In our case, the interaction is enhanced by the relative bulk velocity of the plasma of the zones that inevitably boosts the radiation density of each zone as seen in the frame of the other. Based on the parameters of Table~\ref{table:steady}, we estimate that if the two zones (possibly re-collimation shocks) are separated along the jet's axis by at least $r_0>1.5 \times 10^{17}$\,cm the interaction can be safely neglected. Such a separation, equivalent to $\lesssim10^{-1}$\,pc, is realistic and well below the scale of the jet and the distance between the stationary features detected in Mrk~421 \citep{2019ApJ...877...26H, 2016AJ....152...12L}. The detailed quantification is presented in Appendix~\ref{sec:zones_interaction}.\par

The parameters for the ``slow'' zone are only evolved on daily timescale according to Tab.~\ref{table:steady}. This is inline with the variability noticed in the radio/optical/MeV-GeV bands (Fig.~\ref{light_curve_flare}). Differently, the rapid X-ray \& VHE spectral changes imply intranight variations of the parameters in the ``fast'' zone. We try to reproduce the X-ray \& VHE data by evolving $p$ and $l_e$. The variations are applied on timescales equal to the light-crossing time, which is $R/c/\delta_{\rm fast} \sim 6$\,min in the observer's frame. The procedure is described below.

\subsubsection{Time-dependent intraday SED modeling}
\label{sec:p_le_variations_setup}

For each day, the $p$ and $l_e$ variations in the ``fast'' zone are applied with all other model parameters fixed to the values of Tab.~\ref{table:steady}. We build time evolution curves of $p$ and $l_e$ in the observer's frame, which are then transformed in the jet's rest-frame by multiplying by $\delta_{\rm fast}$ and then are smoothed with a quadratic interpolation so that the derivatives are handled correctly when solving the stiff kinetic equations. We remind that the code outputs model curves with temporal steps equal to the light-crossing time, which is $R/c/\delta_{\rm fast} \sim 6$\,min in the observer's frame. To remove potential artifacts originating from the initial conditions, we let the system reach a steady state (by evolving the system for 0.1\,days, equivalent to $\approx25$ light-crossing times) before we start applying the variations in model parameters.\par

We follow a two-step approach when applying the simultaneous $p$ and $l_e$ variations. We start by building light curves by varying only $p$. To determine the time series of $p$ we directly exploit the X-ray data (using \textit{Swift}-XRT and \textit{NuSTAR}) as this energy regime of the SED originates from synchrotron radiation, whose spectrum brings a direct constraint on the electron slope at any time throughout the flare. The details on the determination of the $p$ are provided in Appendix~\ref{sec:p_evolution_curve_determination}. In summary, we approximate the \textit{NuSTAR} data with a simple power-law model, which is used to estimate the photon index $\Gamma_{\rm X-ray}$ in the soft X-ray band that is radiated by electrons far below the high-energy (exponential) cut-off. The slope $p$ of the injected electrons is then derived using the relation $\Gamma_{\rm X-ray} = (p+1)/2$ for a synchrotron-uncooled population of electrons \citep{1979rpa..book.....R}, which is the case for all stationary states derived in the previous section. In fact, by balancing the escape timescale, $R/c$, with the synchrotron and inverse-Compton cooling timescales, the cooling break is located around \mbox{$\log_{10}{\gamma}\sim 6.2$,} beyond the cut-off Lorentz factor $\gamma_{\rm max}$ of the injected distribution. \par

After running the code with the $p-$variations, the X-ray and VHE flux from the model differs significantly from that of the data, since the observed variations are significantly more intense than those generated with a $p-$only evolution in the electron population. That is expected, given the observed multi-band amplitude of variations. For the aforementioned reasons, we correct for the unaccounted flux-variations by directly evolving $l_e$. To do so, we need to know at each time $t$ what is the amount of flux to be added/subtracted, as well as how to translate this flux difference to an electron compactness difference. As in \citet{polkas21}, we assume a power-law dependence of $l_e$ on the flux of a certain band, and we select here the $3-7$\,keV 15-min light curve from \textit{NuSTAR} as our reference band:
\begin{equation}
l_e(t) = l_{e,0}\times\left(\frac{F_{3-7\rm keV}(t)}{F_{3-7 \rm keV,0}}\right)^{1/\sigma_x}.
\label{eq:transle}
\end{equation}
We find the value $\sigma_x \approx 1$ yields a good description for all days.  Note that for the SSC scenario, the compactness of synchrotron (here also X-ray) photons $l_{\rm syn}$ is expected to have a linear/square-root dependence on $l_e$ in the slow/fast synchrotron-dominated/fast SSC-dominated cooling regime \citep[Eq.\,A7 in][]{polkas21}. However, this does not take into account the dependence of $F_{3-7\rm keV}(t)$ on the spectrum of the photon distribution (that also changes with varying electron compactness) when close to transiting cooling regimes. We compute the residuals $\Delta F = F_{\rm 3-7 \rm keV,data}-F_{3-7 \rm keV,\rm p}$, where $F_{3-7 \rm keV,\rm p}$ stands for the flux for the $p-$variation simulation. Then, for each point, we create a synthetic $l_e$ time curve as $l_e(t) = l_{e,0} \times (1+\Delta F/F_{3-7 \rm keV,p})^{1/\sigma_x} $, where $l_{e,0}$ is the value determined from the stationary state. We then apply this time series, together (values are changed at the same substep in the code) with the initial $p$ light curve on the blob that has reached the quasi-steady state, and produce the final time-dependent electron and photon distributions. The SED of the ``slow'' zone (transformed with $\delta_{\rm slow}=30$) is added to the contribution of the ``fast'' zone using the parameters given in Table~\ref{table:steady} in order to obtain the broadband emission at each time. \par 

\begin{figure}[t!]
\centering
\includegraphics[width=1\columnwidth]{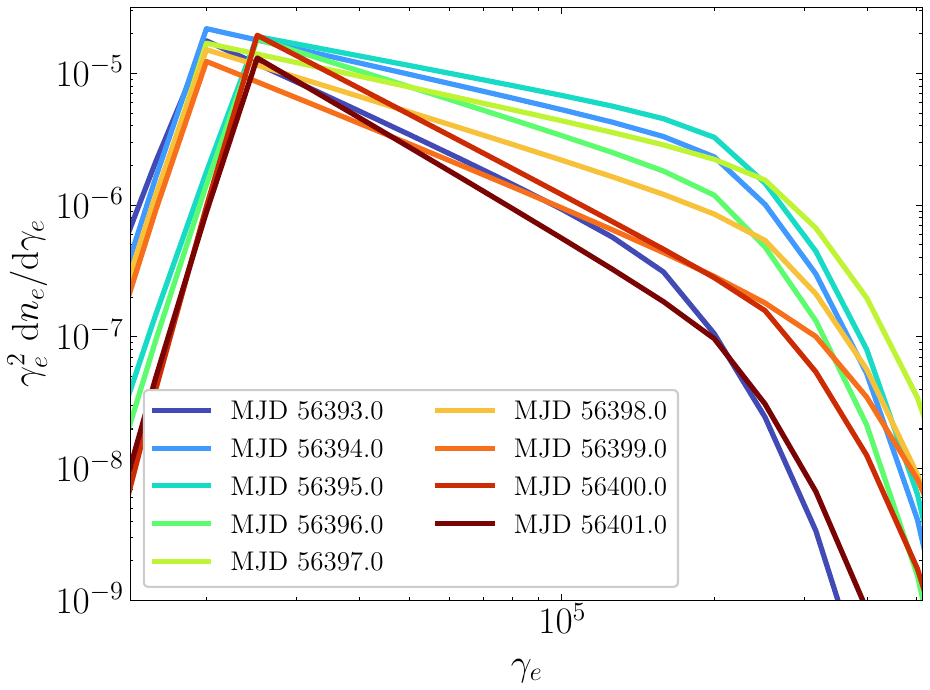}
\caption{Snapshots from the temporal evolution of the electron distribution in the ``fast'' zone, extracted at the time MJD~56XXX.0 from each day of the flare. Here, $n_e$ is defined as the number for particles contained in a volume $\sigma_T \, R$.
\label{representative_EEDs}}
\end{figure}
\begin{figure}[t!]
\centering
\includegraphics[width=1\columnwidth]{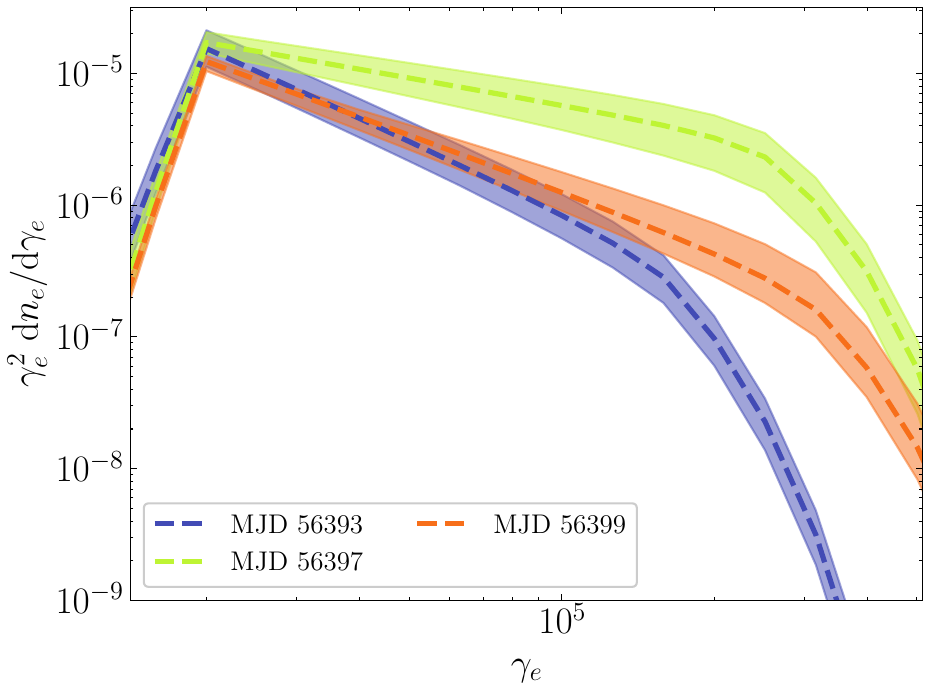}
\caption{Range of values covered by the electron distributions from three selected days that relate to the beginning (MJD~56393), middle (MJD~56397) and end (MJD~56399) of the 9-day flaring activity. The bands represent the minimum and maximum value reached at each Lorentz factor, while the dashed lines depict the average.
\label{range_EEDs}}
\end{figure}

The simultaneous evolution of the injected $p$ and $l_e$ in the ``fast'' zone at each time step is presented in Fig.~\ref{p_le_vs_mjd}. The model is applied over time intervals with \textit{NuSTAR} and VHE exposures. For illustrative purposes, we also plot in Fig.~\ref{representative_EEDs} snapshots of the resulting electron distribution at a fixed time for each day (we arbitrarily select the start of the day in MJD units). To give an idea of the intraday variations of the electron distribution, Fig.~\ref{range_EEDs} presents with bands the range of values covered during three different days.

\begin{figure*}[p!]
\centering
\gridline{\fig{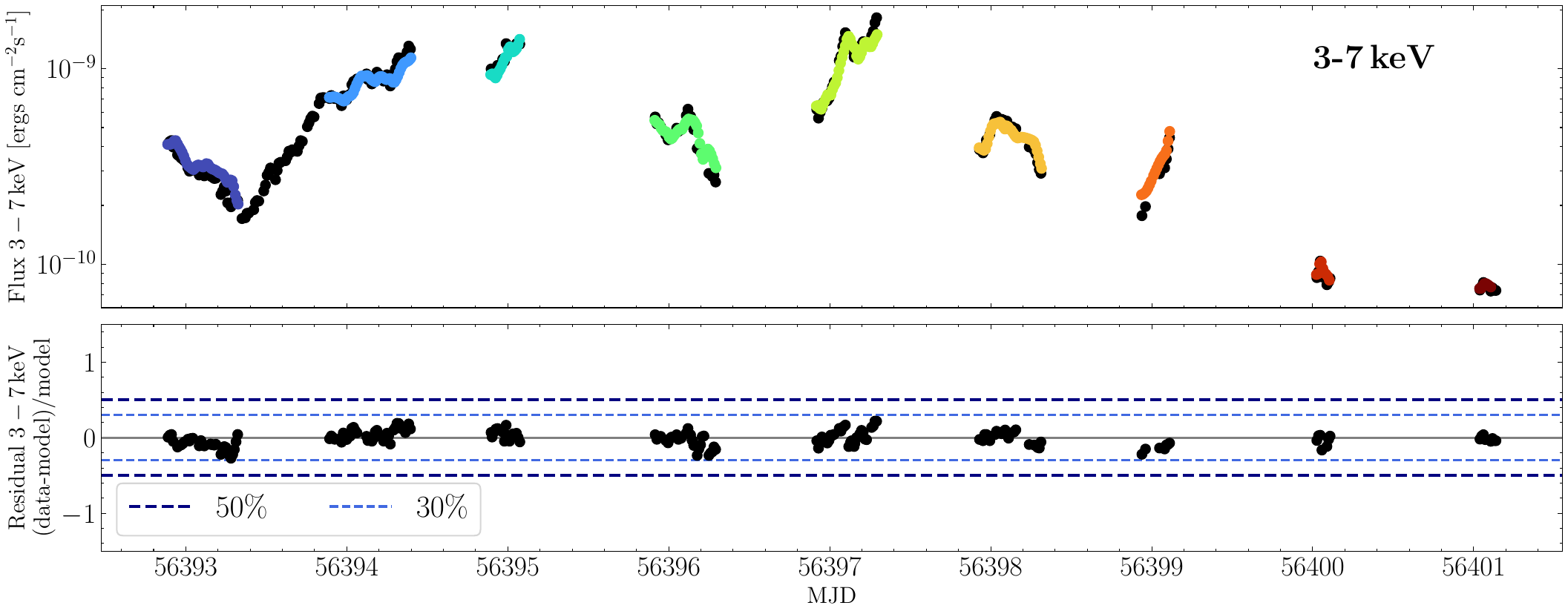}{0.9\textwidth}{(a)}}
\gridline{\fig{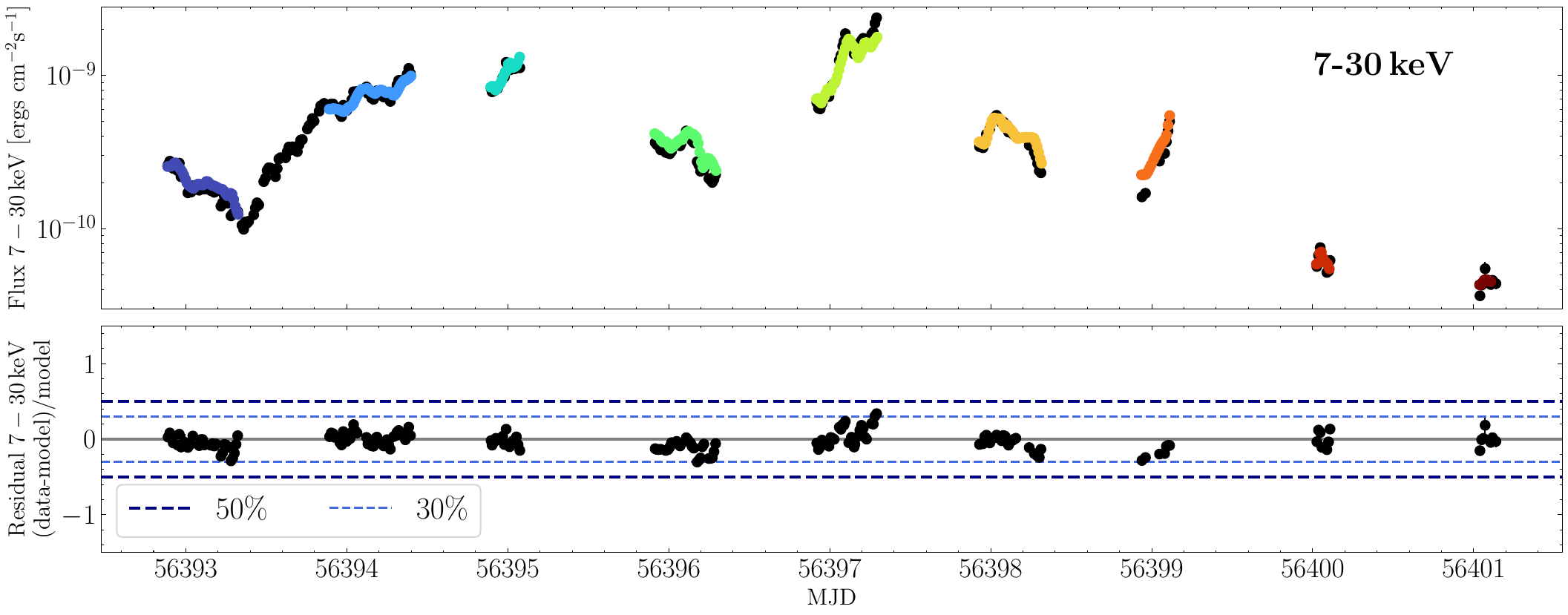}{0.9\textwidth}{(b)}}
\gridline{\fig{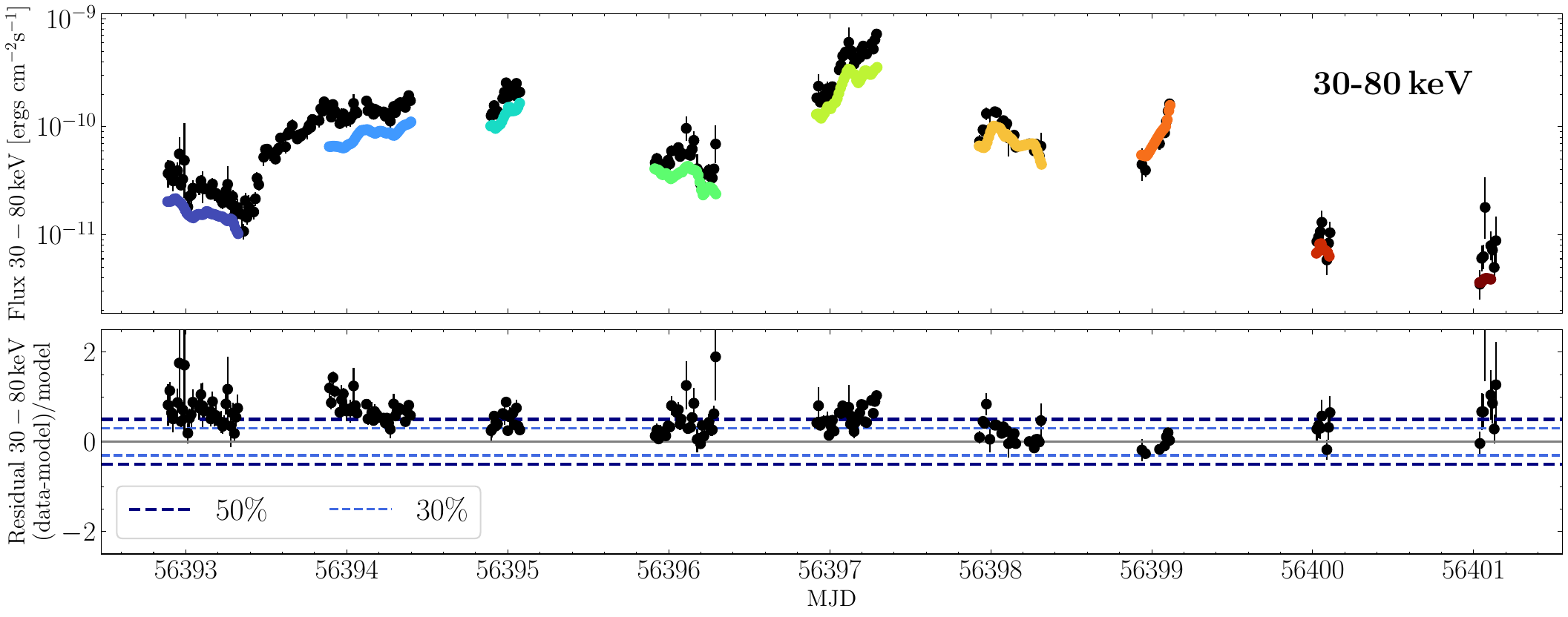}{0.9\textwidth}{(c)}}
\caption{Comparisons between the light curve obtained form observations and light curves provided by the model. Solid colourful markers are the fluxes derived from the model, while dark points represent the data. From top to bottom: (a) $3-7$\,keV band (b) $7-30$\,keV band and (c) $30-80$\,keV band. The bottom panel of each subplot is the residuals defined as $\frac{ (data - model)}{model}$.
\label{xray_model_comp}}
\end{figure*}
\begin{figure*}[p!]
\centering
\gridline{\fig{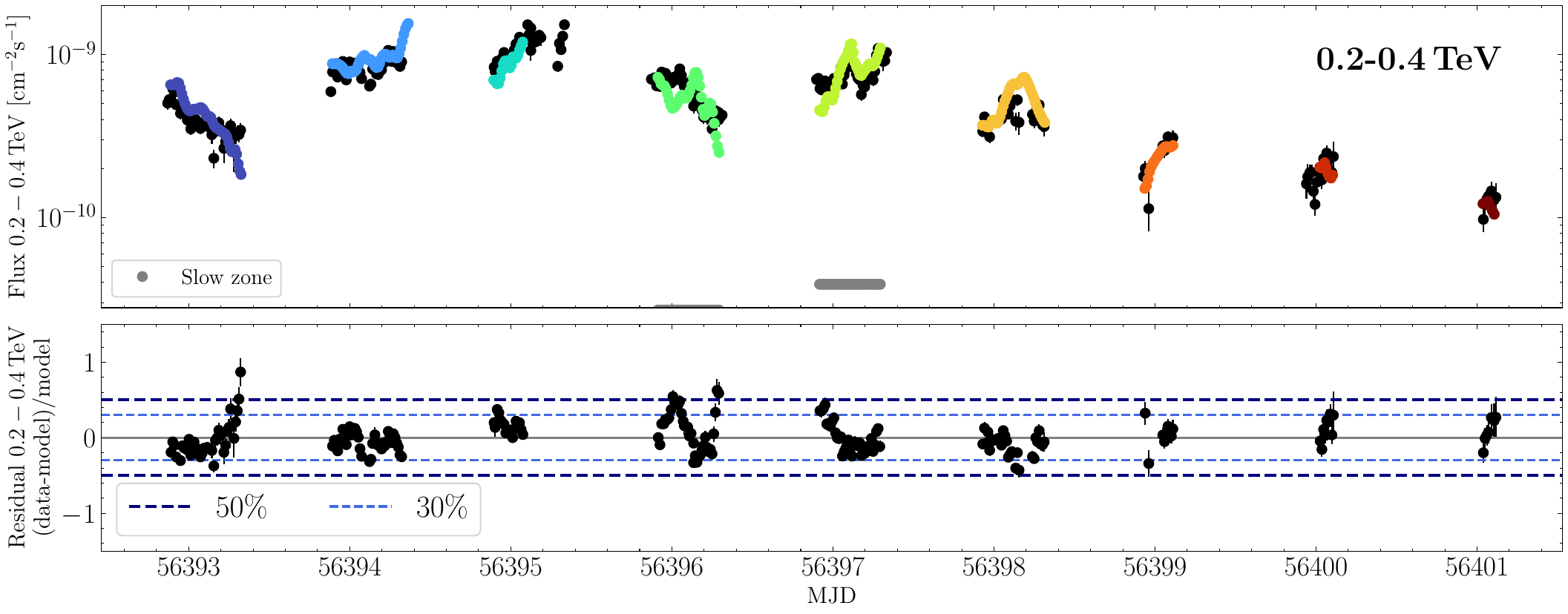}{0.9\textwidth}{(a)}}
\gridline{\fig{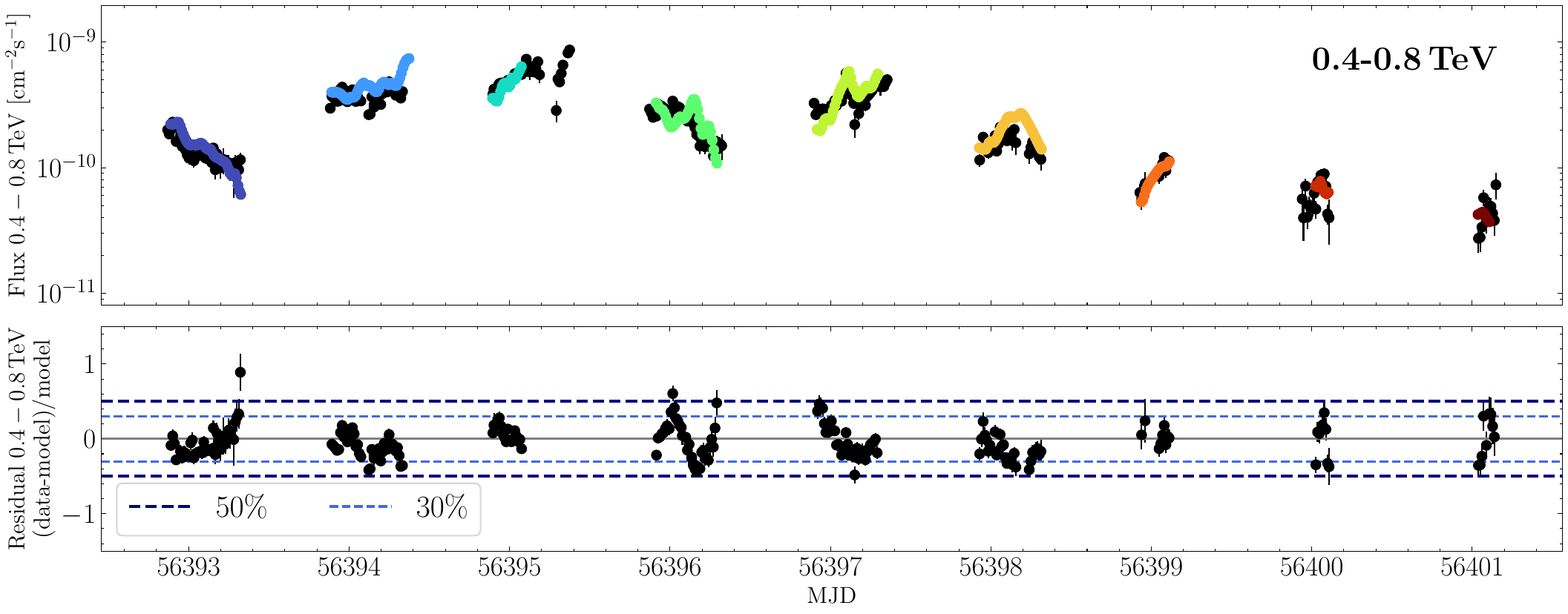}{0.9\textwidth}{(b)}}
\gridline{\fig{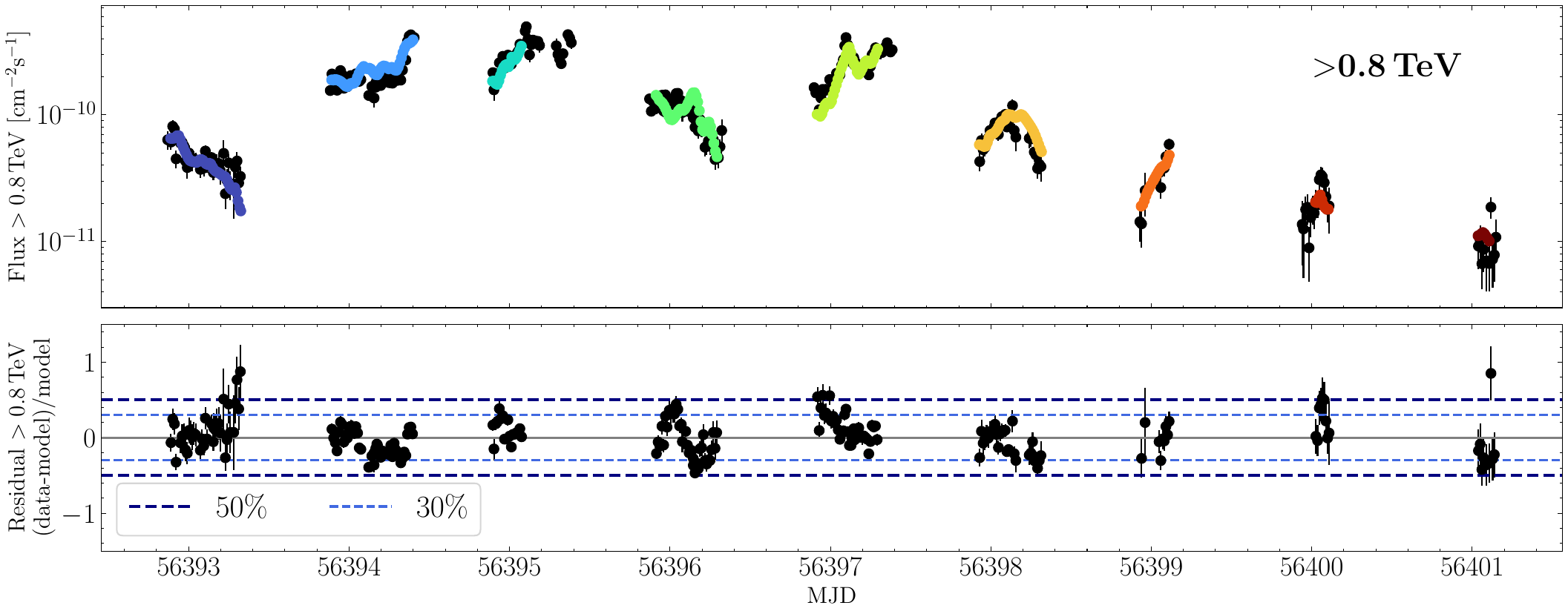}{0.9\textwidth}{(c)}}
\caption{Same as Fig.~\ref{xray_model_comp}, but at (a) $>0.8$\,TeV (b) $0.4-0.8$\,TeV and (c) $0.2-0.4$\,TeV energies. For the $0.2-0.4$\,TeV and $0.4-0.8$\,TeV bands we further show with grey markers the flux from the ``slow'' component, which brings a small contribution to the total flux on MJD~56397. For the other days, as well as for the entire $0.2-0.4$\,TeV and $>0.8$\,TeV light curve, the ``slow'' zone contribution is completely negligible and therefore omitted from the plot.  
\label{vhe_model_comp}}
\end{figure*}

\begin{figure}[h!]
\centering
\gridline{\fig{figures/res_hist_3_7keV.pdf}{0.425\textwidth}{(a)}}
\gridline{\fig{figures/res_hist_7_30keV.pdf}{0.425\textwidth}{(b)}}
\gridline{\fig{figures/res_hist_30_80keV.pdf}{0.425\textwidth}{(c)}}
\caption{Histogram of the residuals defined as $\frac{ (data - model)}{model}$ for the three X-ray bands (a) $3-7$\,keV (b) $7-30$\,keV and (c) $30-80$\,keV.
\label{res_hist_xray}}
\end{figure}
\begin{figure}[h!]
\gridline{\fig{figures/res_hist_02_04teV.pdf}{0.425\textwidth}{(a)}}
\gridline{\fig{figures/res_hist_04_08teV.pdf}{0.425\textwidth}{(b)}}
\gridline{\fig{figures/res_hist_08teV.pdf}{0.425\textwidth}{(c)}}
\caption{Same as for Fig.~\ref{res_hist_xray}, but for the three VHE bands, (a) $0.2-0.4$\,TeV, (b) $0.4-0.8$\,TeV and (c) $>~0.8$\,TeV.
\label{res_hist_vhe}}
\end{figure}

\subsection{Multi-band light curves from the theoretical model and comparison to the measurements}
\label{sec:modelling_results}

The optical and MeV-GeV emissions, which display small and slow variations, are well reproduced by the ``slow'' component, as shown in Fig.~\ref{stationary_states} and Fig.~\ref{stationary_states_2}. Therefore, this section focuses on the data/model comparisons of the light curves in the X-ray and VHE bands, which are dominated by the model parameters related to the ``fast'' component.\par 

From the broadband SEDs derived by our time-dependent theoretical model, we compute fluxes in the \mbox{$3-7$\,keV,} $7-30$\,keV, $30-80$\,keV, $0.2-0.4$\,TeV, $0.4-0.8$\,TeV and $>0.8$\,TeV bands, which are the same energy ranges used in the  analyses of the observational data from \textit{NuSTAR} and MAGIC/VERITAS. The fluxes are extracted by fitting a log-parabola function to the model SED curves (with a curvature $\beta$ let as a free parameter). The best-fit function is then integrated in each energy band to extract the flux in units of $1/\rm{cm}^2/s$ for the VHE, and $\rm{erg}/\rm{cm}^2/s$ for the X-rays. The VHE photon fluxes are extracted after including EBL absorption using the model of \citet{2011MNRAS.410.2556D}, which is the same template used throughout this work, and for the MAGIC spectral analysis (Sect.~\ref{sec:spectral_analysis}). Finally, we average the model fluxes over 15\,min temporal bins as done for the data (the code provides model curves in temporal steps of $\sim6$\,min in the observer's frame).\par

The results are presented in Fig.~\ref{xray_model_comp} and Fig.~\ref{vhe_model_comp} for the X-ray and VHE bands, respectively. The residuals, defined as $\frac{ (data - model)}{model}$, are given on the bottom panel of each light curve. Histograms of the residuals can be found in Fig.~\ref{res_hist_xray} and \ref{res_hist_vhe}.\par 

The model describes the data within 30\% for the vast majority of the bins in the $3-7$\,keV, $7-30$\,keV, $0.2-0.4$\,TeV, $0.4-0.8$\,TeV and $>0.8$\,TeV bands. The $3-7$\,keV and $7-30$\,keV bands are by far the ones that are the best reproduced. They respectively show a data/model mismatch below 15\% for $\approx90$\% and $\approx80$\% of the bins. This is an expected behavior given that the $l_e$ and $p$ variations are applied directly based on the $3-7$\,keV \textit{NuSTAR} measurements, these results provide an important validation of the procedure adopted for the parameter variations. Hence, any mismatch noticed in the other bands is likely caused by more fundamental aspects related to the model itself, like the choice of varying parameters and/or the initial modeling setup.\par 

For the highest X-ray energies, $30-80$\,keV, there is a data/model mismatch at level of 50\% on average during the first 5 days of the flare (MJD~56393 to MJD~56397). For the other days, the $30-80$\,keV emission is reasonably well captured by the model. The mismatch appears as a systematic underestimation of the flux, whereas the relative variations are well preserved by the model. The $30-80$\,keV range probes the most energetic part of the electron distribution, close to the cut-off $\gamma_{\rm max}$ (see Eq.~\ref{eq:pwl_cutoff}). The underestimation likely originates from an inaccurate modeling of the shape of the exponential cut-off. Because these electron energies are probed by the high-energy edge of the \textit{NuSTAR} bandwidth (covered by 1 or 2 spectral points at most), it is hardly possible to constrain a more complex shape of the cut-off than the one use here (see Eq.~\ref{eq:pwl_cutoff}). In Sect.~\ref{sec:discussion} we discuss ways to resolve this systematic difference.\par 

In the VHE bands, the data/model match is mostly within 30\%. Such a mismatch is also comparable to the systematics expected for MAGIC and Cherenkov telescopes in general for a Crab-like spectrum \citep{2016APh....72...76A}. A systematic underestimation of the fluxes is happening for the last few bins of MJD~56393. Over this time window, however, the $3-7$\,keV and $7-30$\,keV emission is well reproduced. Variations of additional parameters (on top of $p$ and $l_e$) are thus necessary to counteract the decrease of the VHE flux. It may also indicate the presence of an unaccounted, subdominant additional emitting region not included in the current scenario. We stress that tuning the parameters of the ``slow'' zone will not help in this regard, since the systematic underestimation is also present at $>0.8$\,TeV energies where the flux of the ``slow'' component is largely subdominant.\par

\begin{figure*}[htb!]
\gridline{\fig{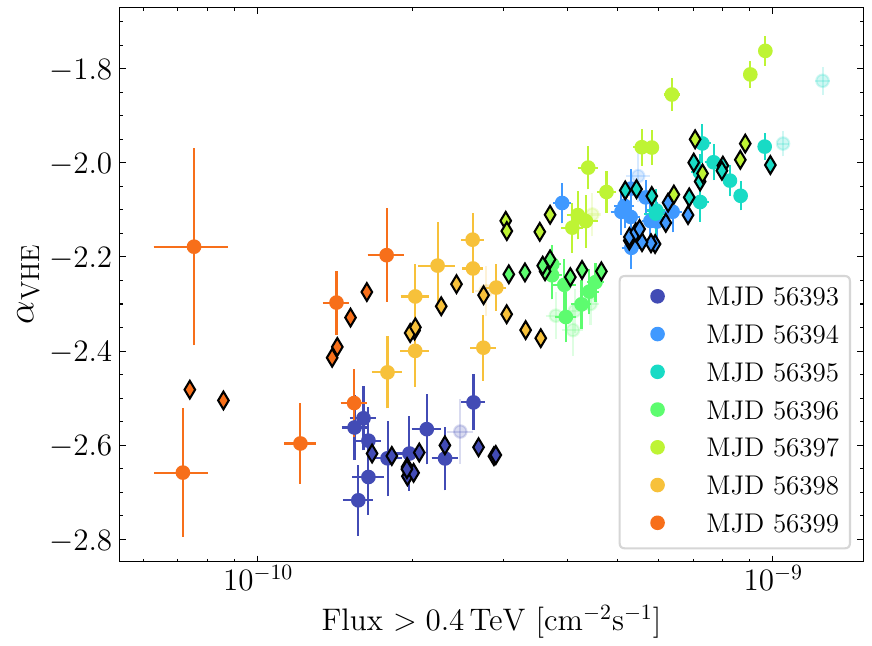}{0.48\textwidth}{(a)}
          \fig{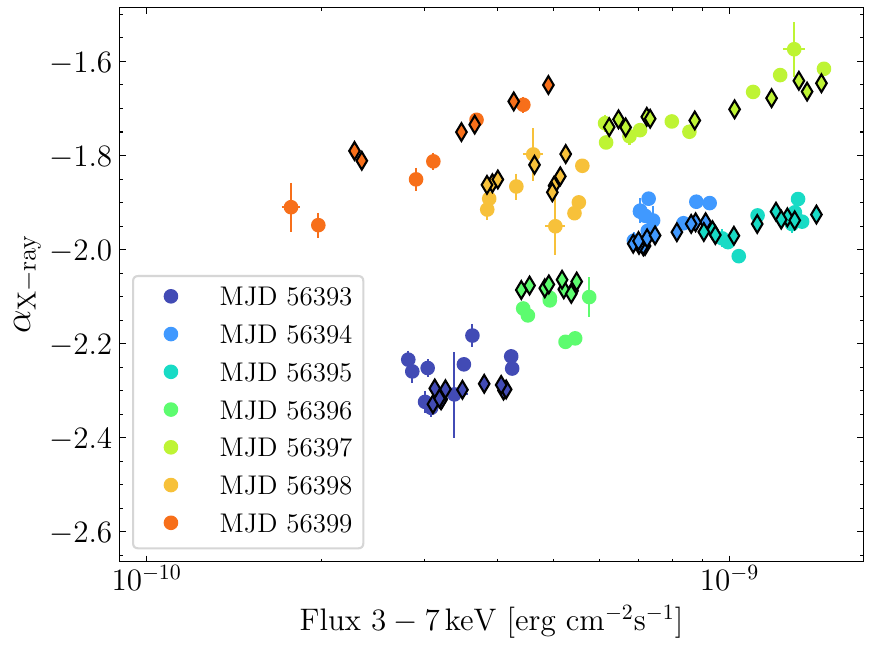}{0.48\textwidth}{(b)}
          }
\caption{Spectral parameter $\alpha$ as function of the flux for (a) MAGIC and (b) \textit{NuSTAR}, as obtained from the data (circle marker) and from the model (solid diamond marker). Transparent markers in the VHE band depict measurements for which MAGIC does not have a strictly simultaneous \textit{NuSTAR} measurements.} 
\label{alpha_vs_flux_model}
\end{figure*}

We now compare the spectral evolution of the model with the results of Sect.~\ref{sec:spectral_analysis}. To do so, we fit the model curves with the same spectral shape applied to the data, i.e. a log-parabola with curvature fixed to $\beta=0.38$ for the X-ray and $\beta=0.40$ for the VHE. At VHE, the fit also takes into account measurement uncertainties by simulating the MAGIC instrument response using the sensitivity curves from \citet{2016APh....72...76A}. Fig.~\ref{alpha_vs_flux_model} shows the photon indices $\alpha_{\rm X-ray}$ and $\alpha_{\rm VHE}$ plotted versus the $3-7$\,keV flux for the X-ray band and versus the $>400$\,GeV flux for the VHE band, respectively, as in Sect.~\ref{sec:spectral_analysis}. The comparison is only performed over simultaneous MAGIC/\textit{NuSTAR} time windows because, at the moment, the publicly available VERITAS results from this April 2013 flare of Mrk\,421 only relate to the multi-band light curves \citep{2017ICRC...35..641B}.\par 

\begin{figure*}[htb!]
\gridline{\fig{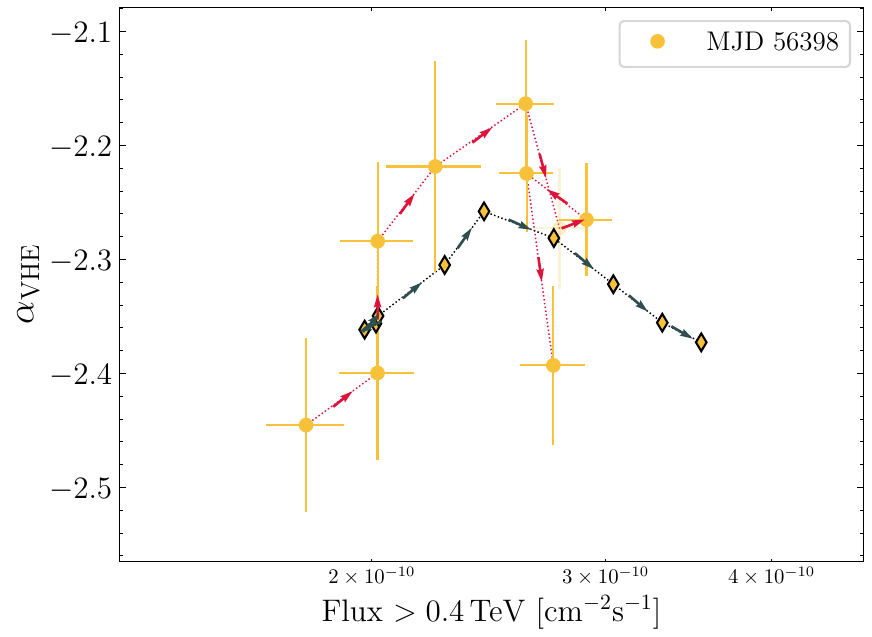}{0.48\textwidth}{(a)}
          \fig{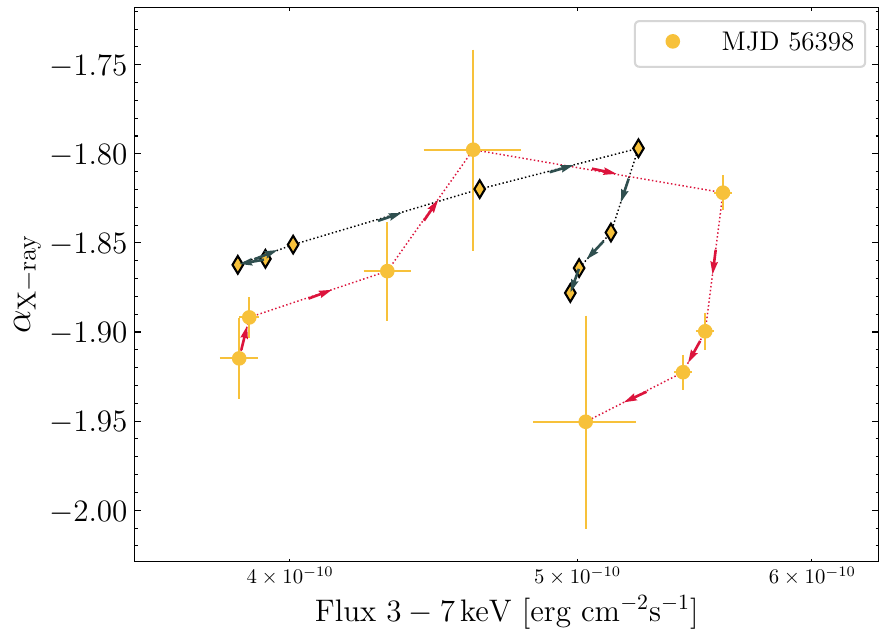}{0.49\textwidth}{(b)}
          }
\caption{Same as Fig.~\ref{alpha_vs_flux_model}, but zoomed on MJD~56398 during which an indication of intraday spectral hysteresis is measured. Red and black arrows show the direction of time for the data and the model, respectively. 
\label{alpha_vs_flux_model_56398zoom}}
\end{figure*}

In the X-rays, the model hardness is in good agreement with the data, and $\alpha$ differs at most by $\sim10$\% from the measurements. In fact, the great majority of the bins agree within 5\%. A similarly good agreement is obtained for the VHE band, except for MJD~56397. For the latter day, VHE spectra from the model are significantly softer than the observations. Panel (b) of Fig.~\ref{alpha_vs_flux_model} shows that the X-ray hardness is well reproduced for all the bins during MJD~56397, implying that the mismatch at VHE can hardly be solved by simply modifying the shape of the electron distribution. In Fig.~\ref{alpha_vs_flux_model_56398zoom}, we present a zoom on MJD~56398 during which an indication of an intraday hysteresis loop is visible.\par 

Finally, we compare the model broadband SEDs with observations from the radio to VHE. SEDs in strictly simultaneous bins of 15\,min are extracted from the \textit{NuSTAR}, MAGIC and VERITAS data. The VERITAS SEDs are obtained in three energy bins by making use of the light curves quoted in Paper~1: 0.2-0.4\,TeV \& 0.4-0.8\,TeV \& $>$0.8\,TeV. The SED point in each energy bin is placed at the mean energy weighted by the source spectral shape,  using the spectral shape averaged over the flare\footnote{Since we do not take into account short timescale spectral variability, we risk introducing some bias in the VERITAS SED points. Based on the spectral variability observed in the MAGIC data, we evaluate that this bias is expected to be at most at the level of the statistical uncertainty ($\sim10\%$) of the individual SED points.} ($dN/dE \propto E^{-2.14-0.45\log_{10}{(E/300\rm{GeV})}}$; see Paper~1). The \textit{Fermi}-LAT SED are averaged 12\,hours centered at the VHE measurements. Mrk~421 being relatively faint in the \textit{Fermi}-LAT band, this time binning provides a good trade-off between simultaneity and statistical uncertainty. At lower energies, SED points are computed in the radio, optical and UV bands using the light curves of Fig.~\ref{light_curve_flare}. We select observations that are the closest in time to the VHE/X-ray data, leading to a temporal offset of at most $\sim5$\,hours from the VHE/X-ray measurements, below the observed variability timescale.\par

In total, we obtain in total 310 broadband SEDs that extend over the 9-day flaring activity, namely from April 11 until April 19 (MJD 56393-56401). Out of these 310 SEDs, 240 lie inside time intervals that contain a VHE and X-ray coverage, i.e. the intervals over which we apply our model. For convenience purposes, we only provide in this manuscript snapshots for each of the days in Fig.~\ref{snapshot_timedep_sed} and Fig.~\ref{snapshot_timedep_sed_2} in Appendix~\ref{snapshot_sed_modelling}. The complete set of SEDs is published as online material, and can be retrieved from the following Zenodo repository \url{https://zenodo.org/records/17054582}.

\section{Discussion} \label{sec:discussion}

\subsection{Spectral hysteresis in the temporal evolution of the \mbox{X-ray} and VHE emissions}

The \textit{NuSTAR} data reveal hysteresis loops in clockwise direction (see Sect.~\ref{sec:spectral_analysis}). The loops occur on daily and sub-hour timescales, and are repeating during several days. Similar X-ray spectral behavior has already been reported in Mrk~421 and other HBLs \citep{1996ApJ...470L..89T, 2002ApJ...572..762Z, 2004A&A...424..841R, 2004ApJ...601..165F, 2008AIPC.1085..447S, 2016ApJ...831..102K, 2017ApJ...834....2A}. \citet{2016ApJ...831..102K} investigated spectral hysteresis in X-ray data from {\it Swift}-XRT and {\it NusTAR} during the period January 2013 to June 2013 and reported clockwise and counter-clockwise loops, underlining the complex character of X-ray variability on a wide range of timescales. At VHE, the MAGIC spectra closely follow the patterns observed in the X-rays over the simultaneous time windows (see Fig.~\ref{alphavsflux_magic} and Sect.~\ref{sec:spectral_analysis}). One of the most interesting features is the indication of a simultaneous X-ray/VHE clockwise loop on MJD~56398 (see Fig.~\ref{alphavsflux_magic}). Attempts to detect spectral hysteresis in VHE have been reported in the past \citep[see e.g.][]{2012A&A...539A.149H, 2017ApJ...834....2A}, but with no success, likely due to lack of statistics. To our knowledge, this is the first time that structured spectral trends going beyond the usual harder-when-brighter evolution are detected at VHE in a blazar flare. The similarities in the evolution of the VHE and X-ray hardness indicate the need for a common underlying particle population, hence favoring the SSC model \citep{1992ApJ...397L...5M}. \par

Hysteresis loops in clockwise direction are caused by a lag of the lower energy photons with respect to the higher energy ones. It can be interpreted as a signature of synchrotron cooling \citep{Kirk98}. Under this assumption, $\delta$ and $B$ are constrained as \citep{2002ApJ...572..762Z}:\par 
\begin{equation}
\label{eq:delay}
  B \delta^{1/3} = 209.91 \left( \frac{1+z}{E_l} \right)^{1/3} \left( \frac{1-(E_l/E_h)^{1/2}}{t_{delay}}   \right)^{2/3}
\end{equation}

where $t_{delay}$ is the time delay (in s), $E_l$ and $E_h$ the lower and higher photon energies (in keV) between which the delay is observed, B is the magnetic field strength in Gauss, and $z$ is the redshift. In order to make rough estimates, we set $E_l=3$\,keV and $E_h=15$\,keV, which are respectively the start energy and the logarithmic mean energy used in the \textit{NuSTAR} fits, and $t_{delay}\approx1$\,hr is the timescale of the hysteresis loops observed on MJD~56398. Setting $\delta=100$, as used in the ``fast'' zone, one finds $B\sim0.09$\,G. This is very close to the values adopted in the time-dependent model (see Table~\ref{table:steady}). Nonetheless, for such a magnetic field strength and assuming that electrons escape on timescales $t_{\rm esc} =R/c$ (where $R\sim10^{15}$\,cm is the radius of the ``fast'' zone) the synchrotron cooling break lies at $\log{\gamma_{\rm syn, break}}\sim6.2$, which is beyond $\gamma_{\rm max}$ for all days. Synchrotron cooling therefore does not significantly affect the flux variations in our model, and the evolution of the particle distribution is predominantly dictated by the acceleration/injection and escape process. Note that the particle cooling from inverse-Compton scattering remains sub-dominant.\par  

If the escape timescale is substantially increased to $t_{\rm esc}\sim10\,R/c$, synchrotron cooling may affect the shape of the electron distribution because, under such conditions, $\gamma_{\rm syn, break}\lesssim\gamma_{\rm max}$. In case the particles are contained in a blob of highly turbulent plasma, the escape timescale is related to the diffusion timescale \citep{2006ApJ...647..539B}, which can in principle be an order of magnitude longer than $R/c$. However, as argued in Sect.~\ref{sec:sec_stationary}, the large Doppler factors and the stability of the source parameters throughout the flare favor an emitting region behind a stationary shock. In such a scenario, the escape timescale is given by the advection velocity ($v_{\rm adv}$) of the plasma, leading to $t_{\rm esc} \sim R/v_{\rm adv} \sim R/c$ (since $v_{\rm adv}\sim c$). 

\subsection{Particle dynamics and acceleration processes}

The sub-hour variations in X-ray and VHE bands are well reproduced by varying two parameters: the electron luminosity and the slope of the injected distribution. The fluxes match the data within $\sim 30\%$, and for the vast majority of the time bins the photon index agree within $< 0.2$ (equivalent to $<10$\%; see Fig.~\ref{alpha_vs_flux_model}).\par

We can not exclude that other combinations of parameters in the ``fast'' zone are able to explain the \mbox{X-ray} and VHE variability. Nonetheless, variations of $p$ is a minimal requirement to explain the sub-hour spectral variations discussed in Sect.~\ref{sec:spectral_analysis}, and the changes in $p$ are also implied by the chromatic behavior of the X-ray/VHE correlation (see Paper~1). Furthermore, spectral variability can not be reproduced with synchrotron cooling only because the changes in the spectral indices are not monotonic. Variations of $\gamma_{\rm max}$ may lead to an evolution of the X-ray/VHE photon indices and fluxes \citep[as proposed in][]{1997A&A...320...19M, 2014A&A...571A..83P}, but in our case this is not sufficient. Indeed, from the stationary states (see Table~\ref{table:steady}), one finds $\log \gamma_{\rm max} \sim 5.2-5.4$. Due to the Klein-Nishina suppression effects, changes of $\gamma_{\rm max}$ around such high Lorentz factors will not be able to generate the intraday spectral variation of $\Delta \alpha_{\rm VHE} \approx 0.3-0.4$ measured at VHE (see Fig.\,\ref{alphavsflux_magic}).\par 

The evolution of $l_e$ (or equivalently $L_{\rm e}$) dominates the intraday flux variability in our model, but rapid fluctuations of $B$ may also have a significant contribution since $L_{\rm SSC} \sim L_{\rm syn} \sim B^2$. However, a smaller inverse-Compton variability will be induced compared to the scenario with varying $l_e$ given that $L_{\rm SSC} \sim L^2_{\rm syn} \sim l^2_e$. If only the magnetic field strength varies, it will thus underproduce the observed VHE variability and additional varying parameters would be needed to compensate this. In summary, our investigations favor $p$ and $l_e$ as the minimal combination of parameters able to capture the spectral and flux variations.

\begin{figure}[t!]
\plotone{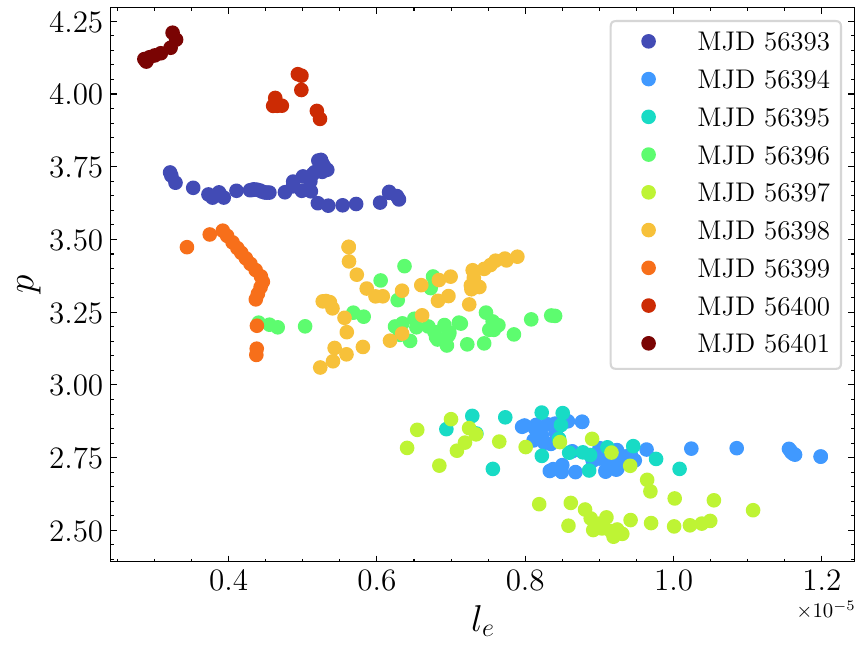}
\caption{Correlation between the $p$ and $l_e$ parameters of the electron distribution injected in the ``fast'' zone for the time-dependent modeling described in Sect.~\ref{sec:modelling}. The data are binned in 15-min intervals.
\label{p_vs_le}}
\end{figure}

We present in Fig.~\ref{p_vs_le} the correlation plot of $p$ and $l_e$. For clarity, the data points are binned over 15\,min similarly to the observed light curves. An evident (anti-)correlation is visible, implying a particle distribution that becomes harder when the electron density increases. This is expected from the general harder-when-brighter trend visible in the hardness versus flux plots presented in Sect.~\ref{sec:spectral_analysis}.\par

Magnetic reconnection is in general favored with respect to shocks to explain fast variability. A fraction of the magnetic energy is transferred to bulk kinetic energy of plasmoids formed in the reconnection layer, leading to the production of plasmoids moving at relativistic speeds with respect to the jet's comoving frame. As a result, the emitting regions have an effective Doppler factor reaching up to $\delta \sim 100$, as used here for the ``fast'' zone \citep{2009MNRAS.395L..29G, 2013MNRAS.431..355G, 2016MNRAS.462.3325P}. \citet{2023A&A...678A.140J} compared predictions of magnetic reconnection using particle-in-cell (PIC) simulations with the flux variability patterns observed in the 2013 flare of Mrk~421. The authors derived a spread in the model photon index of $\Delta \Gamma \approx 0.4$, equivalent to a spread in the electron power-law slope of $\approx0.8$ (see Eq.~\ref{eq:index_to_p}). This amplitude is significantly lower than the one derived in this work necessary to match the spectral changes in the data ($\Delta p=1.8$). We note that \citet{2023A&A...678A.140J} adopted a fixed magnetization\footnote{$\sigma=B^2/4\pi  m n c^2$, where $n$ is the electron number density} of $\sigma=50$, while the power-law slope of particles accelerated in reconnection events scales with the magnetization of the plasma \citep[see e.g.][]{2016MNRAS.462...48S}. Based on 2D PIC reconnection simulations presented in \cite{2014ApJ...783L..21S}, the particle distribution index approaches $p \approx 4$ for a plasma magnetization of $\sigma = 1$ and $p \approx 2$ for $\sigma = 10$. Since $\sigma \propto B^2$, it means that the magnetic field should vary by a factor of a few to explain the amplitude in $p$. In our model, $B$ varies by less than a factor of 2 between the days (see Table~\ref{table:steady}), hence somewhat at odds with a reconnection scenario. Finally, in Paper~1 we showed that to explain some of the broad features like the flux doubling time in the X-rays, $\sigma \approx 10$ had to be assumed (the bulk velocity of the plasmoids scales as $\sim \sqrt \sigma$), which would imply much harder indices $p$ that derived in our model.\par

Relativistic plasma shocks constitute an alternative efficient acceleration mechanism \citep{1999JPhG...25R.163K}. In this case, the slope of the particles sensitively depends on the shock compression ratio $r$, or equivalently the ratio $t_{\rm acc} / t_{\rm esc}$, where $t_{\rm acc}$ is the acceleration timescale \citep[][]{Kirk98, 2005PhRvL..94k1102K, 2005A&A...439..461V}. Using equation (2) from \citet{2005A&A...439..461V}, which is a good approximation of the index generated by relativistic parallel shocks of finite thickness, one finds that variations of $r$ by only a factor of $\sim 2$ can easily lead to modification of the slope from $p\approx 2.5$ to $p\approx4$ as in Fig.~\ref{p_vs_le}. From this point of view, diffusive shock acceleration is favored over magnetic reconnection.\par

As mentioned in Sect.~\ref{sec:sec_stationary}, the model parameters suggest emission regions from quasi-stationary features, which may be indicative of recollimation shocks. To stay subdominant in the optical band (as suggested by the absence of X-ray and optical correlation) and to satisfactorily describe the SED, the ``fast'' zone necessitate a minimum Lorentz factor of $\gamma_{\rm min} \sim 10^{4}$. As discussed in \citet{2021A&A...654A..96Z}, such large values of $\gamma_{\rm min}$ are in general possible in case of electron-proton co-acceleration at shocks, but is puzzling in the specific case of recollimation shocks. Indeed, in the latter scenario the minimum electron Lorentz factor is approximated as $\gamma_{\rm min}~\approx~1200\, \Gamma_{\rm j}/\delta_{\rm fast}$ \citep{2021A&A...654A..96Z}, where $\Gamma_{\rm j}$ is the jet Lorentz factor. This implies $\Gamma_{\rm j} \sim 10^3$, in tension with constraints from very-long-baseline interferometry (VLBI) observations that typically yield $\Gamma_{\rm j}$ of a few tens. We note however, that \cite{2021ApJ...923...67H} found extreme rare cases with $\Gamma_{\rm j} \sim 10^2-10^3$ on parsec scales in some AGN jets with MOJAVE and 2\,cm Survey programs at 15\,GHz. These extreme cases are mostly found among flat spectrum radio quasars, while BL Lac type objects seem to saturate at $\Gamma_{\rm j} \lesssim 10^2$.\par  

To relax these extreme jet requirements and at the same time explain the high $\gamma_{\rm min}$, a possibility is to invoke pre-acceleration processes. For instance, a sequence of standing shocks may successively increase $\gamma_{\rm min}$ as long as particles do not cool efficiently (and produce detectable radiation) when traveling between each shock \citep{2021A&A...654A..96Z}. \cite{2019ApJ...877...26H} also investigated this scenario to interpret the repetitive X-ray flaring episode in Mrk~421. Interestingly, the latter analysis constrained jet parameters to values consistent with the ``fast'' zone (i.e. $\Gamma_b \in [43 - 66], \; \delta>32$).

\subsection{Caveats and short comings of the model}

The model assumes two distinct, non-interacting emitting regions, possibly associated with stationary shocks. As argued in Sect.~\ref{sec:sec_stationary}, if the two regions are separated by more than a few $\sim 10^{16}$\,cm ($\sim10^{-2}$\,pc) it is well-justified to neglect in each component the interaction of particles with the radiative field from the other region. Therefore, such a configuration effectively leads to two ``one-zone'' SSC models. As shown in \cite{2016MNRAS.456.2374T}, large Doppler factors (higher than VLBI constraints) and low magnetic field are general features of one-zone models. A possible solution to solve this ``Doppler crisis'' is to consider a stratified region, where the plasma flows with a higher Lorentz factor close to the jet base and then drops further downstream the jet (as in a decelerating flow). The region may also be radially stratified and consist of a central component of the jet that possesses a fast plasma flow (the spine) surrounded by a slower layer (the sheath). In both cases, the interaction and the relative motion between the emitting regions could result in a lower $\delta$ and higher $B$, while capturing satisfactorily the synchrotron and inverse-Compton components. This was extensively discussed in earlier works, e.g. in \citet{2003ApJ...594L..27G} and \citet{2005A&A...432..401G}.\par

A time-dependent model with interacting emitting regions is beyond the scope of this paper, and it demands a significantly higher degree of complexity. Indeed, the particle distributions in the different components need to be evolved simultaneously after incorporating the mutual interplay of their radiation fields. Assuming that an interacting stratified scenario indeed allows for a larger $B$ compared to our model, the synchrotron cooling timescale may be sufficiently shortened to impact the evolution of the particle distribution, thus modifying the conclusions from the previous section regarding the origin of the hysteresis loops (which we attribute to the acceleration/injection process). An increase of $B$ by a factor of $\sim 3$ (while keeping $R\sim 10^{15}$\,cm and $t_{\rm esc} \sim R/c$), is sufficient to obtain $\gamma_{\rm syn, break}\lesssim\gamma_{\rm max}$. \citet{2020ApJS..247...16A} showed such difference in $B$ is possible between homogeneous and spine-sheath scenarios. However, a dedicated modeling is necessary to precisely quantify to which extend $\delta$ and $B$ can be modified with respect to Table~\ref{table:steady}.\par

We anticipate that a stratified/interacting model will only marginally affect most of our conclusions. First, the correlation between $l_e$ and $p$ in Fig.~\ref{p_vs_le} will remain largely unchanged, as well as the need of $\gamma_{\rm min} \sim 10^4$ in the ``fast'' zone to stay subdominant in the optical band. Moreover, assuming very conservatively that a stratified region allows for a reduction of $\delta_{\rm fast}$ by an order of magnitude, an emitting region would travel on $\Delta d \sim \frac{\Gamma_{\rm b, fast}^2 \delta_{\rm fast} c \Delta T_{\rm obs}}{1+z} \sim 10^{18}$\,cm over the flaring episode, equivalent to $\sim$\,pc scales, which is again at odds with the stability of the parameters of the stationary states. Hence, recollimation shocks would still be favored as opposed to a blob-in-jet scenario. Finally, an order of magnitude lower Doppler factor demands $\Gamma_{\rm j}\sim10^2$ to obtain $\gamma_{\rm min} \sim 10^4$ ($\gamma_{\rm min}~\approx~1200\, \Gamma_{\rm j}/\delta_{\rm fast}$) for electron-proton co-acceleration on a recollimation shock, which is still higher than VLBI observations.\par 

On MJD~56397, the VHE hardness predicted by the model is clearly softer than in the data (see left panel of Fig.~\ref{alpha_vs_flux_model}). Since the X-ray spectrum is well reproduced (see right panel of Fig.~\ref{alpha_vs_flux_model}), the mismatch at VHE can not be resolved by only modifying the parameters of the electron distribution. Instead, this may be indicative of the emergence of short-lived additional emitting region filled with highly energetic electrons, which is currently ignored in our model. The origin of these extra components remain however unclear. Ultra-fast plasmoids (that are short-lived for the observer) can be easily produced in a magnetic reconnection scenario \citep{2009MNRAS.395L..29G, 2013MNRAS.431..355G, 2016MNRAS.462.3325P}. On the other hand, as already mentioned, our model parameters (and their stability over time) tends to favor shocks as acceleration mechanism.\par 

We conclude this section by discussing a general short coming of our model: a systematic underestimation of the 30-80\,keV fluxes from \textit{NuSTAR}. Although the relative flux variations are well preserved by the model, the fluxes are lower than the observations by roughly 50\% on average from MJD~56393 to MJD~56397. Given that the X-ray photon indices are well reproduced by the model (within 10\%; see Fig.~\ref{alpha_vs_flux_model}), this mismatch points towards a non-optimal description of the electron distribution close to $\gamma_{\rm max}$. Although we deem the systematic flux offset relatively mild in light of the complexity of the emission patterns during the flare and that the spectral shapes and the rest of the fluxes in the other sub-energy bands are well reproduced, these \textit{NuSTAR} observations constitute one of the very few data sample providing a precise measurement of the 30-80\,keV sub-hour variability in HBLs. Hence, we investigated alternative parametrization of the exponential cut-off to see if the discrepancy can be resolved.\par 

To enhance the synchrotron emission above 30\,keV, we softened the sharpness of the cut-off by lowering $a$ in the electron distribution ($\propto \gamma^{-p} \, \exp{\left(-(\gamma / \gamma_{\rm max})^{a}\right)}$, see Eq.~\ref{eq:pwl_cutoff}). We find that $a=1.5$ instead of $a=2$ provides a reasonable description of the 30-80\,keV fluxes together with the other X-ray and VHE bands. However, the X-ray spectral index $\alpha_{\rm X-ray}$ becomes systematically harder by $\Delta \alpha_{\rm X-ray} \approx 0.15$. Thus, $a=1.5$ results in an overall degradation of the X-ray spectra compared to the $a=2$ model (see Fig.~\ref{alpha_vs_flux_model}). The VHE spectra remained unchanged, mostly due to the Klein-Nishina suppression of the inverse-Compton scattering from electrons close to the cut-off. Additionally, we enhanced the \mbox{30-80\,keV} emission after increasing $\gamma_{\rm max}$ by 0.1\,dex (being the particle grid resolution used in the code). As in the previous case, the 30-80\,keV fluxes can be well captured but the X-ray spectra becomes significantly harder than the data.\par 

In conclusion, an optimal description of the \mbox{30-80\,keV} fluxes (i.e. within $\lesssim 50\%$) can not be achieved with the current parameters of our model if one wants to preserve the good description of the \textit{NuSTAR} photon indices. A more complex shape of the high-energy cut-off in the electron distribution may be needed, but since \mbox{the 30-80\,keV} energy range is only covered by 1 or 2 \textit{NuSTAR} SED points, a model with more degrees of freedom than Eq.~\ref{eq:pwl_cutoff} can hardly be constrained. Nonetheless, it is interesting to note that our model prefers $a\approx2$ rather than $a\approx1$. This is in agreement with results from \citet{2024PhRvD.109j3039B} that fitted a sample of \textit{NuSTAR} observations from Mrk~421.

\section{Conclusions} \label{sec:summary}

The TeV blazar Mrk~421 exhibited substantial flaring activity in April 2013, marking one of the largest outbursts recorded to date. This event was extensively monitored over nine consecutive days (April 11–19) by MAGIC, VERITAS, and \textit{NuSTAR}, in conjunction with observations from \textit{Swift}, as well as optical and radio telescopes. The intraband correlations and variability were thoroughly analyzed in Paper~1. In the study reported in this paper, we extended our work by characterizing the simultaneous VHE and X-ray spectral evolution. We subsequently modeled the broadband evolution on a 15-minute scale using a time-dependent approach to accurately track the history of the particle distribution.

The high flux and rich multiwavelength coverage enabled the construction of 240 broadband SEDs throughout the 9-day flaring period, on which the model was applied. We stress that, in most of the previous works on blazar flares that were modelled with a time-independent radiation code, the time variations were described on daily or longer timescales \citep[see e.g.][]{2017ApJ...834....2A, 2020A&A...633A.162H, 2020A&A...637A..86M, 2021A&A...647A.163M}. The key findings from our work are as follows: 

\begin{itemize}

     \item The \textit{NuSTAR} data in the X-ray band reveal spectral hysteresis loops in a clockwise direction at various intervals during the flare, indicating a behavior that exceeds a simple harder-when-brighter correlation.

    \item In the VHE band, MAGIC spectra extend to energies exceeding 10\,TeV. The evolution of spectral hardness in the VHE band closely follows that in the \mbox{X-ray} band. For the first time, we report evidence of concurrent clockwise VHE spectral hysteresis alongside the X-ray emission. The simultaneous spectral trends in the X-ray and VHE bands strongly support leptonic models for the origin of the gamma-ray emission.

    \item Our time-dependent leptonic modeling demonstrates that the majority of the sub-hour variations in flux and spectral characteristics can be explained by changes in the luminosity and the slope of the injected electron distribution.

    \item Our model indicates that the magnetic field remains stable, varying by less than a factor of two, while the size of the emitting region remains constant. The small variations of the source environment support an emitting region behind a standing shock within the jet, such as recollimation shocks.

    \item The model's derived variations in the slope of the electron distribution could occur within a shock-acceleration scenario if the shock compression ratio changes by a factor of approximately 2. In contrast, a magnetic reconnection model would necessitate significant changes in plasma magnetization, which contradicts the stable magnetic field characteristics observed in our model.

     \item To accurately reproduce the VHE data at the highest energies, a Doppler factor $\delta$ close to 100 is required. In a standing shock scenario, as indicated by the model parameters, this necessitates a jet Lorentz factor of $\Gamma_{\rm j} \sim 10^3$ to achieve the necessary $\gamma_{\rm min} \sim 10^4$. This large $\gamma_{\rm min}$ is suggested by the absence of strong variability in the optical/UV bands and their lack of correlation with the highly variable \mbox{X-rays.} The required high jet velocity of $\Gamma_{\rm j} \sim 10^3$ significantly exceeds typical values obtained from VLBI observations. While the Doppler factor could be reduced by assuming interactions between the radiation fields of compact and extended regions, even in this scenario, a Lorentz factor of at least $\Gamma_{\rm j} \sim 100$ would still be necessary within the standing shock framework to achieve $\gamma_{\rm min} \sim 10^4$. 
    
\end{itemize}


\begin{acknowledgments}
We would like to thank the Instituto de Astrof\'{\i}sica de Canarias for the excellent working conditions at the Observatorio del Roque de los Muchachos in La Palma. The financial support of the German BMFTR, MPG and HGF; the Italian INFN and INAF; the Swiss National Fund SNF; the grants PID2019-107988GB-C22, PID2022-136828NB-C41, PID2022-137810NB-C22, PID2022-138172NB-C41, PID2022-138172NB-C42, PID2022-138172NB-C43, PID2022-139117NB-C41, PID2022-139117NB-C42, PID2022-139117NB-C43, PID2022-139117NB-C44, CNS2023-144504 funded by the Spanish MCIN/AEI/ 10.13039/501100011033 and "ERDF A way of making Europe; the Indian Department of Atomic Energy; the Japanese ICRR, the University of Tokyo, JSPS, and MEXT; the Bulgarian Ministry of Education and Science, National RI Roadmap Project DO1-400/18.12.2020 and the Academy of Finland grant nr. 320045 is gratefully acknowledged. This work has also been supported by Centros de Excelencia ``Severo Ochoa'' y Unidades ``Mar\'{\i}a de Maeztu'' program of the Spanish MCIN/AEI/ 10.13039/501100011033 (CEX2019-000920-S, CEX2019-000918-M, CEX2021-001131-S) and by the CERCA institution and grants 2021SGR00426 and 2021SGR00773 of the Generalitat de Catalunya; by the Croatian Science Foundation (HrZZ) Project IP-2022-10-4595 and the University of Rijeka Project uniri-prirod-18-48; by the Deutsche Forschungsgemeinschaft (SFB1491) and by the Lamarr-Institute for Machine Learning and Artificial Intelligence; by the Polish Ministry Of Education and Science grant No. 2021/WK/08; and by the Brazilian MCTIC, the CNPq Productivity Grant 309053/2022-6 and FAPERJ Grants E-26/200.532/2023 and E-26/211.342/2021.
Axel Arbet-Engels and David Paneque acknowledge support from the Deutsche Forschungs gemeinschaft (DFG, German Research Foundation) under Germany’s Excellence Strategy – EXC-2094 – 390783311. Maria Petropoulou acknowledges support from the Hellenic Foundation for Research and Innovation (H.F.R.I.) under the ``2nd call for H.F.R.I. Research Projects to support Faculty members and Researchers'' through the project UNTRAPHOB (Project ID 3013).   
\end{acknowledgments}

\section*{Data availability} \label{sec:data_avail}

The broadband SEDs (from radio to TeV) in the 15-minute intervals throughout the entire flare are made available at the Zenodo repository \url{https://zenodo.org/records/17054582}. The repository also includes the SED models and the comparison with the data for all 15-minute bins together with an animation illustrating the temporal evolution.\par

All the MAGIC data used in this work are further released in the Data Level 3 (DL3) format in order to be analyzed with the \texttt{gammapy} open source software \citep{2023A&A...678A.157D}. The DL3 files are available at the Zenodo repository \url{https://zenodo.org/records/17064461} and they are released as part of the ``ACME'' MAGIC Open Data-Analysis School\footnote{\url{https://acme-magic-odas.sciencesconf.org/}}. Tutorials to process the DL3 files can be found in the following GitHub repository \url{https://github.com/magic-telescopes/acme_magic_odas/blob/master/3_light_curve_variable_source.ipynb}.

\section*{Author contributions} \label{sec:author_contrib}
A.~Arbet-Engels: project leadership, MAGIC, \textit{Fermi}-LAT \& \textit{NuSTAR} analysis, modeling, paper drafting; A.~Babi\'c: MAGIC analysis crosscheck; D.~Paneque: organization of the observations, coordination of the multiwavelength data reduction, paper drafting; M.~Petropoulou: modeling, paper drafting; M.~Polkas: modeling, paper drafting. The rest of the authors have contributed in one or several of the following ways: design, construction, maintenance, and operation of the instrument(s) used to acquire the data; preparation and/or evaluation of the observation proposals; data acquisition, processing, calibration, and/or reduction; production of analysis tools and/or related MC simulations; overall discussions about the contents of the draft, as well as related refinements in the descriptions.

%

\facilities{MAGIC, VERITAS, \textit{NuSTAR}, \textit{Swift}-XRT, \textit{Swift}-UVOT, GASP-WEBT, OVRO, IRAM 30-m telescope, Mets\"ahovi}


\software{MARS \citep{zanin2013},
          XSpec \citep{1996ASPC..101...17A},
          NuSTARDAS,
          XRTDAS,
          Fermitools (\url{https://fermi.gsfc.nasa.gov/ssc/data/analysis/})
          }



\appendix

\section{Observations and data processing} \label{sec:appendix_analysis}

In the radio, the observations were performed by the OVRO, Mets\"ahovi and IRAM 30-m telescopes. In the optical, the data were taken in the R-band by telescopes belonging to the GLAST-AGILE Support Program \citep[GASP, e.g.][]{villata2008, villata2009, 2017MNRAS.472.3789C} of the Whole Earth Blazar Telescope\footnote{http://www.oato.inaf.it/blazars/webt/} \citep[WEBT, e.g.][]{villata2002, villata2006, rai2007,raiteri2017}. The measurements in the UV regime were performed by the Swift UV and Optical Telescope \citep[UVOT,][]{2005SSRv..120...95R} in the UVW1, UVM2 and UVW2 spectral bands. The corresponding fluxes were directly taken from Paper~1, and the data reduction method was reported in \citet{2016ApJ...819..156B}. \par  

The soft X-rays (0.3-10\,keV) measurements are from the \textit{Swift} X-ray Telescope \citep[XRT;][]{2005SSRv..120..165B} and the data processing is described in \citet{2016ApJ...819..156B}. 
In the hard X-ray band (3-79\,keV), \textit{NuSTAR} observed for about 70\,hours during this flare. With respect to Paper~1, the data were reprocessed with the NuSTARDAS software (version 2.1.1) using an updated CALDB version 20220215, which includes the major update in the effective area calibration released in October 2021. As described in \citet{2021arXiv211011522M}, this update leads to a systematic increase of the flux by $\sim5-10\%$ with respect to older CALDB versions. The events were screened from the effect of the south Atlantic anomaly using the options \texttt{tentacle=yes} and \texttt{saamode=optimized} in the \texttt{nupipepline} run. The source region was defined as a circular shape with a radius of $\sim$100\,arcsec centered at the target. The background was extracted from an annulus around the target with an inner radius of $\sim$120\,arcsec and outer radius of $\sim$220\,arcsec. To build broadband SEDs and calculate the fluxes, both \textit{Swift}-XRT and \textit{NuSTAR} data were independently fitted in \texttt{Xspec} \citep{1996ASPC..101...17A} with a log parabola model and a column density $N_{\rm H}=1.92\times10^{20}$\,cm$^{-2}$ \citep{2005A&A...440..775K}. The spectra from the two \textit{NuSTAR} focal plane modules (FPMA \& FPMB) were modeled simultaneously adopting a cross-calibration normalization factor as a free parameter in the \texttt{Xspec} model. The resulting cross-normalization factor was almost exclusively in the order of a few percents, firmly within the expected systematics \citep{2013ApJ...770..103H}. We fixed the pivot energy to 1\,keV, as commonly adopted for \textit{Swift}-XRT analysis, and to ease the comparison of the spectral parameters with recent works \citep[e.g.][]{2021A&A...655A..89M}. We have verified that a pivot energy closer to the center of the \textit{NuSTAR} bandwidth (for instance 10\,keV) does not change the \textit{NuSTAR} fitting results. \par 

The Large Area Telescope (LAT) on board the \textit{Fermi Gamma-ray Space Telescope} \citep[\textit{Fermi}-LAT;][]{2009ApJ...697.1071A,2012ApJS..203....4A} was used to characterize the emission in the MeV-GeV band. The \textit{Fermi}-LAT data were reduced in a similar fashion as in Paper~1, using the same analysis cuts and selecting the same region of interest (ROI) of $10^\circ$ around Mrk~421. We adopted an updated version of the instrument response functions, P8R3\_SOURCE\_V3 instead of P8R3\_SOURCE\_V2 that was used in Paper~1. The diffuse Galactic and isotropic extragalactic background models\footnote{http://fermi.gsfc.nasa.gov/ssc/data/access/lat/\\BackgroundModels.html} were included using the files gll\_iem\_v07.fits and iso\_P8R3\_SOURCE\_V3\_v1.txt, respectively. For the SEDs, we computed spectral points in each energy bin reaching a test statistics \citep[TS;][]{1996ApJ...461..396M} above 5, and calculated an upper limit at a 95\% confidence level otherwise. The data were reduced with the analysis software \texttt{FERMITOOLS}\footnote{https://fermi.gsfc.nasa.gov/ssc/data/analysis/} (version 2.0.8).\par 

At VHE, the MAGIC \citep{2016APh....72...76A} observations were analized using the standard analysis tools from the MAGIC Analysis and Reconstruction Software (MARS) package \citep{zanin2013, 2016APh....72...76A}. More details on the MAGIC observations and analysis can be found in Paper~1. In this work, we simply extend the analysis of Paper~1 to extract SED points and spectral parameters at VHE. The spectra were fitted using a log-parabola model:
\begin{equation}
\label{eq:logparabola_MAGIC}
    \frac{dN}{dE} = f_0 \left(\frac{E}{E_0}\right)^{\alpha-\beta \log_{10}{\left(\frac{E}{E_0}\right)}}
\end{equation}
where $f_0$ is the normalization constant, $\alpha$ the photon index and $\beta$ the curvature parameter. The normalization energy $E_0$ was set to 500\,GeV. All the SEDs and spectral parameters were evaluated after correcting for the extragalactic background light (EBL) absorption effects using the template described in \citet{2011MNRAS.410.2556D}. Results of the spectral fits are presented in Section~\ref{sec:spectral_analysis}.\par 

In order to extend and maximize the temporal coverage at VHE, we also make use of the VERITAS \citep{2008AIPC.1085..657H} data published in Paper~1. Thanks to their different geographical locations, VERITAS and MAGIC observe one after another with a small overlap. This provides continuous VHE coverage over up to $\sim10$\,hours per day. As detailed in Paper~1, an energy-dependent systematic offset reaching up to 25-30\% was found between the MAGIC and VERITAS fluxes. Such an offset is within the expected instrumental systematics that are dominated by the uncertainty on the absolute energy scale. In fact, by assuming a shift by 20\% of the MAGIC absolute energy scale the SEDs from both instruments are compatible. To correct for the offset, the VERITAS fluxes in Paper~1 in each energy band were renormalized according to scaling factors determined assuming the 20\% shift in the energy scale. In this work, we use the VERITAS light curves using the identical correction method.\par 

\begin{figure*}[ht!]
\gridline{\fig{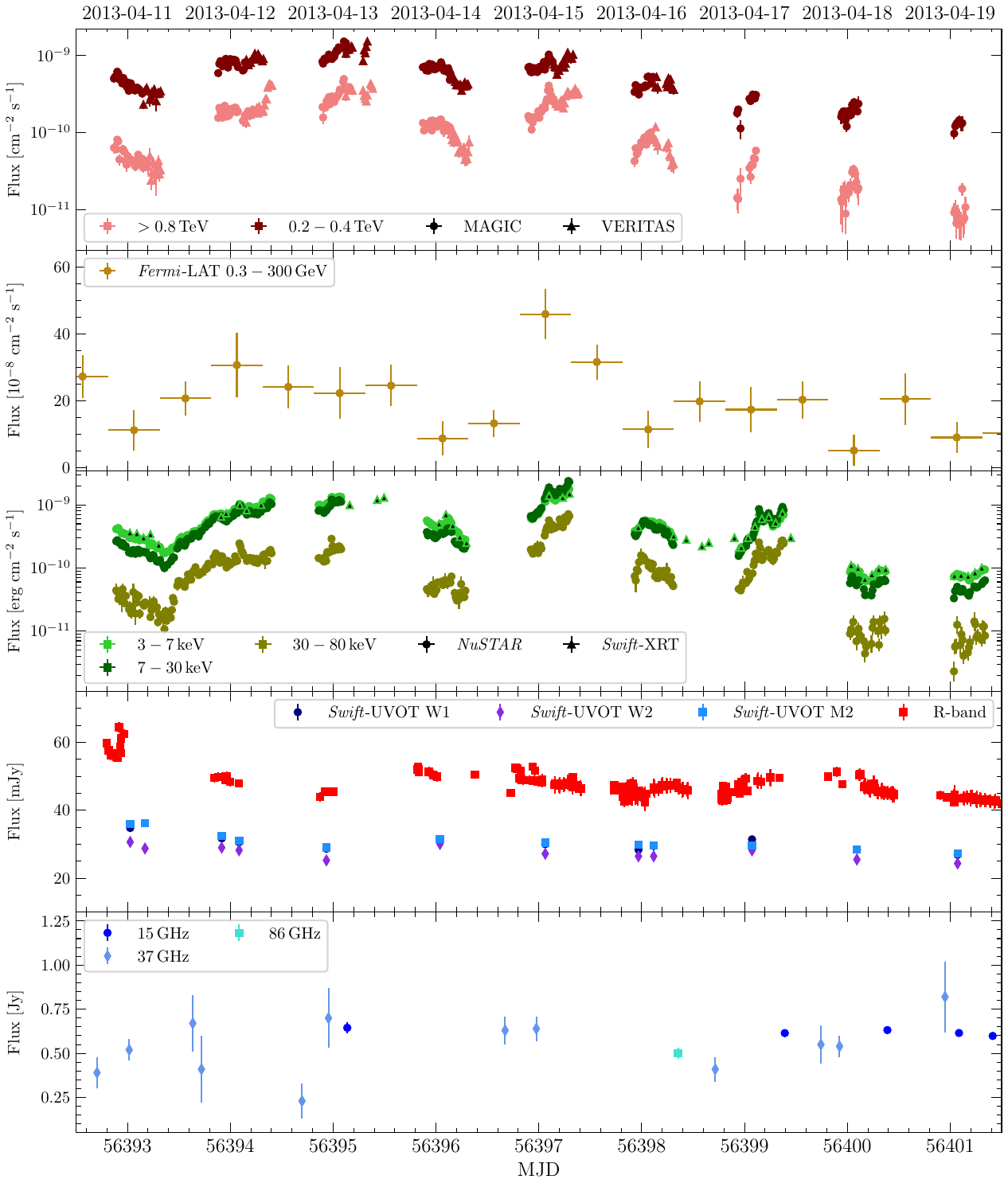}{0.9\columnwidth}{}}
\caption{Multiwavelength light curves from VHE to radio frequencies during the April 2013 flare. Fluxes from MAGIC, VERITAS, \textit{Swift}-XRT, \textit{Swift}-UVOT, optical telescopes (R-band) and radio telescopes (15\,GHz, 37\,GHz and 86\,GHz) are directly taken from Paper~1. \textit{Fermi}-LAT and \textit{NuSTAR} fluxes have been updated using more recent instrument response functions with respect to Paper~1 (see text in Sect.~\ref{sec:appendix_analysis} for more details). The MAGIC, VERITAS, \textit{NuSTAR}  and \textit{Swift}-XRT light curves are binned in 15-minute intervals. The \textit{Fermi}-LAT light curve is sampled homogeneously in identical 12-hours intervals. This binning differs slightly from the one adopted in Paper~1, where the data were first divided into 12-hours intervals centered exactly at the VHE observations and then complemented by intervals where there are no VHE observations. The rest of the light curves are binned observation-wise. \label{light_curve_flare}}
\end{figure*}

The flux behavior was extensively discussed in Paper~1. For completeness, we display in Fig.~\ref{light_curve_flare} the multiwavelength light curves to provide a comprehensive view of the flux evolution during the flare.\par

\section{VHE/X-ray spectral behavior evolution - complementary information and figures}
\label{sec:VHE_spectral_analysis_app}

\begin{figure}[h!]
\gridline{\fig{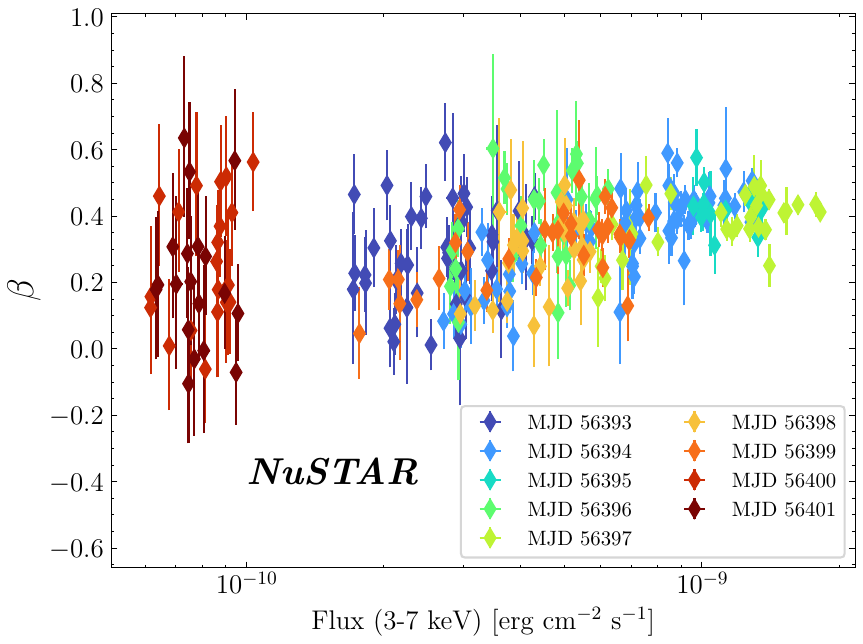}{0.495\textwidth}{(a)}
          \fig{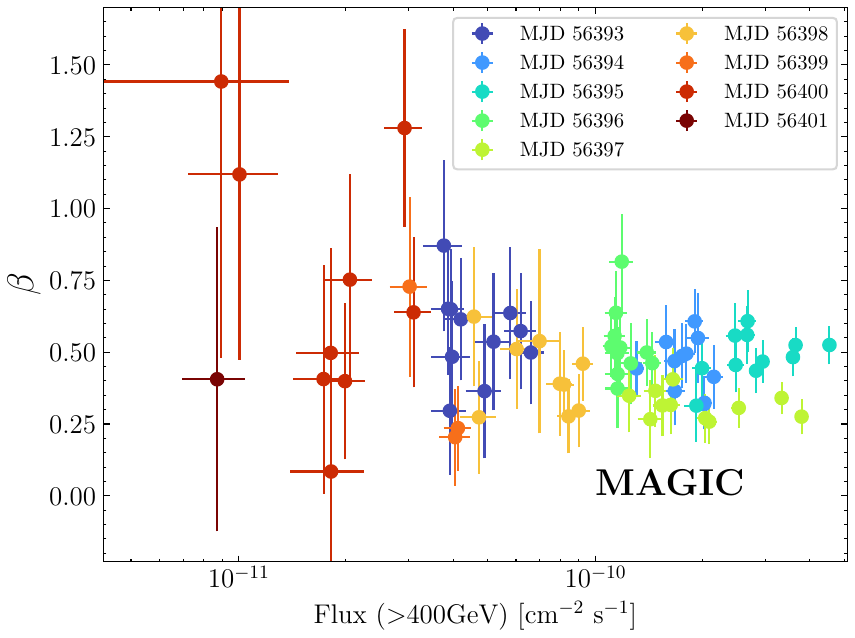}{0.489\textwidth}{(b)}
          }
\caption{Spectral parameter $\beta$ (curvature) from log-parabola fits versus flux for (a) \textit{NuSTAR} and (b) MAGIC. The spectral fits are performed in 15-minute bins for \textit{NuSTAR} and 30-minute for MAGIC. The markers are color-coded according to the respective day of the observation.
\label{betavsflux}}
\end{figure}

In Sect.~\ref{sec:spectral_analysis}, the \textit{NuSTAR} and MAGIC spectra were fitted using a log-parabola model in which the curvature parameter $\beta$ was fixed (see Eq.~\ref{eq:logparabola_MAGIC}). As in such a model the correlation between $\beta$ and the spectral index $\alpha$ is cancelled, $\alpha$ directly quantifies the hardness evolution. In order to determine the fixed value of $\beta$, we first ran a series of fits using a ``$\beta$-free'' model. We then fixed $\beta$ to the average value of all time bins, yielding $\overline{\beta}_{\rm X-ray}=0.38$ and $\overline{\beta}_{\rm VHE}=0.40$ for \textit{NuSTAR} and MAGIC, respectively. Fig.~\ref{betavsflux} presents $\beta_{\rm X-ray}$ versus the 3-7\,keV flux (left), and $\beta_{\rm VHE}$ versus the $>400$\,GeV flux (right). The average is plotted with an black dashed line. For both energy ranges, $\beta$ shows no clear correlation with the flux. Based on a $\chi^2$ test, we find that for both MAGIC and \textit{NuSTAR} the ``$\beta$-fixed'' model is consistent with the data within $3\sigma$ for all time bins, except two bins (out of 286) in the \textit{NuSTAR} data. \par

\begin{figure}[t!]
\epsscale{0.5}
\centering
\gridline{\fig{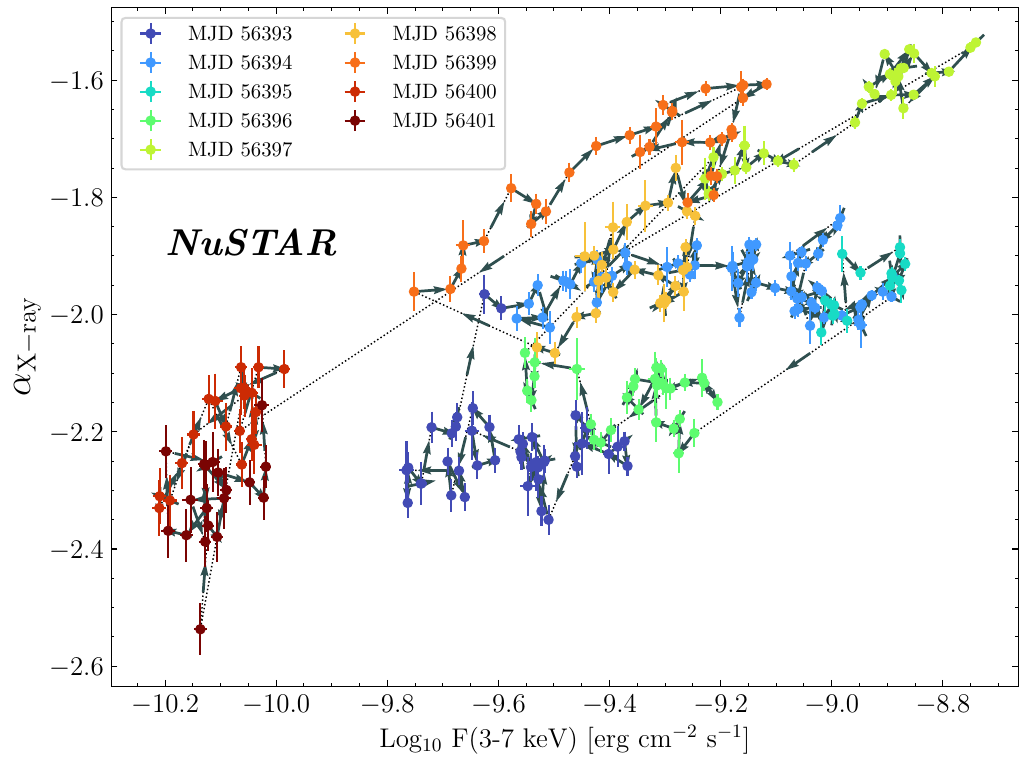}{0.485\textwidth}{(a)}
          \fig{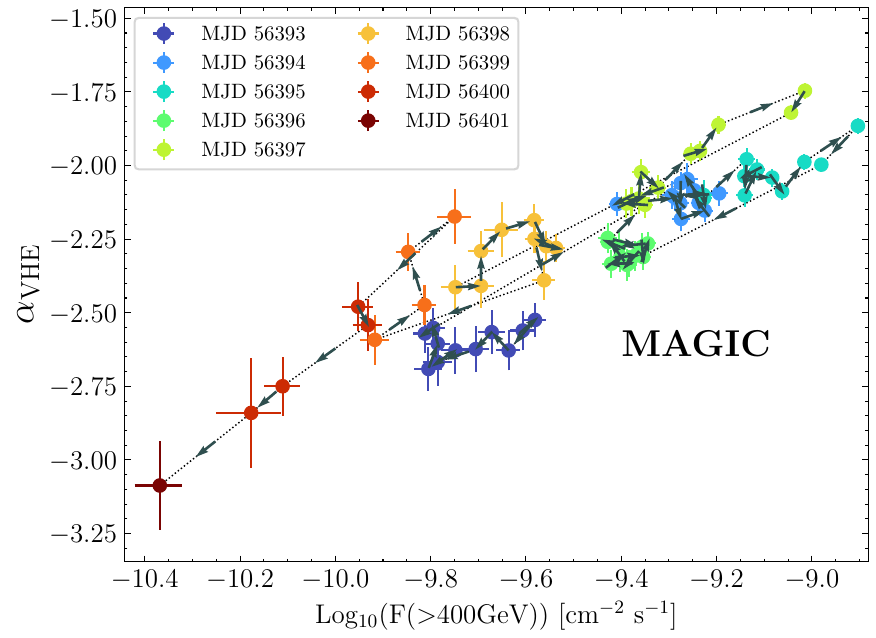}{0.495\textwidth}{(b)}
          }
\caption{Spectral parameter $\alpha$ versus flux for (a) \textit{NuSTAR} and (b) MAGIC for the entire flare, up to MJD~56401. The \textit{NuSTAR} fluxes are evaluated in the 3-7\,keV bands, while for MAGIC they are computed above 400\,GeV. The fits are performed in 15-minute bins for \textit{NuSTAR} and 30-minute for MAGIC. The markers are color-coded according to the respective day of the observation.
\label{alpha_vs_flux_nustar_alldays}}
\end{figure}

For clarity purposes and to focus on the days displaying the most complex spectral behavior, the dependence of $\alpha$ on the flux was only shown between MJD~56393 and MJD~56399 in Sect.~\ref{sec:spectral_analysis}. We now present in Fig.~\ref{alpha_vs_flux_nustar_alldays} the evolution of $\alpha$ for MAGIC and \textit{NuSTAR}, including MJD~56400 and MJD~56401. The last two days show a significantly dimmer flux as well as a softer VHE spectrum with respect to the rest of the flare.

\section{Stationary state modeling with lower Doppler factor in the ``fast'' zone}
\label{sec:doppler_comp_steadystate}

We argued in Sect.~\ref{sec:modelling_setup} that a Doppler factor of at least $\approx100$ is necessary for the ``fast'' component to capture the X-ray and VHE hardness during the flare. Such a Doppler factor is significantly higher than those typically adopted for Mrk~421 in the literature, which are in the range 20-40 \citep[see e.g.][]{2011ApJ...736..131A, 2016ApJ...819..156B}. A high Doppler factor is required due to the large separation between the SED components and to overcome the Klein-Nishina effect, which softens the spectra at $\sim$TeV energies. A brief analytical justification is performed in Sect.~\ref{sec:modelling_setup}. As a further illustration, we show in this section the stationary state of MJD~56393 and MJD~56395 when modeled with a ``fast'' zone having $\delta_{\rm fast}=50$ and compare it with the one using $\delta_{\rm fast}=100$.\par 

\begin{figure}[t!]
\epsscale{0.5}
\centering
\gridline{\fig{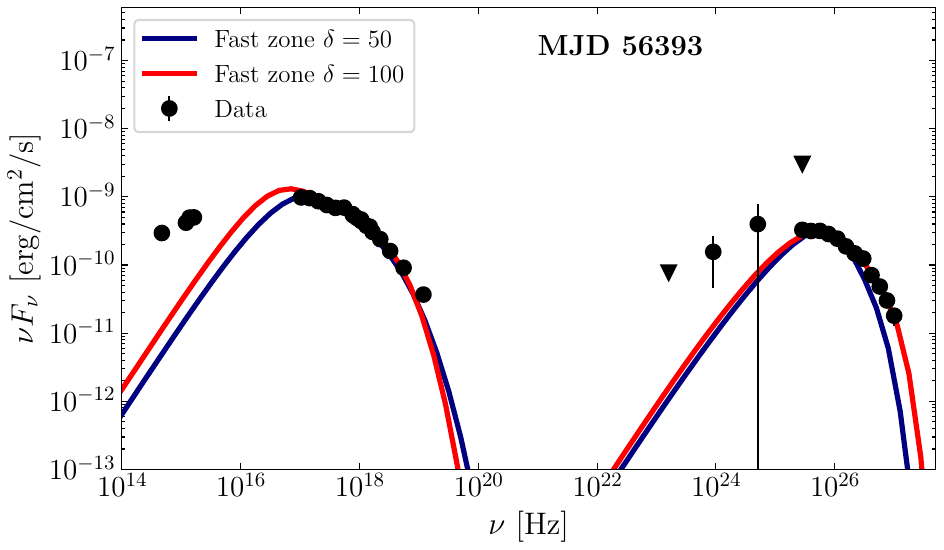}{0.495\textwidth}{(a)}
          \fig{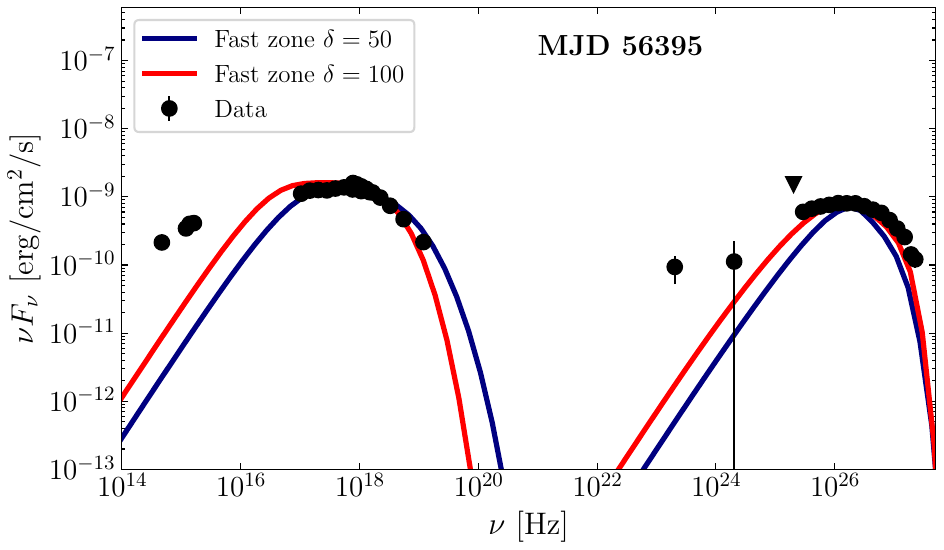}{0.495\textwidth}{(b)}
          }
\caption{Comparison of the ``fast'' component steady state when using $\delta_{\rm fast}=50$ and $\delta_{\rm fast}=100$ for MJD~56393 (left) and MJD~56395 (right). The model parameters for the $\delta_{\rm fast}=100$ case are listed in Table~\ref{table:steady}. Regarding the $\delta_{\rm fast}=50$ case, $\log_{10}l_e=-4.9$, $\log_{10}{\gamma_{\rm min}}=4.4$, $\log_{10}{\gamma_{\rm min}}=5.2$ and $B=0.37$ for MJD~56393, and $\log_{10}l_e=-4.6$, $\log_{10}{\gamma_{\rm min}}=4.6$, $\log_{10}{\gamma_{\rm min}}=5.5$ and $B=0.2$\,G for MJD~56395. The rest of the parameters are unchanged compared to Table~\ref{table:steady}. 
\label{comp_doppler_steadystate}}
\end{figure}

The results are shown in Fig.~\ref{comp_doppler_steadystate}. The blue and red solid lines show the ``fast'' zone with $\delta_{\rm fast}=50$ and $\delta_{\rm fast}=100$, respectively, and the data are plotted with black markers. The model parameters for the $\delta_{\rm fast}=100$ case are listed in Table~\ref{table:steady}. For the $\delta_{\rm fast}=50$ case, $p$ and $R$ were set to the same values as in as in Table~\ref{table:steady}, but we modified $B$, $l_e$, $\gamma_{\rm min}$ and $\gamma_{\rm max}$ to adapt for the change in $\delta_{\rm fast}$. The corresponding parameter values are given in the caption of Fig.~\ref{comp_doppler_steadystate}. $\gamma_{\rm min}$ and $\gamma_{\rm max}$ in the $\delta_{\rm fast}=50$ case were increased as much as the X-ray data allow in order to describe the VHE hardness as close as possible. As one can see, the highest VHE points are underproduced even with those modifications derived with $\delta_{\rm fast}=50$. Furthermore, in the case of MJD~56395, the inverse-Compton component is narrower than that with $\delta_{\rm fast}=100$, hence also causing some tension at the lower VHE points. \par

\section{Spatial separation between the ``fast'' and ``slow'' zones}
\label{sec:zones_interaction}

As discussed in Sect.~\ref{sec:modelling}, our model neglects the interaction between the two emitting zones, i.e. only the synchrotron photons that are locally produced are considered for the inverse-Compton process and the electrons do not interact with the radiation field from the other zone. This simplification holds if the spatial separation between the zones is sufficiently large in which case only the local radiation field dominates. It is particularly relevant in our case since the relative bulk velocity of the plasma of the zones inevitably boosts the radiation density of each zone as seen in the frame of the other one.\par    

The synchrotron energy density from the zones in their respective comoving frame is (single-primed and double-primed quantities are in the rest frame of the ``slow zone'' and ``fast zone'', respectively):
\begin{equation}
    u'_{\rm syn, slow} \approx \frac{\nu_{\rm p, slow} L^{\rm obs}_{\rm syn, slow}(\nu_{\rm p, slow})}{4 \pi c R^2_{\rm slow} \delta^4_{\rm slow}}
\end{equation}
\begin{equation}
    u''_{\rm syn, fast} \approx \frac{\nu_{\rm p, fast} L^{\rm obs}_{\rm syn, fast} (\nu_{\rm p, fast})}{4 \pi c R^2_{\rm fast} \delta^4_{\rm fast}}
\end{equation}

where $L^{\rm obs}_{\rm syn}(\nu_{\rm p})$ the observed synchrotron luminosity at the peak frequency $\nu_{\rm p}$. Assuming that the plasma in the two components travels on the same axis (parallel to the jet), the relative bulk velocity and Lorentz factors are:
\begin{equation}
  \beta_{\rm rel} = \frac{\beta_{\rm slow}-\beta_{\rm fast}}{1 - \beta_{\rm fast}\beta_{\rm slow}} 
\end{equation}
\begin{equation}
  \Gamma_{\rm rel} = \Gamma_{\rm b, fast} \Gamma_{\rm b,slow} (1-\beta_{\rm fast}\beta_{\rm slow})
\end{equation} 
Our model uses $\delta_{\rm b, fast}=100$ and $\delta_{\rm slow}=30$, which implies $\Gamma_{\rm b, fast}=100$ and $\Gamma_{\rm b, slow}=15$ in case the viewing angle relative to the jet's axis is $\Theta= 1/\Gamma_{\rm fast}$. With these choice of parameters, $\beta_{\rm rel}=-0.95$  and $\Gamma_{\rm rel}=3.4$.

Following \citet{2014A&A...571A..83P}, the radiation of the ``slow zone'' as observed in the comoving frame of the ``fast zone'' is given by:
\begin{equation}
    u''_{\rm syn, slow} = \frac{u'_{\rm syn, slow}}{2} \Gamma^2_{\rm rel} \int^1_{\mu_{0}} d\mu' (1-\beta_{\rm rel} \mu')^2 
    = \frac{u'_{\rm syn, slow}}{2} \Gamma^2_{\rm rel} \left( 1- \mu_0 - \beta_{\rm rel} (1-\mu^2_0) + \frac{\beta^2_{\rm rel}}{3} (1-\mu^3_0) \right)
\end{equation}
where $\mu_0 = r_0 / \sqrt{r_0^2+R^2_{\rm slow}}$, with $r_0$ being the separation between the two regions.\par 
Requiring $u''_{\rm syn, slow}~<~0.1\, u''_{\rm syn, fast}$ ensures that the synchrotron photon field of the ``slow zone'' entering the ``fast zone'' is largely subdominant with respect to one being locally produced. This condition translates into a lower limit $r_0 > 1.5 \times 10^{17}$\,cm using the parameters of Table~\ref{table:steady}. Repeating the above exercise but for the ``slow zone'', i.e. finding the minimal $r_0$ for which $u'_{\rm syn, fast}~<~0.1\, u'_{\rm syn, slow}$, yields $r_0 > 0.9 \times 10^{15}$\,cm. We thus conclude that neglecting the interaction holds in a realistic configuration where the two regions are separated by a few $10^{17}$\,cm, equivalent to $\sim 10^{-1}$\,pc, orders of magnitude smaller than the jet's size.

\section{Determination of the time evolution curve of the power-law slope for the injected electrons in the ``fast'' zone}
\label{sec:p_evolution_curve_determination}

Using the X-ray observations, we determine the time evolution of the power-law slope $p$ of the electrons (injected inside the ``fast'' zone). The synchrotron nature of the X-ray photons allows in principle a direct prediction of the electron distribution slope based on the \textit{NuSTAR} best-fit spectral model. We find however that the \textit{NuSTAR} indices (in the $3-30$\,keV band) are significantly softer than the ones measured with \textit{Swift}-XRT in simultaneous time bins (in the $0.3-10$\,keV band). It indicates that \textit{NuSTAR} covers (at least partially) a high-energy cut-off in the electron spectrum (or is probing a radiative cooling break). Hence, the \textit{NuSTAR} photon indices alone prevent us to properly constrain the uncooled electron distribution (or well below the high-energy cut-off) injected in the ``fast'' zone. The photon indices determined closer to the \textit{Swift}-XRT band are required for the modeling. We stress that in our scenario the ``fast'' zone dominates not only in the \textit{NuSTAR} range, but also in the \textit{Swift}-XRT band. Unfortunately, the poor temporal sampling from \textit{Swift}-XRT throughout the flare compared to \textit{NuSTAR} does not allow a determination of the evolution of the X-ray photon index on $15$\,min timescales down to the $0.3-10$\,keV band.\par 

In order to address that issue, we first perform a series of spectral fits in each of the \textit{NuSTAR} $15$\,min time bins (following the same data reduction recipe described in Sect.~\ref{sec:analysis}) as well as in all available \textit{Swift}-XRT observations assuming a simple power-law model ($dN/dE \propto E^{-\Gamma_{X-ray}}$). In a second step, we correlate the \textit{NuSTAR} and \textit{Swift}-XRT slopes using strictly simultaneous measurements. This allows us to find an empirical, linear relationship between the indices of the two instruments. We used the obtained best-fit linear relation to convert all the \textit{NuSTAR} photon indices obtained in $15$\,min bins into a photon index ``light curve'' as it would be measured in the \textit{Swift}-XRT band. Finally, this synthetic photon index light curve ($\Gamma_{X-ray, synthetic}(t)$) is transformed into a time evolution curve of the electron power-law slope $p(t)$ using the well-known relationship for synchrotron emission \citep{1979rpa..book.....R}:
\begin{equation}
\label{eq:index_to_p}
\Gamma_{X-ray, synthetic}(t) = \frac{p(t)+1}{2},
\end{equation}
where $\Gamma_{X-ray}$ is the photon power-law index and $p$ the electron slope. We highlight that although a log-parabola model is significantly preferred in general to describe the \textit{NuSTAR} and \textit{Swift}-XRT spectra (see Sect.~\ref{sec:spectral_analysis}), our aim here is not to characterize with accuracy the X-ray data, but to obtain an estimate of the injected electron power-law slope using Eq.~\ref{eq:index_to_p}. Having a preliminary $p(t)$ evolution curve, we apply some final fine-tuning by hand (at the level of 10\% at most) to optimize the data/model match in the modeling process. Lastly, we apply some Gaussian smoothing on MJD~56393 \& MJD~56394 in order to remove a few statistical fluctuations in the measured photon index that led to artificial short timescale variations in $p(t)$ and thus also in the model fluxes. The other days did not require such smoothing, so we leave them untouched. The final $p(t)$ curve used for the model is shown with black markers in Fig.~\ref{p_le_vs_mjd}.\par

\section{Snapshot of the time-dependent model SEDs}
\label{snapshot_sed_modelling}

\begin{figure*}[htb!]
\centering
\gridline{\fig{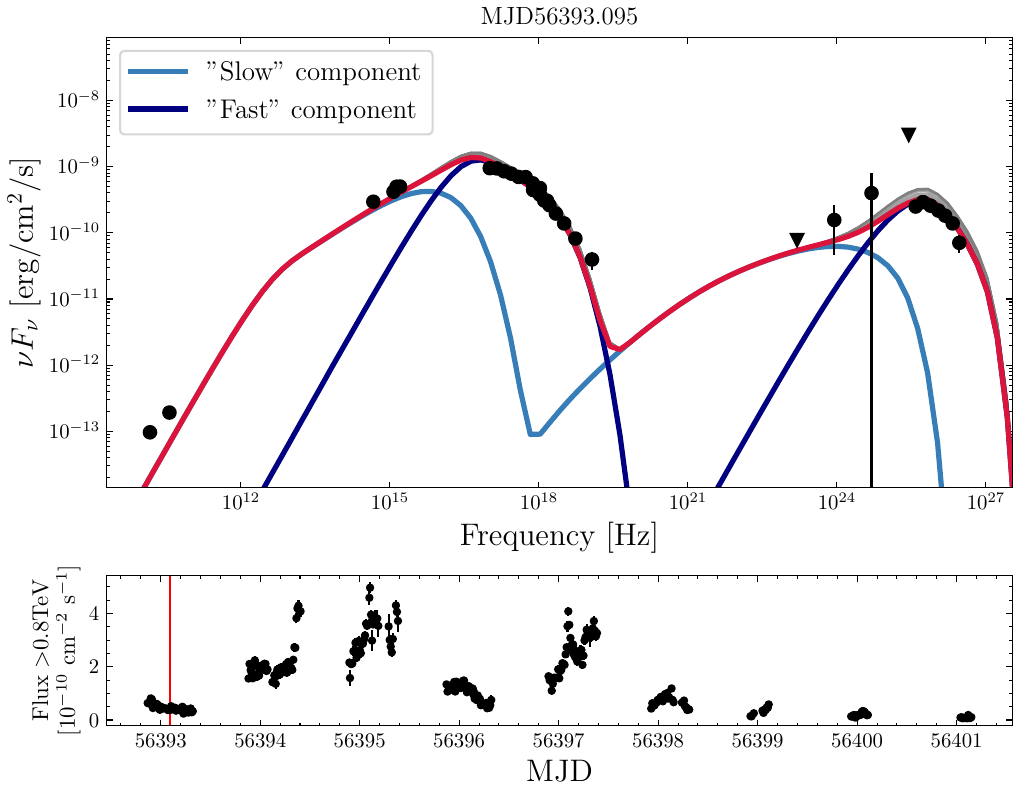}{0.48\textwidth}{(a) MJD~56393}
          \fig{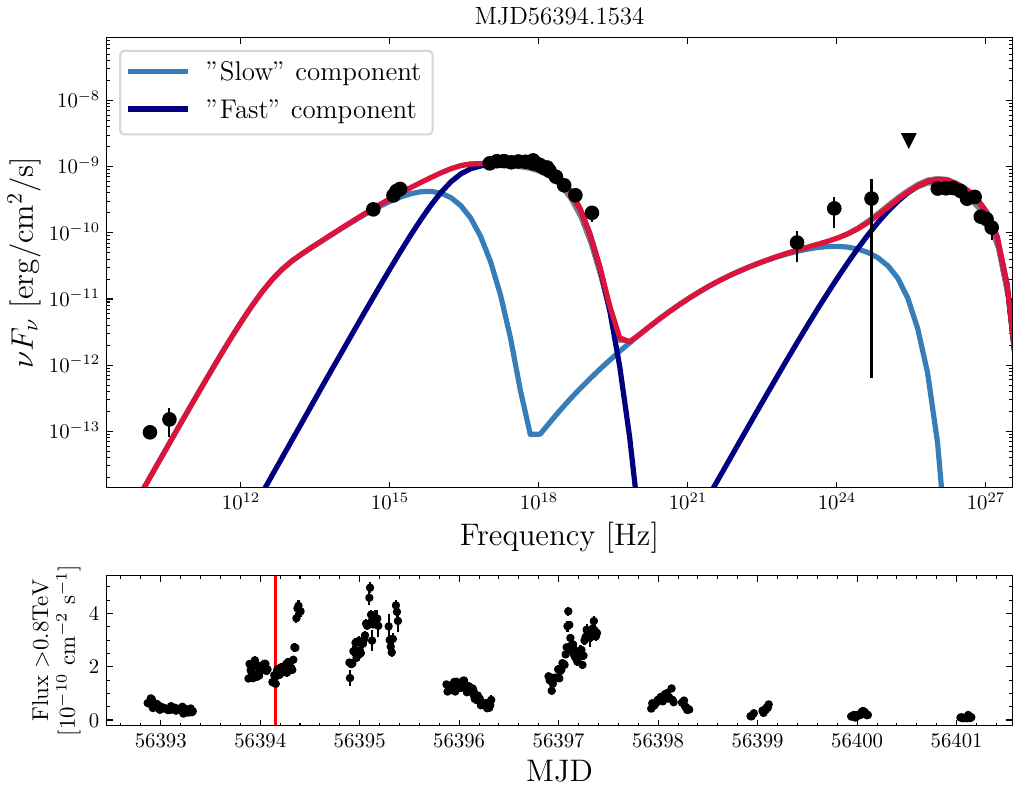}{0.48\textwidth}{(b) MJD~56394}}
\gridline{\fig{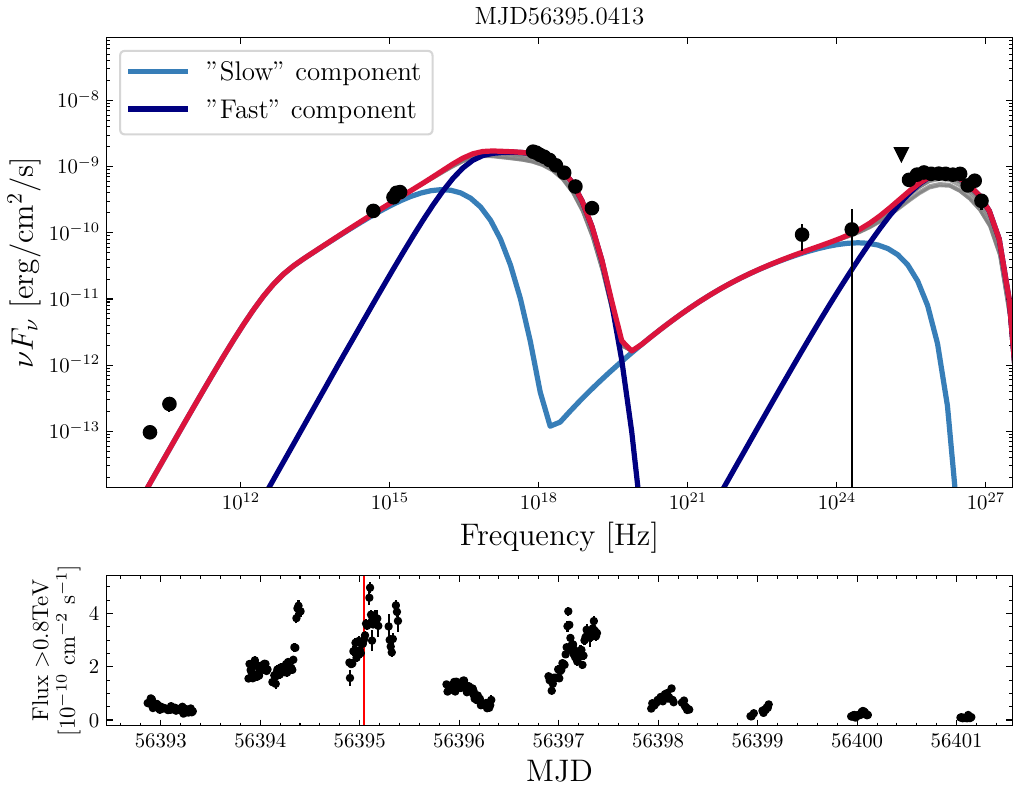}{0.48\textwidth}{(c) MJD~56395}
          \fig{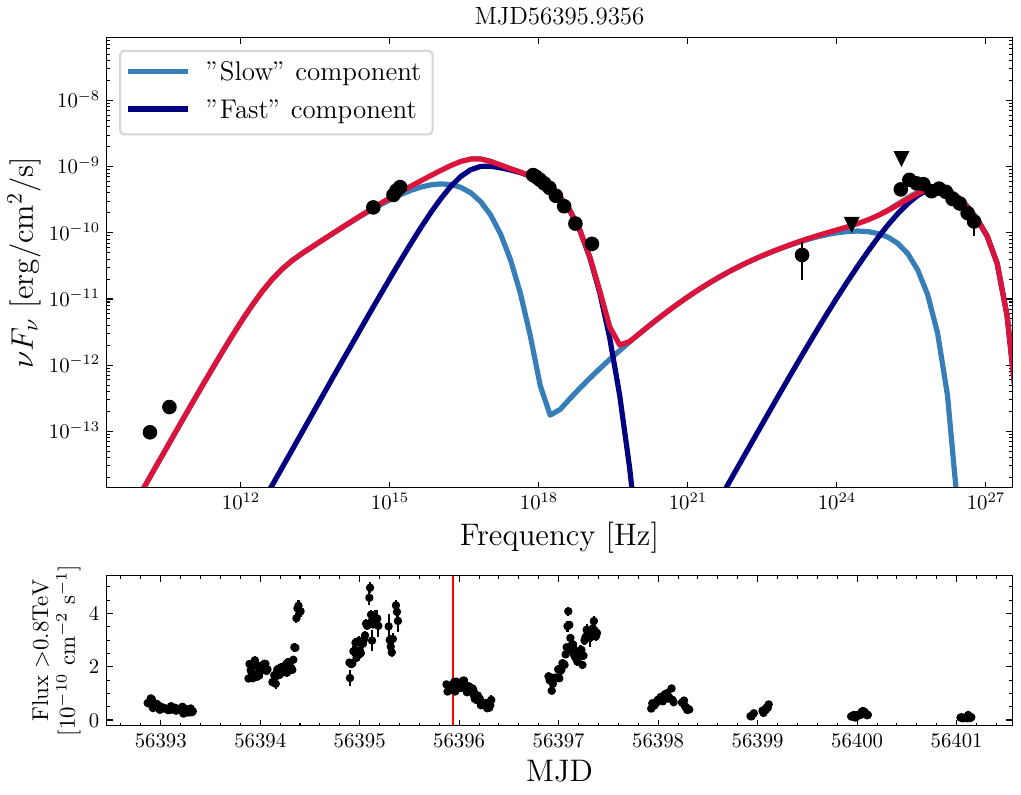}{0.48\textwidth}{(d) MJD~56396}}
\gridline{\fig{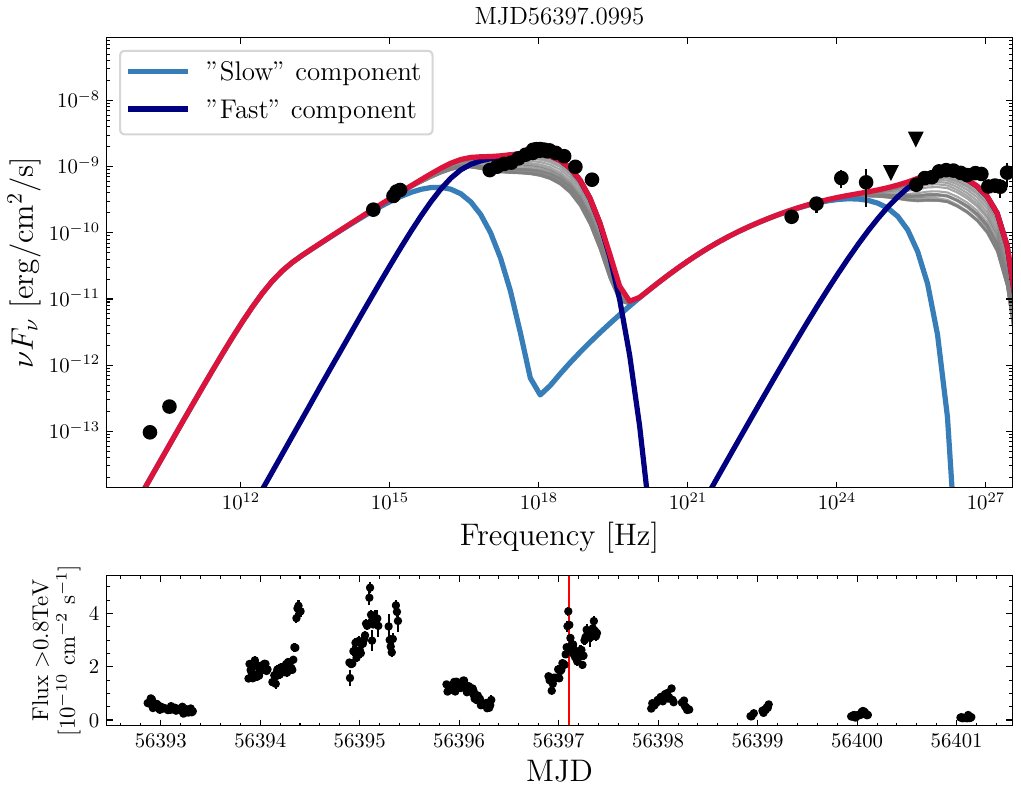}{0.48\textwidth}{(e) MJD~56397}
          \fig{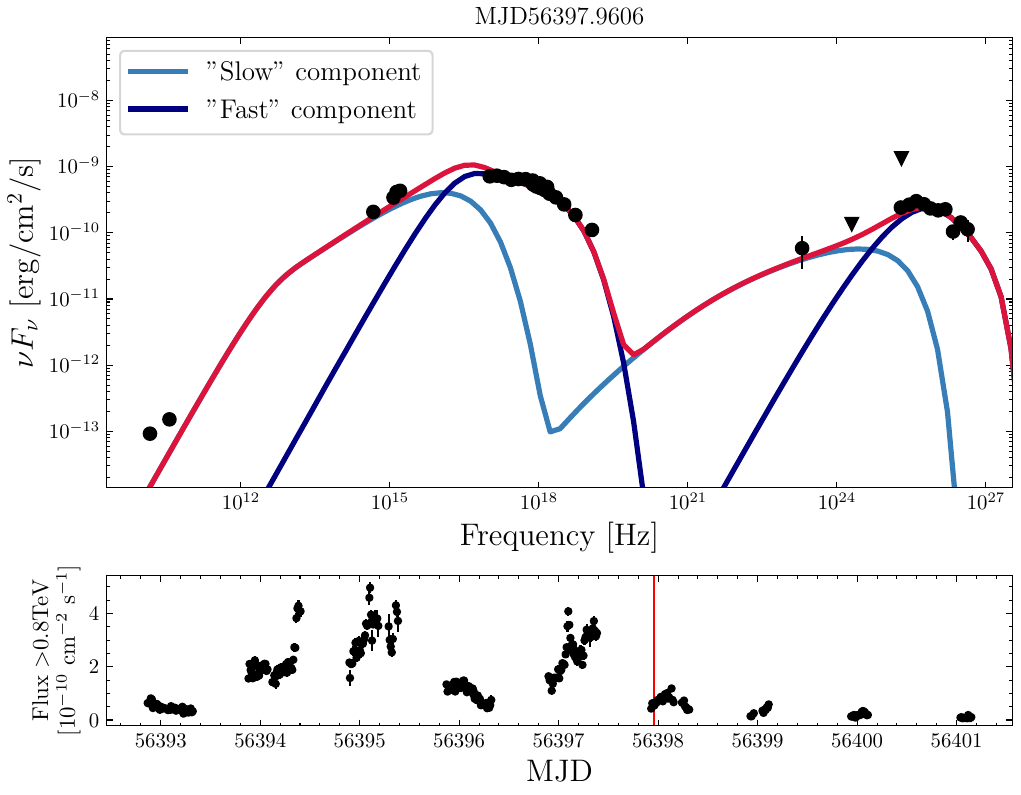}{0.48\textwidth}{(f) MJD~56398}}
\caption{Snapshots from the time-dependent model in one of the 15-minute intervals of each day. The light blue line represents the emission from the ``fast'' zone, the dark blue line is the emission from the ``slow'' zone, and the solid red line is the sum of the two components. Data are depicted with dark points. We show with grey lines the model curves up to 80 light-crossing time prior to the time of each snapshot to illustrate the variability. 
\label{snapshot_timedep_sed}}
\end{figure*}

\begin{figure*}[htb!]
\centering
\gridline{\fig{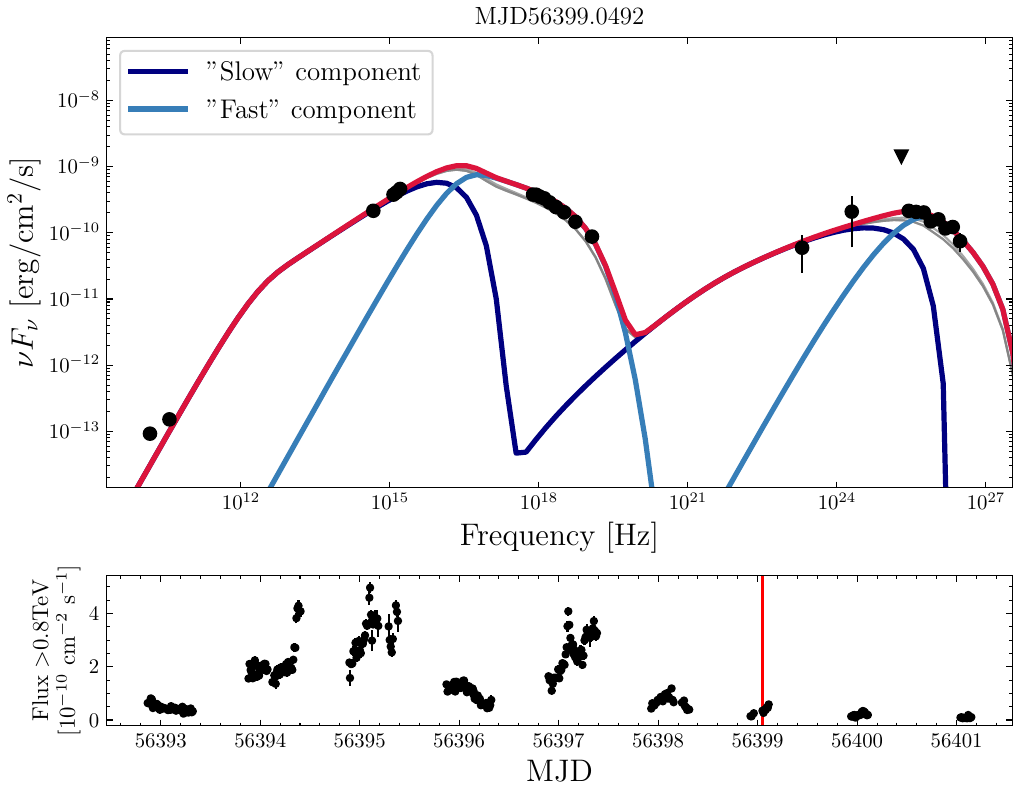}{0.48\textwidth}{(g) MJD~56399}
          \fig{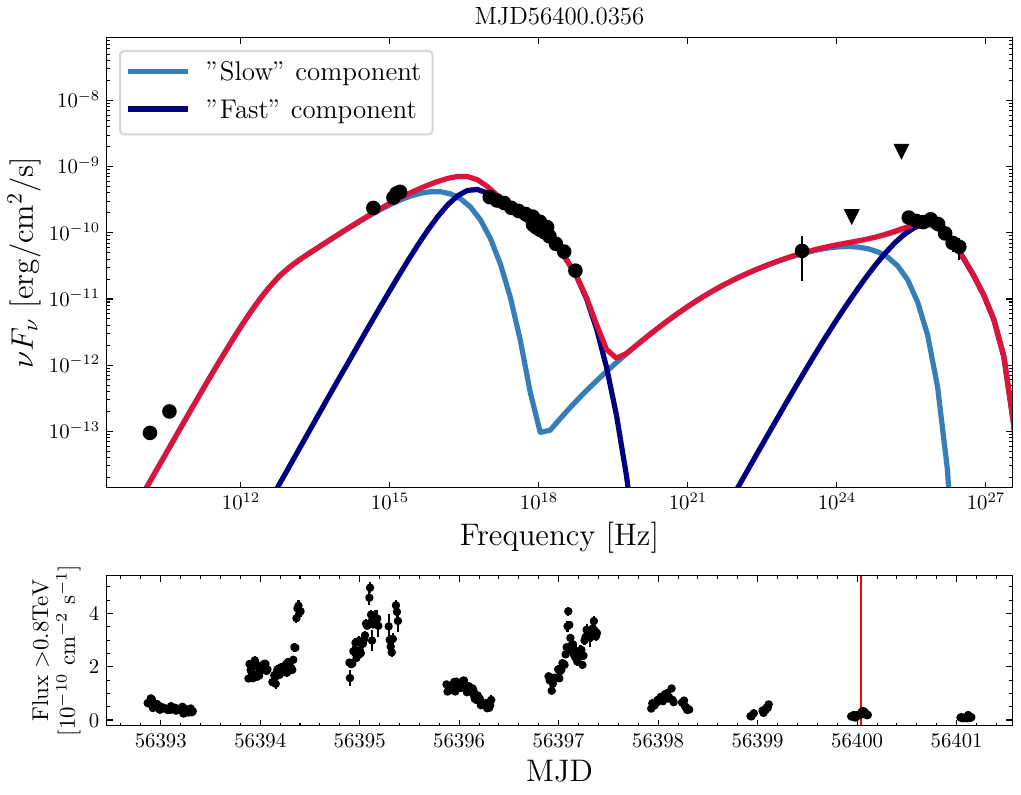}{0.48\textwidth}{(h) MJD~56400}}
\gridline{\fig{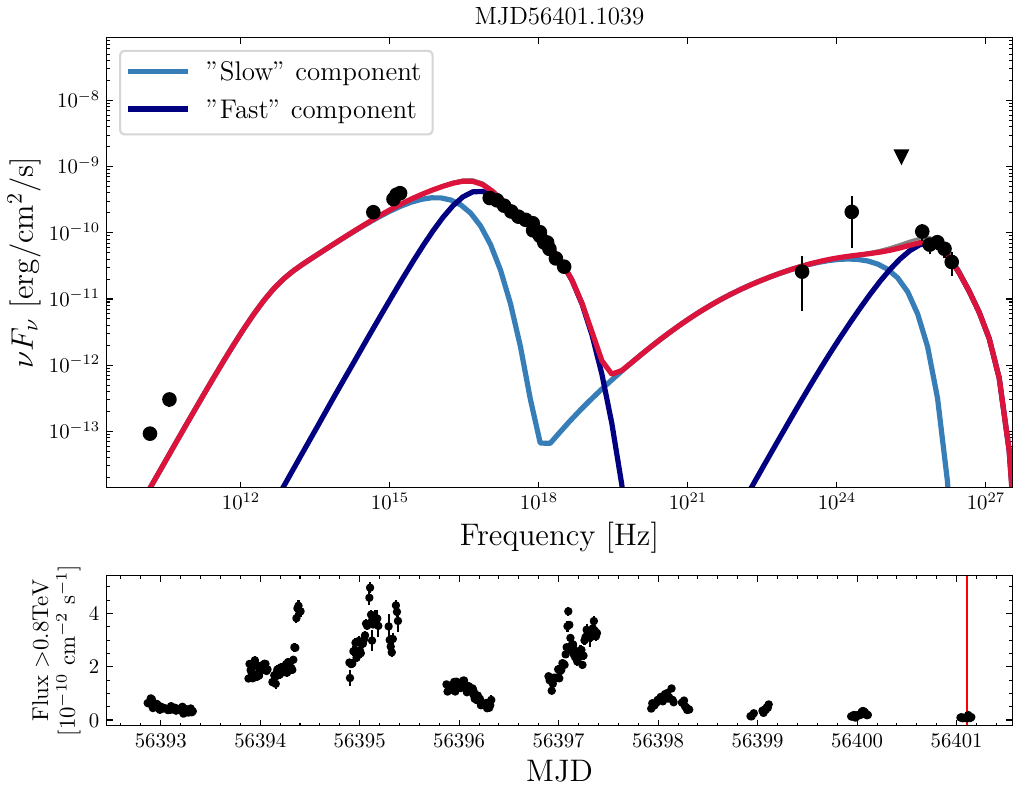}{0.48\textwidth}{(i) MJD~56401}}
\caption{Same as Fig.~\ref{snapshot_timedep_sed}, for the last three days of the 9-day outburst (MJD~56399-56401).}
\label{snapshot_timedep_sed_2}
\end{figure*}

As illustrative examples, we present the SED model in one of the 15-minute intervals of each day. They are shown in Fig.~\ref{snapshot_timedep_sed} and Fig.~\ref{snapshot_timedep_sed_2}. The complete set of SEDs is published as online material, and can be retrieved from the Zenodo repository \url{https://zenodo.org/records/17054582}.


\newpage


\bibliographystyle{aasjournal}

\bibliography{bibliography}



\end{document}